\documentclass[twocolumn]{aastex63}

\newcommand{\eg}{e.g.,\ }

\newcommand{\Msun}{$M_{\odot}$}

\newcommand{\kms}{km~s$^{-1}$}

\newcommand{\OI}{O~{\sc i}}

\newcommand{\CII}{C~{\sc ii}}
\newcommand{\CI}{C~{\sc i}}

\newcommand{\MgII}{Mg~{\sc ii}}

\newcommand{\SiII}{Si~{\sc ii}}
\newcommand{\SiIII}{Si~{\sc iii}}

\newcommand{\SII}{S~{\sc ii}}
\newcommand{\CaII}{Ca~{\sc ii}}
\newcommand{\TiII}{Ti~{\sc ii}}

\newcommand{\FeII}{Fe~{\sc ii}}
\newcommand{\FeIII}{Fe~{\sc iii}}
\newcommand{\CoII}{Co~{\sc ii}}
\newcommand{\CoIII}{Co~{\sc iii}}
\newcommand{\NiII}{Ni~{\sc ii}}

\newcommand{\Nifs}{$^{56}$Ni}

\newcommand{\sBV}{s$_{BV}$}

\newcommand{\mic}{$\mu$m}
\newcommand{\DmB}{\Delta{\rm m_{15}(B)}}

\newcommand{\ved}{$v_{edge}$}
\newcommand{\iBmax}{t$^{i-B}_{max}$}

\newcommand{\SCSN}{2003fg-like SNe}
\newcommand{\red}[1]{\textcolor{black}{#1}}

\usepackage{comment}

\received{\today}
\revised{}
\accepted{}

\submitjournal{ApJ}

\shorttitle{CSP 2003fg-like SNe}
\shortauthors{Ashall et al.}

\begin{document}

\title{\textit{Carnegie Supernova Project:} The First Homogeneous Sample of  ``Super-Chandrasekhar Mass"/2003fg-like Type Ia Supernova} 

\correspondingauthor{Chris Ashall}
\email{Chris.Ashall24@gmail.com}

\author[0000-0002-5221-7557]{C. Ashall}
\affil{Institute for Astronomy, University of Hawaii, 2680 Woodlawn Drive, Honolulu, HI 96822, USA}

\author[0000-0002-3900-1452]{J. Lu}
\affil{Department of Physics, Florida State University, 77 Chieftan Way, Tallahassee, FL 32306, USA}

\author[0000-0003-1039-2928]{E. Y. Hsiao}
\affil{Department of Physics, Florida State University, 77 Chieftan Way, Tallahassee, FL 32306, USA}

\author[0000-0002-4338-6586]{P. Hoeflich}
\affil{Department of Physics, Florida State University, 77 Chieftan Way, Tallahassee, FL 32306, USA}

\author[0000-0003-2734-0796]{M. M. Phillips}
\affil{Carnegie Observatories, Las Campanas Observatory, Casilla 601, La Serena, Chile}

\author[0000-0002-1296-6887]{L. Galbany}
\affil{Institute of Space Sciences (ICE, CSIC), Campus UAB, Carrer de Can Magrans, s/n, E-08193 Barcelona, Spain.}

\author[0000-0003-4625-6629]{C.~R.~Burns}
\affiliation{The Observatories of the Carnegie Institution for Science, 813 Santa Barbara Street, Pasadena, CA 91101, USA}

\author[0000-0001-6293-9062]{C.~Contreras}
\affil{Carnegie Observatories, Las Campanas Observatory, Casilla 601, La Serena, Chile}

\author[0000-0002-6650-694X]{K. Krisciunas}
\affiliation{George P. and Cynthia Woods Mitchell Institute for Fundamental Physics \& Astronomy, Texas A\&M University, Department of Physics and Astronomy, 4242 TAMU, College Station, TX 77843}

\author[0000-0003-2535-3091]{N.~Morrell}
\affil{Carnegie Observatories, Las Campanas Observatory, Casilla 601, La Serena, Chile}

\author[0000-0002-5571-1833]{M.~D.~Stritzinger}
\affiliation{Department of Physics and Astronomy, Aarhus University, Ny Munkegade 120, DK-8000 Aarhus C, Denmark}

\author[0000-0002-8102-181X]{N.~B.~Suntzeff}
\affiliation{George P. and Cynthia Woods Mitchell Institute for Fundamental Physics \& Astronomy, Texas A\&M University, Department of Physics and Astronomy, 4242 TAMU, College Station, TX 77843}
 
\author[0000-0002-2387-6801]{F. Taddia}
\affiliation{Department of Physics and Astronomy, Aarhus University, Ny Munkegade 120, DK-8000 Aarhus C, Denmark}
 
\author{J. Anais}
\affil{Carnegie Observatories, Las Campanas Observatory, Casilla 601, La Serena, Chile}

\author[0000-0001-5393-1608]{E. Baron}
\affiliation{Homer L. Dodge Department of Physics and Astronomy, University of Oklahoma, 440 W. Brooks, Rm 100, Norman, OK 73019-2061, USA}
\affiliation{Hamburger Sternwarte, Gojenbergsweg 112, D-21029 Hamburg, Germany}

\author[0000-0001-6272-5507]{P.J.~Brown}
\affiliation{George P. and Cynthia Woods Mitchell Institute for Fundamental
 Physics and Astronomy, Department of Physics and Astronomy, Texas
 A\&M University, College Station, TX 77843, USA}

\author{L. Busta}
\affil{Carnegie Observatories, Las Campanas Observatory, Casilla 601, La Serena, Chile}

\author{A. Campillay}
\affil{Departamento de F´ısica, Universidad de La Serena, Cisternas 1200, La Serena, Chile}

\author{S. Castell\'{o}n }
\affil{Carnegie Observatories, Las Campanas Observatory, Casilla 601, La Serena, Chile}


\author{C. Corco}
\affil{Carnegie Observatories, Las Campanas Observatory, Casilla 601, La Serena, Chile}
\affil{SOAR Telescope, Casilla 603, La Serena, Chile}

\author[0000-0002-2806-5821]{S. Davis}
\affil{Department of Physics and astronomy, University of California, 1 Shields Avenue, Davis, CA 95616-5270, USA}

\author{G. Folatelli}
\affil{Facultad de Ciencias Astron\'{o}micas y Geof\'{i}sicas, Universidad Nacional de La Plata, Instituto de
Astrof\'{i}sica de La Plata (IALP), CONICET, Paseo del Bosque S/N, B1900FWA La Plata,
Argentina}

\author[0000-0003-3459-2270]{F. F\"{o}rster}
\affil{Millennium Institute of Astrophysics, Casilla 36-D,7591245, Santiago, Chile}
\affil{Departamento de Astronom\'{i}a, Universidad de Chile, Casilla 36-D, Santiago, Chile}

\author[0000-0003-3431-9135]{W.~L.~Freedman}
\affil{Department of Astronomy and Astrophysics, University of Chicago, 5640 S. Ellis Avenue, Chicago, IL 60637, USA}

\author{C. Gonzal\'{e}z}
\affil{Carnegie Observatories, Las Campanas Observatory, Casilla 601, La Serena, Chile}

\author[0000-0001-7981-8320]{M.~Hamuy}
\affil{Universidad de Chile, Departamento de Astronom\'{i}a, Casilla 36-D, Santiago, Chile}

\author[0000-0002-3415-322X]{S. Holmbo}
\affiliation{Department of Physics and Astronomy, Aarhus University, Ny Munkegade 120, DK-8000 Aarhus C, Denmark}

\author[0000-0002-1966-3942]{R.~P.~Kirshner}
\affil{Gordon and Betty Moore Foundation, 1661 Page Mill Road, Palo Alto, CA 94304, USA}
\affil{Harvard-Smithsonian Center for Astrophysics, 60 Garden Street, Cambridge, MA 02138, USA}

\author[0000-0001-8367-7591]{S. Kumar}
\affiliation{Department of Physics, Florida State University, Tallahassee, FL 32306, USA}

\author{G.~H.~Marion}
\affil{University of Texas at Austin, 1 University Station C1400, Austin, TX 78712-0259, USA}

\author{P. Mazzali}
\affil{Astrophysics Research Institute, Liverpool John Moores University, IC2, Liverpool Science Park, 146  Brownlow Hill, Liverpool L3 5RF, UK}

\author[0000-0001-7449-4814]{T. Morokuma}
\affil{Institute of Astronomy, Graduate School of Science, The University of Tokyo, 2-21-1 Osawa, Mitaka, Tokyo 181-0015, Japan}

\author[0000-0002-3389-0586]{P. E. Nugent}
\affil{Lawrence Berkeley National Laboratory, Department of Physics, 1 Cyclotron Road, Berkeley, CA 94720, USA}
\affil{Astronomy Department, University of California at Berkeley, Berkeley, CA 94720, USA}

\author[0000-0003-0554-7083]{S.~E.~Persson}
\affil{The Observatories of the Carnegie Institution for Science, 813 Santa Barbara Street, Pasadena, CA 91101, USA}

\author[0000-0001-6806-0673]{A.~L.~Piro}
\affil{The Observatories of the Carnegie Institution for Science, 813 Santa Barbara Street, Pasadena, CA 91101, USA}

\author{M. Roth}
\affil{Carnegie Observatories, Las Campanas Observatory, Casilla 601, La Serena, Chile}
\affil{GMTO Corporation, Presidente Riesco 5335, Of. 501, Nueva Las Condes, Santiago}

\author{F. Salgado}
\affil{Carnegie Observatories, Las Campanas Observatory, Casilla 601, La Serena, Chile}

\author[0000-0003-4102-380X]{D.~J.~Sand}
\affil{Steward Observatory, University of Arizona, 933 North Cherry Avenue, Rm.~N204, Tucson, AZ 85721-0065, USA}

\author{J. Seron}
\affil{Carnegie Observatories, Las Campanas Observatory, Casilla 601, La Serena, Chile}
\affil{Cerro Tololo Inter-American Observatory, Casilla 603, La Serena, Chile}

\author[0000-0002-9301-5302]{M.~Shahbandeh}
\affil{Department of Physics, Florida State University, 77 Chieftan Way, Tallahassee, FL 32306, USA}

\author[0000-0003-4631-1149]{B.~J.~Shappee}
\affil{Institute for Astronomy, University of Hawaii, 2680 Woodlawn Drive, Honolulu, HI 96822, USA}

\begin{abstract}
We present a multi-wavelength photometric and spectroscopic analysis
of thirteen ``Super-Chandrasekhar Mass"/2003fg-like type Ia Supernova (SNe~Ia). Nine of these objects
were observed by the Carnegie Supernova Project. \SCSN\ have slowly
declining light curves ($\DmB<$1.3~mag), and peak absolute $B$-band magnitudes
between $-19<M_{B}<-21$~mag. Many \SCSN\ are located in the same part of the luminosity width relation as normal SNe~Ia. In the optical $B$ and $V$
bands, \SCSN\ look like normal SNe~Ia, but at redder wavelengths they diverge. Unlike other luminous SNe~Ia, \SCSN\  generally have only one $i$-band maximum which peaks after the epoch of $B$-band maximum, while their NIR light curve rise times can be $\gtrsim$40 days longer than those of normal SNe~Ia. They are also at least one magnitude brighter in the NIR bands than normal SNe~Ia, peaking above $M_H < -19$~mag, and generally have negative Hubble residuals,
\red{which may be the cause of some systematics in dark energy experiments. }
Spectroscopically, \SCSN\ exhibit peculiarities such as unburnt carbon well past maximum light, a large spread (8000--12000~\kms) in \SiII\ $\lambda$6355  velocities at maximum  light with no rapid early velocity decline, and no clear $H$-band break at +10~d, e. 
We find that SNe with a larger pseudo equivalent width  of \CII\ at maximum light have lower \SiII\ $\lambda$6355 velocities and slower declining light curves.
There are also multiple factors that contribute to the peak luminosity of \SCSN. The explosion of a C-O degenerate core inside a carbon-rich envelope is consistent with these observations.
Such a configuration may come from the core degenerate scenario. 
\end{abstract}


\section{Introduction} 
\label{sect:intro}

Type Ia supernovae (SNe~Ia) originate from the thermonuclear  explosion of at least one C-O White Dwarf (WD) in a binary system \citep[\eg][]{Whelan73,Livne90, Iben84,Hoeflich:Khokhlov:96}. These luminous events follow empirical observational relationships  which are fundamental  for  the use of SNe~Ia as extra galactic distance indicators \citep{Pskovskii84,Phillips93,Phillips99}. This led to the  discovery of  the accelerating expansion of the cosmos, or dark energy \citep[\eg][]{Riess98,Perlmutter99}. 

To date, there have been many sub-types of SNe~Ia discovered including: 1991bg-like SNe \citep[\eg][]{Filippenko:91bg:1992,Leibundgut93}, transitional SNe  \citep[\eg][]{Hsiao15,Gall18}, 2002cx-like SNe \citep[\eg][]{Li03,Foley13}, 2002es-like SNe \citep[\eg][]{Ganeshalingam12}, 1991T-like SNe \citep[\eg][]{Filippenko92,Phillips92} and  2003fg-like SNe \citep[\eg][]{Howell06,Hicken07}. 2003fg-like SNe, which are also known as  ``super-Chandrasekhar-mass" SNe~Ia, are among the most rare sub-type of SNe~Ia. It was previously thought that all \SCSN\  were over-luminous and require more \Nifs\ to power the light curve than could be produced in a detonation of a non-rotating Chandrasekhar-mass ($M_{Ch}$) C-O WD \citep{Howell06, Hicken07}. Hence, they were named ``super-Chandrasekhar mass''.
However, it has since become evident that not all \SCSN\ are over-luminous and their properties are more diverse \citep[\eg][]{Taubenberger19,Lu21}. 
Therefore, in this work, we follow the convention of naming the sub-type after the first SN discovered in the group, SN~2003fg \citep{Howell06}.   

There have only been a handful of \SCSN\ discovered and their observational traits are varied \citep{Howell06,Hicken07,Yamanaka09,Scalzo10,Yuan10,Silverman11,Taubenberger11,Chakradhari14,Taubenberger19,Hsiao20,Lu21}. 
They do, however, share a few characteristics: They generally have broad light curves, slow expansion velocity gradient before maximum light, and a very strong $\lambda$6580\ \CII\ absorption feature which lasts well past $B$-band maximum.  
They also peak in the $i$-band well after the time of $B$-band maximum \citep{Ashall20}. 
Furthermore,  \SCSN\ do not show a distinct $H$-band break at +10~d past $B$-band maximum which is seen in normal SNe~Ia \citep{Taubenberger11, Hsiao19, Lu21}. 
This $H$-band break is directly linked to the distribution and bulk of \Nifs\ in the explosion \citep{Wheeler98}. 
SN~2007if and SN~2009dc show low continuum polarizations which suggest spherical explosions \citep{Tanaka10, Cikota19}.
SN~2012dn and LSQ14fmg show evidence of dense circumstellar medium (CSM) \citep{Nagao17,Hsiao20}.
The majority of \SCSN\ occur in low metallicity,  low surface brightness galaxies with high specific star formation rates \citep{Childress11,Hsiao20,Lu21,Galbany21}.

There are several theoretical models that have been proposed for \SCSN. An early suggestion is the violent merger of two WDs that exceed the $M_{ch}$ \citep{Howell06,Scalzo10}. Alternatively, others have suggested these bright SNe experience  interaction with dense CSM \citep{Hachinger12,Noebauer16}.  This is also referred to as an envelope model \citep{Hoeflich:Khokhlov:96}.  Such an explosion may occur from the explosion of a degenerate core of an  Asymptotic giant branch (AGB) star in the core degenerate scenario \citep{Kashi11,Hsiao20, Lu21} or from the disruption of a C-O WD with surrounding circumstellar dust \citep{Nagao17,Nagao18}.  Finally,  the explosion of a C-O WD which exceeds the classical non-rotating  $M_{ch}$ limit due to rapid rotation or high magnetic fields may also be a viable model \citep{Yoon05,Das13}. 
The current dataset of \SCSN\ is sparse, and it has not been possible to disentangle the effects predicted by these scenarios.

The \textit{Carnegie Supernova project}-I \& II (CSP-I \& II)  ran two
observing campaigns between 2004 and 2015, during which we obtained
optical and NIR spectra and photometry of over 300 SNe~Ia
\citep{Krisciunas17,Phillips19,Hsiao19}. Nine \SCSN\ were followed in
these  two campaigns. 
In this work, we combine this data set  
with data from the literature to produce and analyze the first statistical and homogeneous sample \SCSN.
In Section \ref{sect:sample} we present the observational sample, followed by 
 the data reduction in Section \ref{sect:reduc}. Host galaxy extinction is discussed in Section \ref{sect:Extinction}. Section \ref{sect:phot} presents the photometric observations, followed by the spectroscopic observations in Section \ref{sect:spec}. Important correlations and their implications are discussed in Section \ref{sect:correlations}.  Finally the discussion of possible explosion models is given in Section \ref{sect:discussion}, followed by the conclusions in Section \ref{sect:conclusion}.

\section{Sample Characteristics}
\label{sect:sample}

The vast majority of the \SCSN\ followed by CSP were observed during CSP-II (7 out of 9). 
This reflects one of the main differences between the CSP-I and CSP-II campaigns: 
while nearly 90\% of the SNe~Ia followed up by CSP-I came from targeted searches, 83\% of those followed up by CSP-II came from untargeted searches. 
As \SCSN\ preferentially explode in low-luminosity hosts, untargeted surveys have the advantage in detecting them. 
While we strove for a complete and unbiased sample in CSP-II, the strategy also contributes to our success at following up on a statistically significant and uniform sample of \SCSN, the only sample of its kind.
Furthermore, the sample optical light curves are obtained with nearly nightly cadence and are placed on a single well-understood photometric system of the Swope Telescope.
All of the \SCSN\ observed by CSP-II came from untargeted searches.
Only 3 (SNe~2006gz, 2009dc, and 2012dn) out of the 13 \SCSN\ presented here were discovered by targeted searches. The discovery information on the objects not previously published can be see in Appendix \ref{sec:Discoveryinfo}.

\SCSN\ are generally characterized by:
\begin{itemize}
    \item[$\bigstar$]  A primary $i$-band maximum after that of the $B$-band maximum
    \item[$\bigstar$] A lack of $H$-band break at +10\,d in the NIR spectra
     \item[$\bigstar$] A low ionization state in nebular phase spectra
    \item  A broad light curve
    \item A strong  \CII\ \red{$\lambda \lambda$6578,6582} feature past maximum light
    \item A lack of or weak $i$-band secondary maximum
    \item Low ejecta velocity gradients before maximum light
     \item A lack of \TiII\ in the maximum light spectra
\end{itemize}
However, not all of these features are observed in every 2003fg-like SN, and only the timing of the $i$-band primary, the lack of an $H$-band break, and the low ionization nebular phase appear to be ubiquitous \citep[see \eg][]{
Howell06,Silverman11,Taubenberger11,Taubenberger19,Chen19,Hsiao19,Ashall20,Lu21}.

In this work the \SCSN\ were identified in the CSP-I and CSP-II through photometric criteria. As mentioned previously, \SCSN\ have distinct photometric properties from the normal population as well as other peculiar sub-types \citep{Gonzalez14}. 
Here we adopt the method of \citet{Ashall20} to identify \SCSN: 
For an object to be chosen for the sample, it must have its $i$-band primary maximum occurring after that in $B$ band. 
Furthermore, the object must have slowly declining light curves as indicated by $s_{BV}\gtrsim0.8$ or $\Delta \rm{m}_{15}(B)\lesssim1.3$\,mag 
\footnote{See section \ref{sect:LWR} for the definition of $s_{BV}$ and $\Delta \rm{m}_{15}(B)$.}.

These photometric selection criteria are then used in conjunction with the examination of its optical spectrum near maximum to look for the identifying properties of \SCSN\ described above. 
The spectroscopic criteria eliminate the peculiar SN~2006bt, as it contains \TiII\ features. 
Note that using spectra alone can lead to misleading results, such as those of SN 2011hr \citep{Zhang16} and LSQ12gdj \citep{Scalzo14}.
Note that CSS140126 is a border-line case between a 1991T-like and a \SCSN, it has only one low S/N optical spectrum which is featureless, a primary $i$-band maximum which peaks after that of $B$ band, but displays a secondary $i$-band maximum. 
We chose to keep it in the sample, however  due to poor spectral temporal coverage it is difficult to ascertain if it is a true 2003fg-like SN. Nine SNe in the CSP samples and a further four in the literature were found to meet these criteria. 

\begin{figure*}[htb]
\centering
 \includegraphics[width=.95\textwidth]{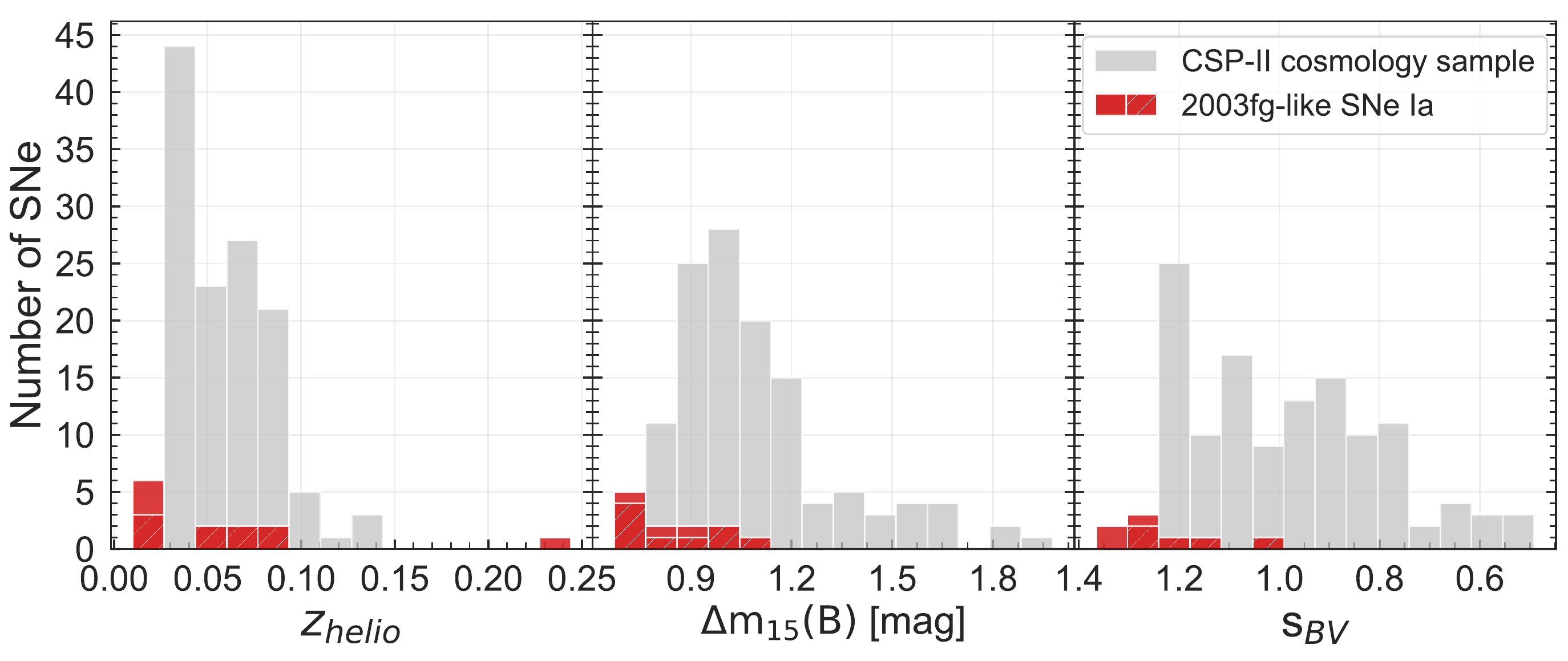}
\caption{Histograms of $z_{helio}$ (left panel), $\DmB$ (middle panel) and $s_{BV}$ (right panel) of the \SCSN\ sample compared to the CSP-II SNe~Ia cosmology samples from \citet{Phillips19}. Nine \SCSN\ followed up by CSP (including CSP-I  and CSP-II) are marked with shaded red bars, and four not followed by CSP are stacked on top with non-stripped red bars. The SNe~Ia from  \citet{Phillips19} are the gray shaded bars. }
\label{fig:hist}
\end{figure*}

Table~\ref{table:prop} contains the basic information of all of the SNe used in this analysis, and Table~\ref{table:propphot} summarizes their photometric properties.
For ten objects $z_{helio}$ was determined using integral field spectroscopy (IFS) data which will be presented in \citep{Galbany21}. For the other objects $z_{helio}$  was determined from a spectrum of the host galaxy, or, in the case of previously published \SCSN\ objects, it was taken from the literature. 
In Fig.~\ref{fig:hist}, the sample characteristics are compared to that of the ``Cosmology'' SN~Ia sample of CSP-II \citep{Phillips19}, as 96\% of the SNe were discovered by untargeted searches.
The majority of the objects from the sample are in the Hubble flow, similar to the CSP-II Cosmology sample.
As our selection criteria dictate, the \SCSN\ are slow decliners as indicated by $s_{BV}$ and $\Delta \rm{m}_{15}(B)$.
However, it should be noted that there is a wide range of light-curve properties that overlap with those of the normal populations.

\begin{deluxetable*}{ c c c c l c c}
\tablewidth{\textwidth}
\tablecaption{The properties of the SNe in the sample.  \label{table:prop}}
\tablehead{
\colhead{SN}&
\colhead{$z_{helio}$}&
\colhead{RA}&
\colhead{DEC}&
\colhead{$\mu$\tablenotemark{a}}&
\colhead{$E(B-V)_{MW}$}&
\colhead{Discoverer}
\\
\colhead{}&
\colhead{}&
\colhead{}&
\colhead{}&
\colhead{Mag}&
\colhead{Mag}&
\colhead{}}
\startdata
2003fg	                &	0.2440	&	14:16:18.8	&	+52:14:53.66	&	40.37	$\pm$	0.01	                &	0.011 & \citet{Howell06}		\\
2006gz	                &	0.0237	&	18:10:26.3	&	+30:59:44.40	&	34.95	$\pm$	0.09	                &	0.055 & \citet{Puckett06gz}		\\
2012dn	                &	0.0100	&	20:23:36.3	&	$-$28:16:43.40	&	33.28	$\pm$	0.21\tablenotemark{b}	&	0.052	&   \citet{12dndisc} \\	
ASASSN-15pz	                &	0.0148	&	03:08:48.4	&	+35:13:50.89	&	33.89	$\pm$	0.14                	&	0.015 & \citet{15pzdisc}		\\
2007if\tablenotemark{c}	&	0.0742	&	01:10:51.4	&	+15:27:39.90	&	37.51	$\pm$	0.03                	&	0.072	&\citet{Akerlof07if}	\\
2009dc\tablenotemark{c}	&	0.0214	&	15:51:12.1	&	+25:42:28.50	&	34.79	$\pm$	0.09                	&	0.060	&\citet{09dcdisc}	\\
LSQ12gpw\tablenotemark{c}	&	0.0506	&	03:12:58.2	&	$-$11:42:40.13	&	36.65	$\pm$	0.04	                &	0.062&	\citet{LSQ}	\\
2013ao\tablenotemark{c}	&	0.0435	&	11:44:44.7	&	$-$20:31:41.10	&	36.39	$\pm$	0.04                	&	0.034	&\citet{13aodisc}	\\
CSS140126\tablenotemark{c~d}	&	0.0772	&	12:03:06.9	&	$-$01:01:31.70	&	37.67	$\pm$	0.03	                &	0.021 &\citet{Drake09}	    \\	
CSS140501\tablenotemark{c~e}	&	0.0797	&	17:04:13.7	&	+17:48:39.40	&	37.74	$\pm$	0.03	                &	0.066	&\citet{Drake09}	\\
LSQ14fmg\tablenotemark{c}	&	0.0649	&	22:16:46.1	&	+15:21:14.13	&	37.24	$\pm$	0.03	                &	0.046	&\citet{LSQ}	\\
2015M\tablenotemark{c}	&	0.0231	&	13:00:32.3	&	+27:58:41.00	&	35.04	$\pm$	0.08	                &	0.009	&\citet{kiss15ndisc}	\\
ASASSN-15hy\tablenotemark{c}&	0.0185	&	20:10:02.4	&	$-$00:44:21.31	&	34.33	$\pm$	0.11	                &	0.130	& \citet{15hydisc}	\\
\enddata
\tablenotetext{a}{Corrected to the CMB rest frame and calculated using $H_{0}$=73\,km\,s$^{-1}$\,Mpc$^{-1}$, $\Omega_{m}$=0.27 and $\Omega_{\Lambda}$=0.73, which is used throughout this work.}
\tablenotetext{b}{Corrected for the infall towards the Virgo cluster and the Great Attractor (recession velocity =3300\kms; \citealt{Mould00}).}
\tablenotetext{c}{SN observed by CSP.}
\tablenotetext{d}{For convenience we shorten CSS140126:120307-010132 to CSS140126.}
\tablenotetext{e}{For convenience we shorten CSS140501-170414+174839 to CSS140501.}

\end{deluxetable*}

\begin{deluxetable*}{ c c c c c c c }
\tablewidth{\textwidth}
\tablecaption{The basic light curve parameters of the \SCSN. 
All parameters were obtained from direct measurements to the Gaussian process interpolations to the data. It should be noted that the \iBmax\ here has not been K-corrected. \red{These values have been corrected for foreground but not host galaxy extinction.}  \label{table:propphot}}
\tablehead{
\colhead{SN}&
\colhead{$T(B)_{max}$}&
\colhead{$B_{max}$}&
\colhead{$\DmB$}&
\colhead{\sBV}&
\colhead{$(B-V)_{Bmax}$}&
\colhead{\iBmax}
\\
\colhead{}&
\colhead{Days}&
\colhead{Mag}&
\colhead{Mag}&
\colhead{}&
\colhead{Mag}&
\colhead{Days}}
\startdata
2003fg	&	2452760.18	$\pm$	0.87	&	$\cdots$			    &	$\cdots$			    &	$\cdots$			    &	$\cdots$			    &	$\cdots$		        \\
2006gz	&	2454021.84	$\pm$	0.10	&	15.86	$\pm$	0.06	&	0.84	$\pm$	0.02	&	1.37	$\pm$	0.03	&	0.03	$\pm$	0.02	&	4.60	$\pm$	1.65	\\
2012dn	&	2456132.63	$\pm$	0.98	&	14.16	$\pm$	0.03	&	0.87	$\pm$	0.03	&	1.25	$\pm$	0.15	&	0.02	$\pm$	0.03	&	2.04	$\pm$	0.61	\\
ASASSN-15pz	&	2457307.47	$\pm$	0.93	&	14.18	$\pm$	0.03	&	0.67	$\pm$	0.09	&	1.37	$\pm$	0.09	&	$-$0.02	$\pm$	0.02	&	0.04	$\pm$	0.81	\\
2007if	&	2454349.97	$\pm$	1.09	&	17.55	$\pm$	0.02	&	0.88	$\pm$	0.09	&	1.26	$\pm$	0.15	&	0.01	$\pm$	0.03	&	3.03	$\pm$	3.24	\\
2009dc	&	2454947.07	$\pm$	0.60    &	15.09	$\pm$	0.02	&	0.70	$\pm$	0.05	&	1.29	$\pm$	0.07	&	$-$0.03	$\pm$	0.01	&	2.52	$\pm$	0.34	\\
LSQ12gpw	&	2456269.75	$\pm$	0.46	&	17.35	$\pm$	0.01	&	0.70	$\pm$	0.03	&	$\cdots$	            &	0.01	$\pm$	0.01	&	$\cdots$            	\\
2013ao	&	2456362.53	$\pm$	0.26	&	16.87	$\pm$	0.01	&	0.99	$\pm$	0.03	&	1.02	$\pm$	0.13	&	0.10	$\pm$	0.01	&	1.80	$\pm$	0.26	\\
CSS140126	&	2456668.48	$\pm$	0.41	&	18.23	$\pm$	0.01	&	0.73	$\pm$	0.05	&	$\cdots$	            &	$-$0.06	$\pm$	0.01	&	3.97	$\pm$	0.73	\\
CSS140501	&	2456787.70	$\pm$	1.60	&	18.09	$\pm$	0.04	&	1.05	$\pm$	0.18	&	$\cdots$	            &	0.03	$\pm$	0.02	&	2.58	$\pm$	2.85	\\
LSQ14fmg	&	2456939.24	$\pm$	0.72	&	17.35	$\pm$	0.01	&	1.04	$\pm$	0.09	&	1.20	$\pm$	0.08	&	0.09	$\pm$	0.01	&	1.51	$\pm$	1.05	\\
2015M	    &	2457169.00	$\pm$	0.82	&	15.54	$\pm$	0.03	&	0.82	$\pm$	0.04	&	$\cdots$	            &	0.14	$\pm$	0.01	&	1.85	$\pm$	2.22	\\
ASASSN-15hy	&	2457151.63	$\pm$	0.40	&	15.19	$\pm$	0.01	&	0.73	$\pm$	0.03	&	1.24	$\pm$	0.18	&	0.19	$\pm$	0.01	&	7.28	$\pm$	0.47	\\
\enddata
\end{deluxetable*}

\section{Data reduction}
\label{sect:reduc}

\subsection{Photometry}

\begin{figure*}
	\setlength\arraycolsep{0pt}
	\renewcommand{\arraystretch}{0}
	\centering$
  \begin{array}{ccc}
    \includegraphics[width=.35\linewidth]{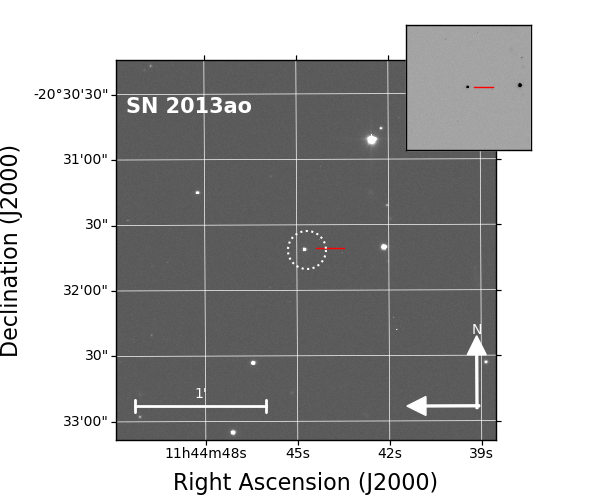} &
    \includegraphics[width=.35\linewidth]{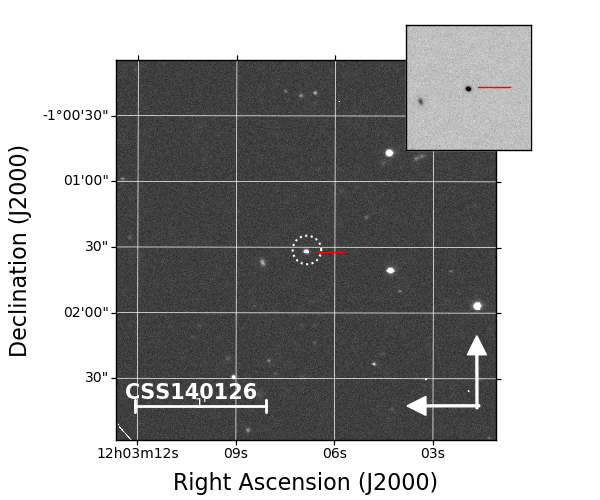}&
    \includegraphics[width=.35\linewidth]{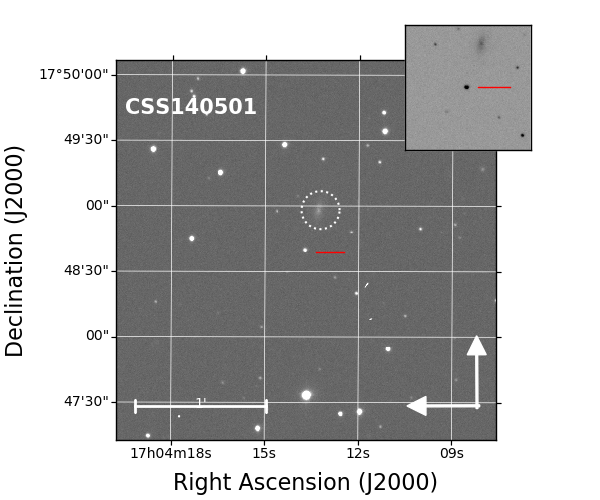}\\
    \includegraphics[width=.35\linewidth]{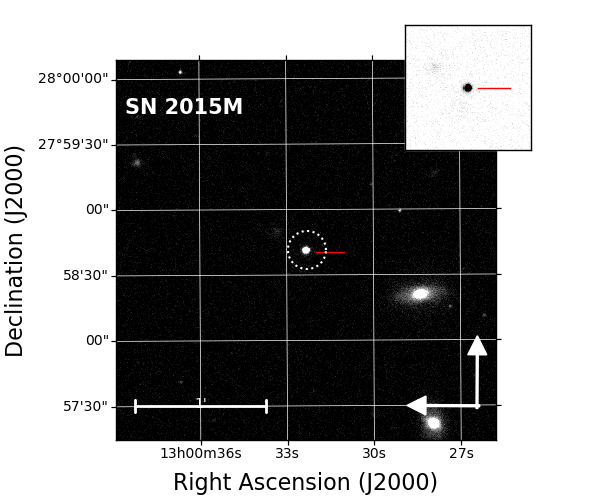}&
    \includegraphics[width=.35\linewidth]{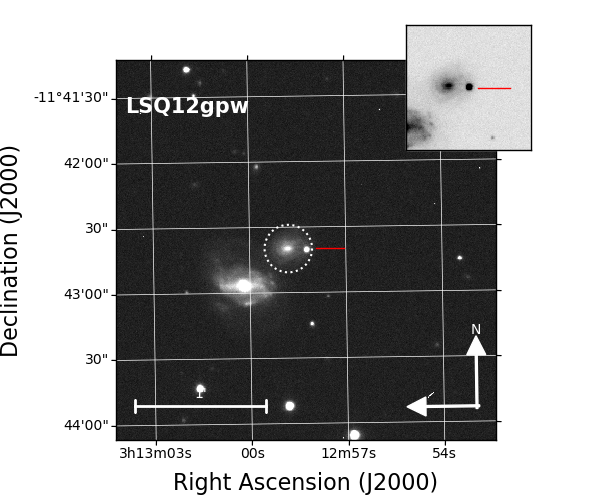}
  \end{array}$
   \caption{The $r$-band finding charts of the five  \SCSN\ observed by CSP-II. The plots are produced with  Swope eV2 images.  An inset of the exact SN location is provided in the top right section of each panel. }
   \label{fig:finders}
\end{figure*}

Optical $uBVgri$-band   imaging was obtained for nine
\SCSN\ using SITe-3 and e2v on the 1-m Swope telescope at Las Campanas
Observatory (LCO). 
For a sub-sample of these, NIR $YJH$-band  imaging was also 
 acquired using  NIR imager called RetroCam which was installed on the Swope telescope during CSP-I and the 2.5-m du~Pont telescope during CSP-II.  All of the photometry presented here is on the well-understood CSP natural system. This allows for systematic differences between SNe to be examined. The reduction and calibration procedures are described in \citet{Krisciunas17} and \citet{Phillips19}, and the final light curves can be found online\footnote{\href{https://csp.obs.carnegiescience.edu/data}{CSP data products.}}. The light curves of four of the nine SNe have been previous published by the CSP: SN\,2007if and SN\,2009dc \citep{Krisciunas17}, LSQ14fmg \citep{Hsiao20}, and ASASSN-15hy \citep{Lu21}. Finder charts of the remaining five \SCSN\ are presented in Fig.~\ref{fig:finders}, and the light curves of all the SNe are presented in the natural system in Fig.~\ref{fig:lightcurves}. These light curves are tabulated in Appendix~\ref{sec:CSP_phot_spec}.

Data from five \SCSN\ that have been published by other groups (SN\,2003fg; \citealt{Howell06}, SN\,2006gz; \citealt{Hicken07} SN\,2009dc; \citealt{Yamanaka09,Tanaka10,Silverman11,Taubenberger11}, SN\,2012dn; \citealt{Chakradhari14,Parrent16,Taubenberger19}, and ASASSN-15pz;  \citealt{Chen19})  are also included in the sample. In the $BVgri$ bands, where possible, S-corrections were applied to transform the photometry to the CSP natural system. \red{Swift  Ultraviolet/Optical Telescope (UVOT) data of SN~2015M was also obtained from the Supernova Archive (SOUSA; \citealt{Brown14a}) via the Swift Supernova website.\footnote{https://pbrown801.github.io/SOUSA/ }}

K-corrections were computed for $BVgri$ light curves using the same method presented in Appendix~B of \citet{Lu21}. In short, the spectral series of SN\,2009dc, SN\,2012dn, ASASSN-15hy, and the Hsiao template were all used independently as the SED to compute the corrections. The SEDs were mangled to match the interpolated observed photometric colors. A comparison was made between the K-correction values between the four SED template spectral series and those computed with the actual observed spectra of the SNe. The template with the smallest average residual K-correction compared to the observed spectra was then selected for each SN. 
This was done because most of the SNe did not have adequate spectral coverage to compute K-corrections directly from the observed spectra. The K-correction process was carried out individually for each SN. As discussed in \citet{Lu21}, the average K-correction uncertainty obtained with this method is smaller than 0.01\,mag, which is consistent with the values obtained for normal SNe~Ia from \citet{Hsiao07}. Due to the lack of spectral data in the UV and NIR, and  hence the ability to understand the SED, no S or K-corrections were applied in these regions.


\subsection{Spectra}

Optical and NIR spectra were obtained of nine \SCSN\ by the CSP I \& II. Twenty-four optical and six NIR spectra are presented here for the first time, which are logged in Appendix~\ref{sec:CSP_phot_spec}.

The majority of these optical spectra were acquired at LCO using B\&C on the 2.5-m du~Pont telescope and LDSS3 and IMACS on the 6.5-m Magellan Baade and Clay Telescopes. Additional spectra were obtained with the ALFOSC on the Nordic Optical Telescope (NOT) at La Palma, EFOSC2 on the New Technology Telescope (NTT) at La Silla, and RSS on the Southern African Large Telescope (SALT) at the South African Astronomical Observatory. The spectra were reduced using the standard \textsc{iraf}\footnote{The Image Reduction and Analysis Facility (\textsc{iraf}) is distributed by the National Optical Astronomy Observatory, which is operated by the Association of Universities for Research in Astronomy, Inc., under cooperative agreement with the National Science Foundation.} packages using the method as described in \citet{Hamuy06} and \citet{Folatelli13}. 

All the NIR spectra were observed with the Folded-port InfraRed Echellette \citep[FIRE;][]{Simcoe13}. The details in observing set up and data reduction are outlined by \citet{Hsiao19}. Along with six NIR spectra of ASASSN-15hy \citep{Lu21}, we have tripled the size of the sample of \SCSN\ NIR spectra, as the previous sample includes only spectra of SN~2009dc \citep{Taubenberger11,Taubenberger13a}.

\begin{figure*}
	\setlength\arraycolsep{0pt}
	\renewcommand{\arraystretch}{0}
	\centering$
  \begin{array}{cc}
    \includegraphics[width=.5\linewidth]{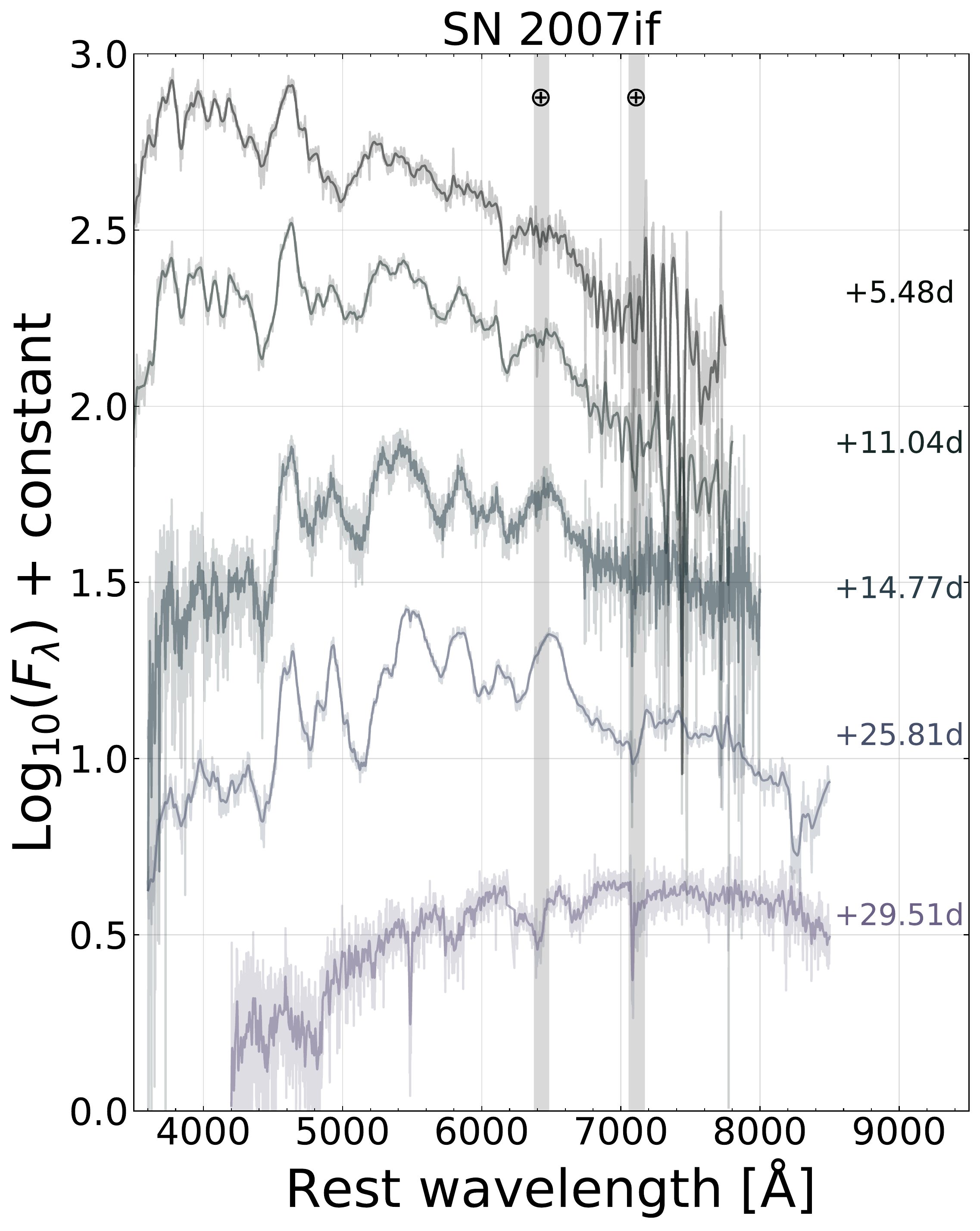} &
    \includegraphics[width=.5\linewidth]{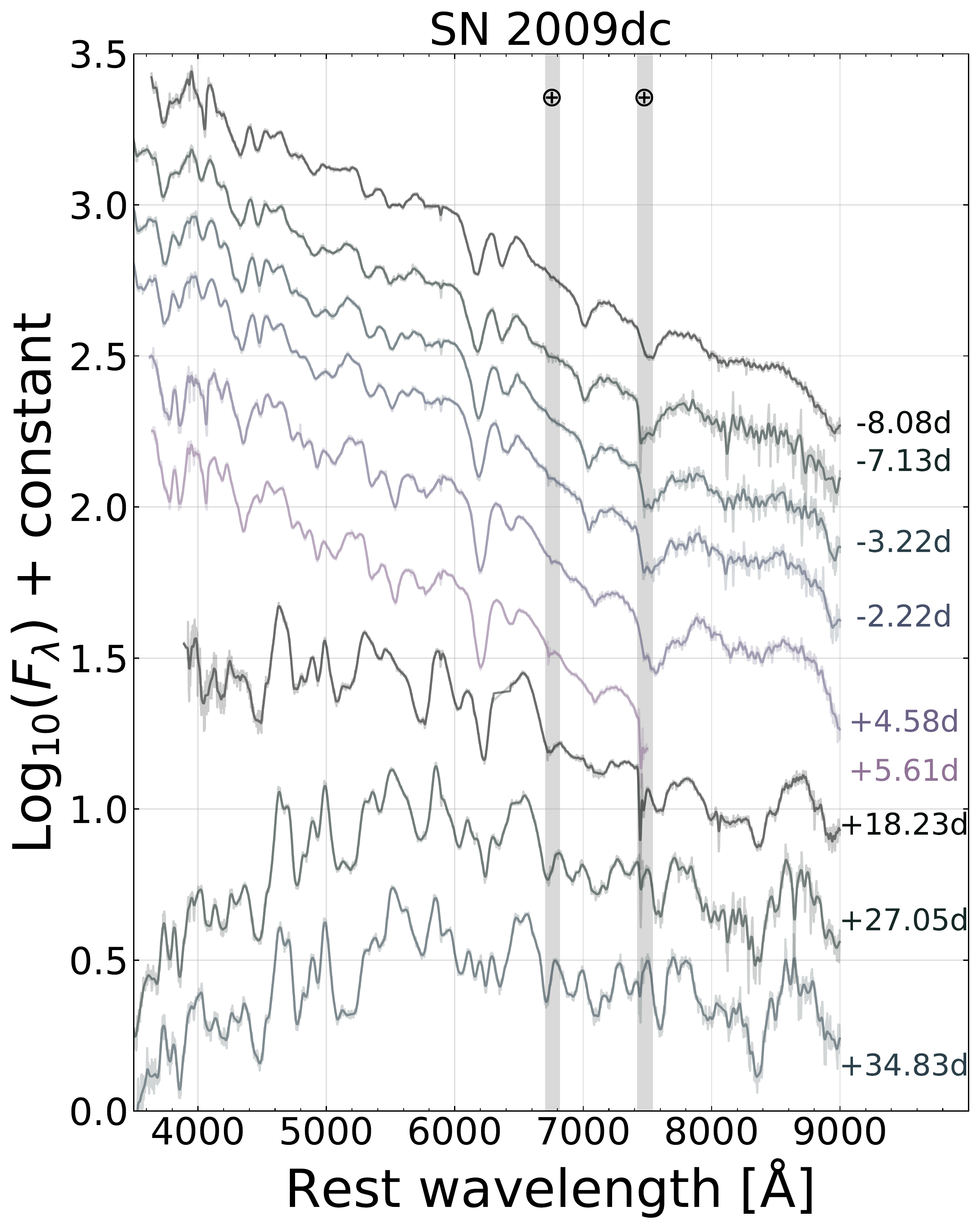} \\
    \includegraphics[width=.5\linewidth]{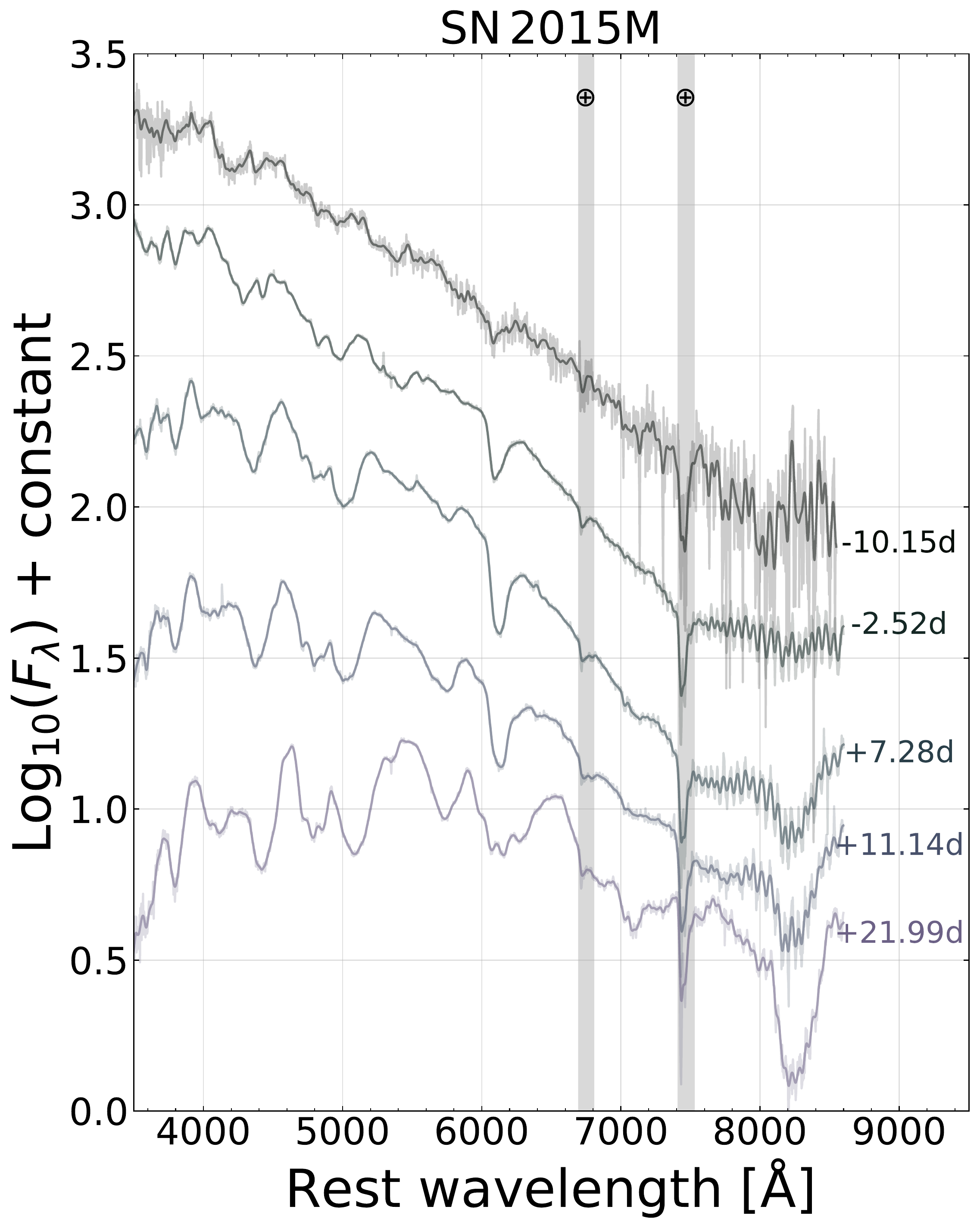} &
    \includegraphics[width=.5\linewidth]{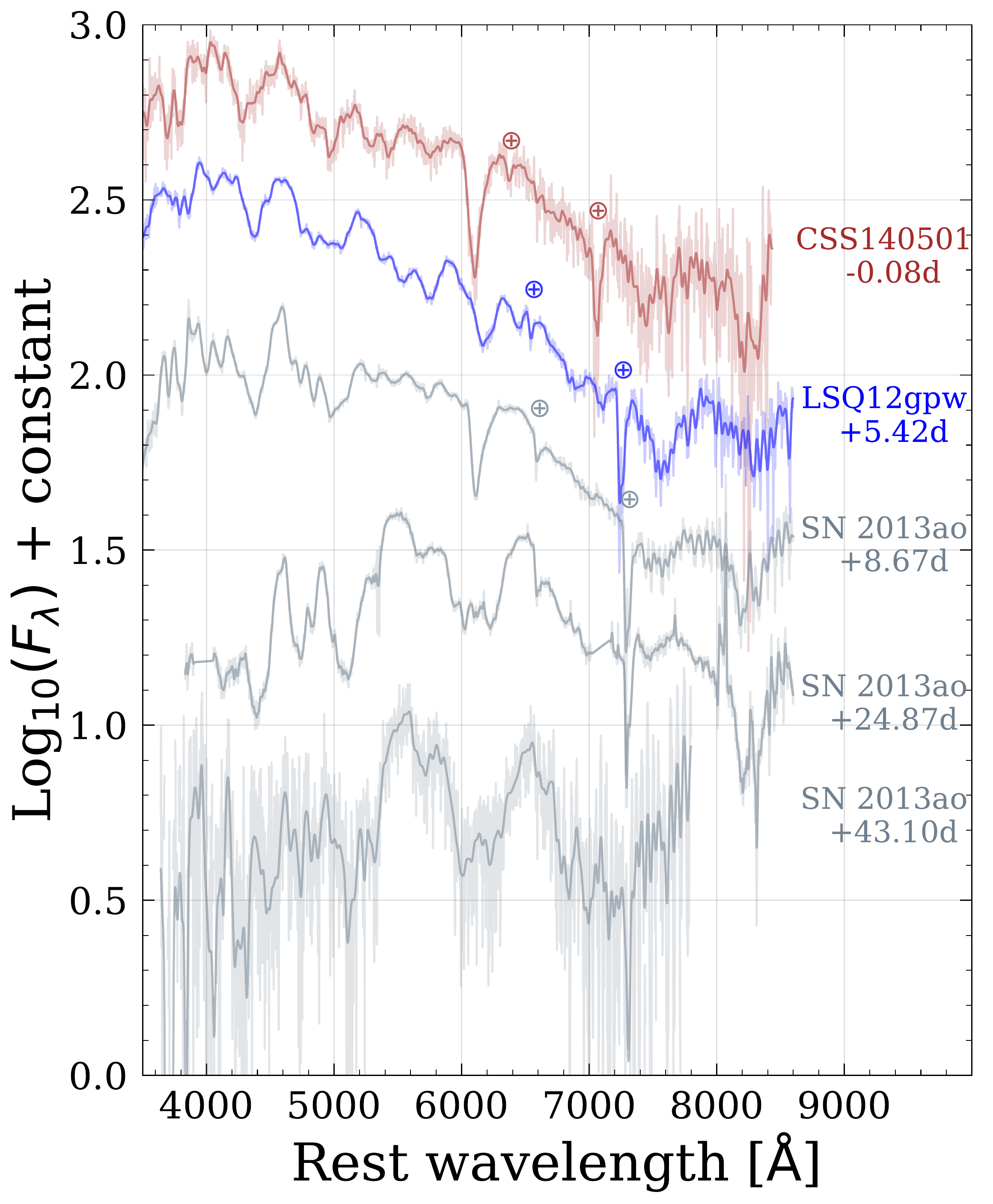} 
  \end{array}$
   \caption{Rest frame optical spectra of all unpublished \SCSN\ from the CSP. Rest frame phases relative to $B$-band maximum are given next to each spectrum. \red{These data have not been corrected for extinction.}  \label{fig:spec}}
   
\end{figure*}

\begin{figure}
\centering
 \includegraphics[width=.5\textwidth]{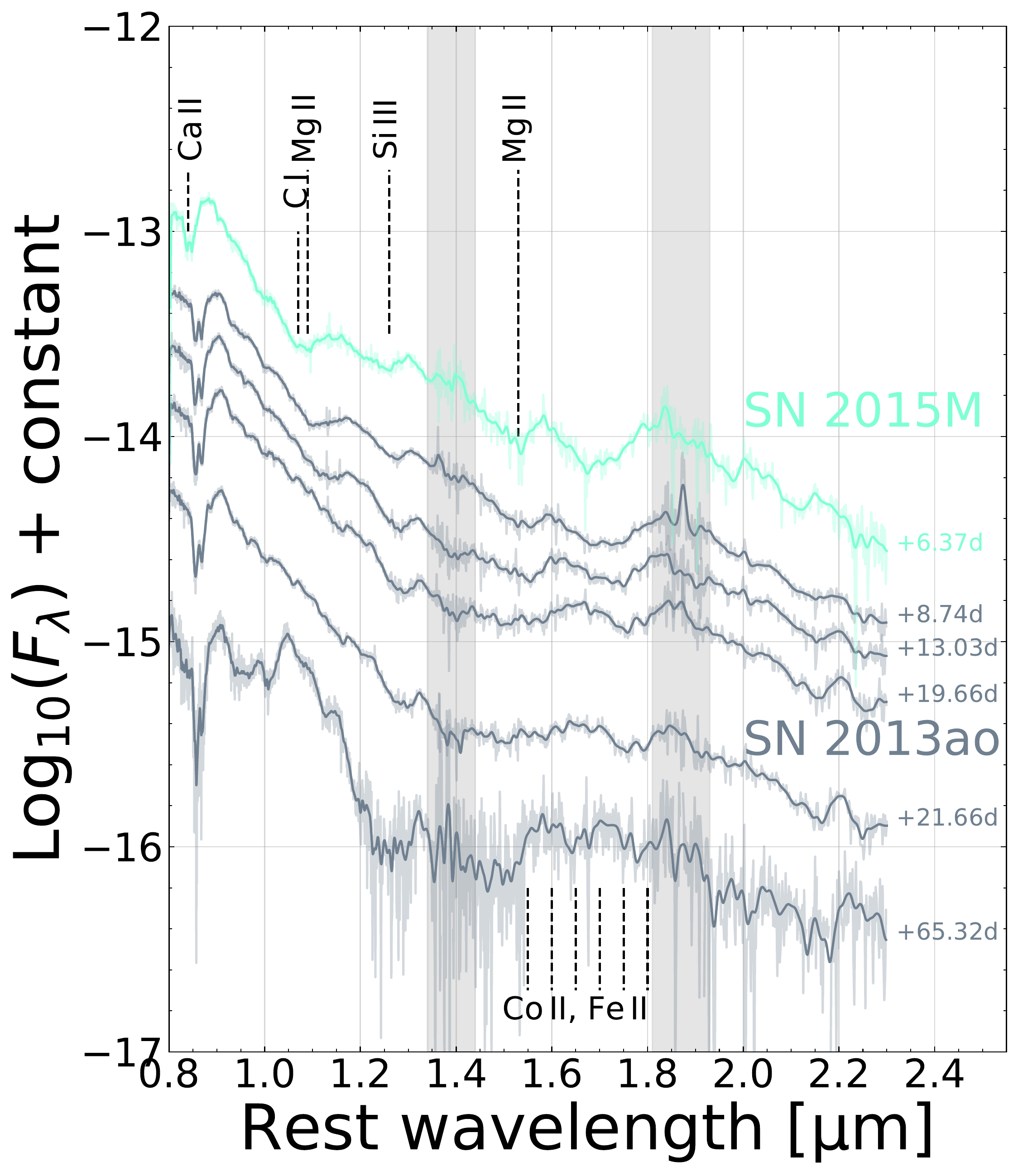}
\caption{Rest frame NIR spectra of all unpublished \SCSN\ from the CSP. Rest frame phases relative to $B$-band maximum are given next to each spectrum. Note that \CI\ is only seen in the  spectra of SN~2015M and not SN~2013ao. \red{These data have not been corrected for extinction.} }
\label{fig:specNIR}
\end{figure}

Plots of the previously unpublished CSP spectra of SN~2007if, SN~2009dc, LSQ12gpw, SN~2013ao, CSS140126, CSS140501, and SN~2015M are presented in Fig.~\ref{fig:spec} and Fig.~\ref{fig:specNIR}. 
All of the spectra for analysis have been made available at the CSP website\footnote{\href{https://csp.obs.carnegiescience.edu/data}{CSP data products.}}.

\begin{figure*}
\centering
 \includegraphics[width=.925\textwidth]{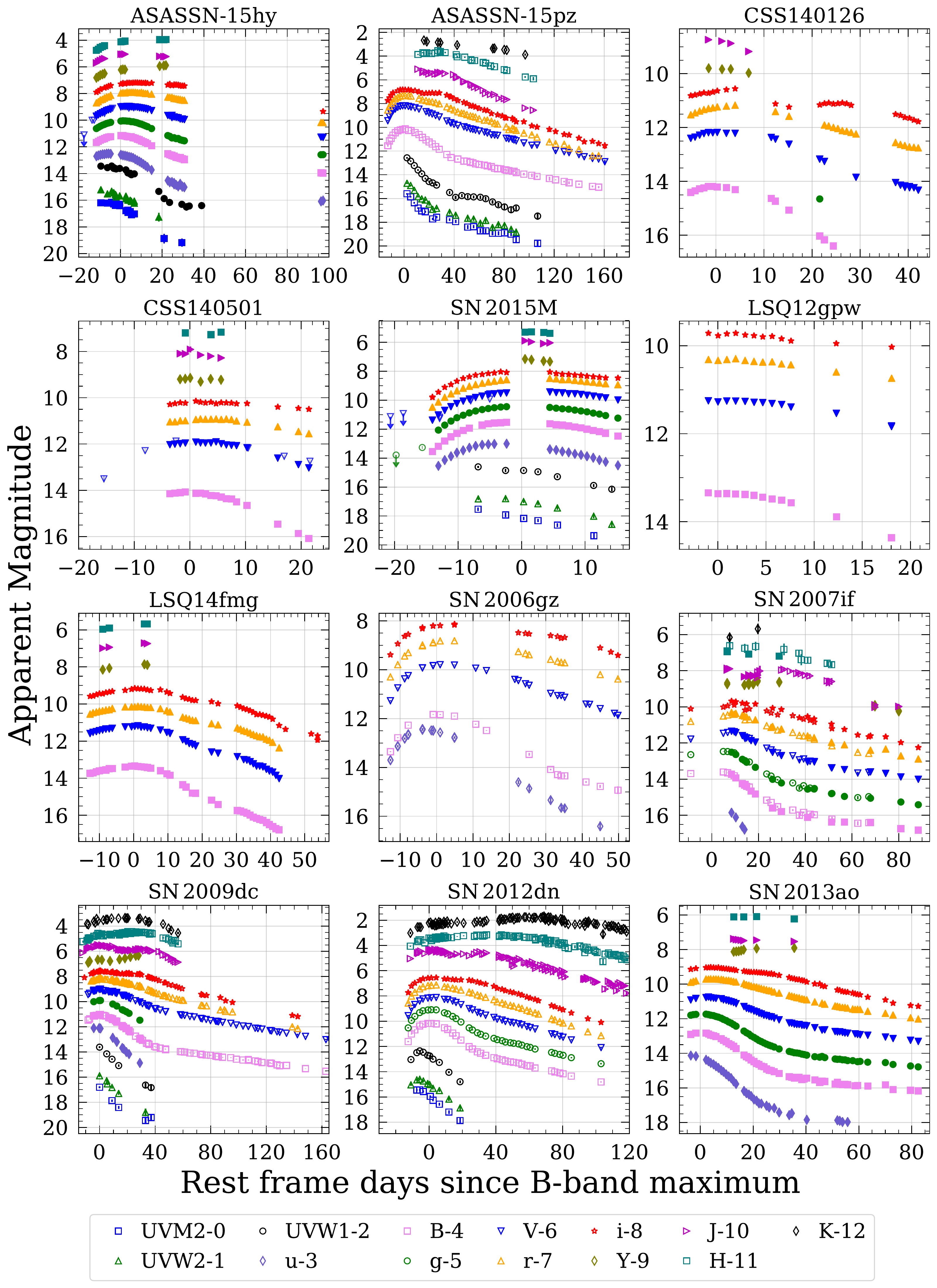}
\caption{Rest-frame UV to NIR photometry of all \SCSN\ in the sample. The photometry in this plot has not been K-corrected or corrected for host galaxy extinction, but have been corrected for foreground extinction. Filled symbols are photometric measurements computed from  Swope or du Pont telescope images and are in the CSP natural system. The open symbols  are from other  sources. SN~2003fg has been excluded from the plot due to poor photometric coverage. Individual light curve plots of these SN can be found in Appendix \ref{sec:CSP_phot_spec}}. 
\label{fig:lightcurves}
\end{figure*}

\section{Extinction}
\label{sect:Extinction}
Throughout this work, the dust map of \citet{Schlafly11} was used to correct for foreground Milky Way (MW) extinction. However, values of the host galaxy extinction of \SCSN\ are still very uncertain due to the fact that their intrinsic brightness and colors are not yet fully understood.

Two methods were adopted to estimate the host extinction for the \SCSN: i)  the equivalent width (EW) of the  host-galaxy Na~{\sc i}~D feature and ii) the Balmer decrements of the host galaxy lines at the SN location using integral field spectroscopy (IFS) data from \citet{Galbany21}.
Both of these techniques suffer from shortcomings,  which we discuss below. It is  established that if a SN has a Na~{\sc i}~D feature it may have some host galaxy extinction, 
and in the Milky Way there is a well-known correlation between the Av and the EW of Na I D (e.g., see Pozananski et al. 2012 and references therein).  However, the scatter in this correlation is $\pm$68\% and a significant fraction of SNe Ia display anomalously high Na I D EWs in comparison with the Av values derived from their colors \citep{Phillips13}. Furthermore, most of the spectra we analyze in this work are of low resolution, making it difficult to detect a weak Na~{\sc i}~D feature. 

If no Na~{\sc i}~D feature can be detected in the observed spectrum we
simulate an upper limit on how much Na~{\sc i}~D absorption could be
hidden in the data. This was done by using the highest resolution
observed spectrum for each SN. An idealized spectrum with infinite
resolution was created by smoothing the observed spectrum via a
Gaussian filter. An artificial absorption line with various widths and
depths simulating the Na~{\sc i}~D feature was then added at the host
redshift. This idealized spectrum was degraded to the resolution of
the observations. The spectrum was re-sampled at the same wavelengths
as the observed spectrum. Finally, random noise was added using the
flux uncertainty measured from the observed spectrum. The EW of the Na~{\sc
  i}~D feature was measured in the idealized spectrum and the
low-resolution spectrum. The strength of the absorption was decreased
until the EW of the low-resolution spectrum equals the EW uncertainty.  At
this point, the EW of the idealized spectrum is taken as the
detection limit. The results are presented in
Table~\ref{table:extinction}. The Na~{\sc i}~D pEW was then converted
to an $E(B-V)$ value using the relation from \citet{Poznanski12}, with a 68\% uncertainty as per \citet{Phillips13}.  

The second method to determine host-galaxy extinction is using the host-galaxy emission lines at the location of the SN. This has been  obtained for ten of the thirteen SNe we analyzed in this work \citep[see][]{Galbany21}. 
The IFS data, which is presented in \citet{Galbany21}, can be used to determine these values. However, this is highly uncertain as it assumes a constant temperature of the gas, and considers the whole column density at the location of the SN, not just in front of it. Only LSQ12gpw has a significant extinction detected at its location using this method. 

Note that the host extinction may be estimated for normal SNe~Ia using the Lira Law \citep{Lira98,Phillips99}, and this was attempted by \citet{Chen19} for  SN~2009dc and ASASSN-15pz. The host galaxy extinction of SN~2009dc has always been uncertain.  \citet{Chen19} found that SN~2009dc, which has a large Na~{\sc i}~D host galaxy feature, had the same $B-V$ color slope as  ASASSN-15pz. It was then pressumed that ASASSN-15pz could be used as an ``unreddened" comparison SN. 
The magnitude offset between the two SNe at the late-time decline was assumed to be caused by extinction.
However, there is no evidence for \SCSN\ following a form of Lira Law, and \citet{Hicken07}
claimed the host color excess derived from Lira relation is most likely not appropriate for \SCSN. Therefore, we choose not to use this relation. 
Furthermore, \citet{Lu21} found that \SCSN\ can look similar after $B$-band maximum during the Lira tail,  but be intrinsically different at early times and therefore have different luminosities demonstrating that for \SCSN\ the Lira law is not reliable. 


In Table~\ref{table:extinction}, the values of extinction obtained with the two different methods are presented.  As all of the methods mentioned above are highly uncertain, we choose to follow the conventional method of using the pEW of the Na~{\sc i}~D feature. 
In the cases where the host $E(B-V)$ measurements based on the Na~{\sc i}~D pEW are consistent with zero, no host extinction correction is employed. 
Furthermore, host galaxy extinction is only applied when explicitly stated in the following analyses.

\begin{deluxetable}{cccccc}
\tablewidth{\textwidth}
\tablecaption{EW values and limits of the Na ID feature, along with the $E(B-V)$ calculated from EW of the Na ID feature as well as  from the ratio of the Balmer lines. \label{table:extinction}}
\tablehead{
\colhead{SN}&
\colhead{Na ID EW}&
\colhead{$E(B-V)$\tablenotemark{a}}&
\colhead{$E(B-V)$\tablenotemark{b}}&
\\
\colhead{}&
\colhead{\AA}&
\colhead{Mag}&
\colhead{Mag}}
\startdata
 \hline
2003fg	&	0.33	$\pm$	0.14 &	0.03	$\pm$	0.03	&	$\cdots$	\\
2006gz	&	0.30	$\pm$	0.11 &	0.03	$\pm$	0.02	&	0.00	\\
2012dn	&	0.27	$\pm$	0.14 &	0.03	$\pm$	0.02	&	$\cdots$	\\
ASASSN-15pz	&	        $\leq$	0.05 &	0.00	$\pm$	0.02	&	$\cdots$	\\
2007if	&		    $\leq$	0.06 &	0.00	$\pm$	0.02	&	$\cdots$	\\
2009dc	&	1.03	$\pm$	0.24 &	0.23	$\pm$	0.22	&	0.00	\\
LSQ12gpw	&	0.23	$\pm$	0.10 &	0.03	$\pm$	0.02	&	0.23	\\
2013ao	&	    	$\leq$	0.08 &	0.00	$\pm$	0.02	&	0.09	\\
CSS140126	&	    	$\leq$	0.42 &	0.00	$\pm$	0.04	&	$\cdots$	\\
CSS140501	&	    	$\leq$	0.14 &	0.00	$\pm$	0.02	&	0.00	\\
LSQ14fmg	&	0.87	$\pm$	0.44 &	0.15	$\pm$	0.20	&	0.00	\\
2015M	    &	    	$\leq$	0.06 &	0.00	$\pm$	0.02	&	$\cdots$	\\
ASASSN-15hy	&	    	$\leq$	0.06 &	0.00	$\pm$	0.02	&	0.00	\\
    \hline
\enddata
\tablenotetext{a}{Calulated using relationship from \citet{Poznanski12} with uncertainties as per \citet{Phillips13}.}
\tablenotetext{b}{Calculated from Balmer line ratio in IFS data from \citet{Galbany21}}
\end{deluxetable}

\section{Photometric properties}
\label{sect:phot}

\subsection{Light curves}

\begin{figure*}
\centering
 \includegraphics[width=0.99\textwidth]{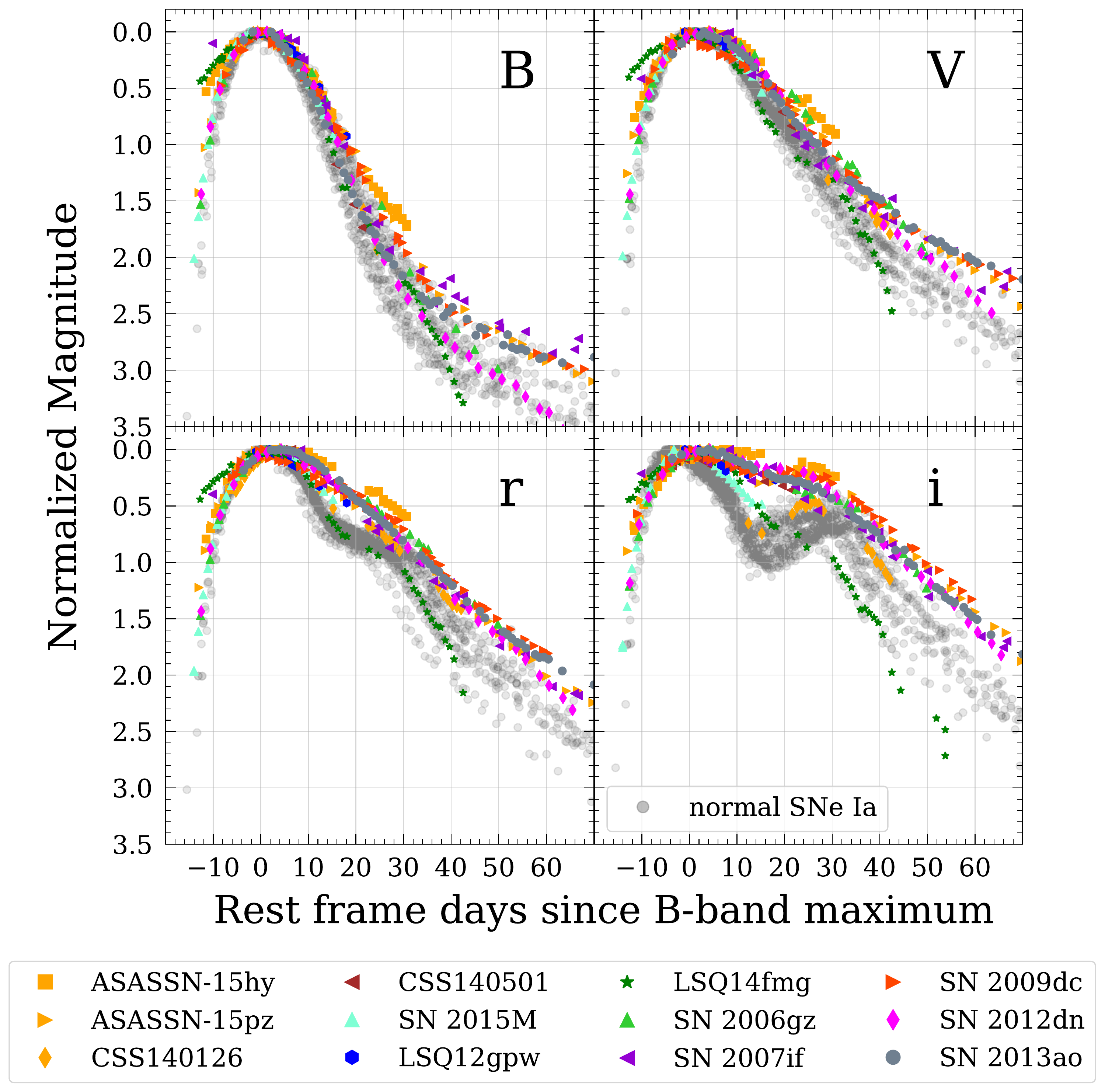}
\caption{Comparison of $BVri$-band light curves of the \SCSN\ and normal SNe~Ia. The light curves are presented relative to $B$-band maximum and in the rest frame. They have also been normalized to the peak of each respective band. Normal CSP SNe~Ia are plotted in light grey for comparison. The selection criteria for the normal SNe are: $\DmB<$1.3~mag and $E(B-V)_{host}<$0.15~mag.}
\label{fig:compLC}
\end{figure*}

Figure~\ref{fig:compLC} presents the K-corrected $BVri$ light curves of the sample compared to a selection of  normal SN~Ia ($\DmB<$1.3~mag), all placed on the CSP natural system. The light curves are normalized to the peak, such that the comparison is in the light-curve shapes. In the $B$ and $V$ bands, the \SCSN\ are largely indistinguishable from normal SNe~Ia, with some noted exceptions. SN~2007if, LSQ14fmg, and ASASSN-15hy have extremely slow rise times.  Furthermore, LSQ14fmg declines rapidly at the start of the decay tail. Similarly rapid declines were also observed but at much later phases in SN~2009dc and SN~2012dn.  CSS140126 also shows a hint of the rapid decline in the $V$ band around the same phase as that of LSQ14fmg.

At redder wavelengths, the light curves of \SCSN\ differ drastically from those of standard SNe~Ia. In the $r$-band, \SCSN\ have a much weaker post-maximum ``knee'', except for LSQ14fmg. In the $i$-band, they have no strong secondary maxima, except in the case of CSS140126. The $i$-band secondary maximum has been suggested to be produced by the recombination of iron group elements in the ejecta \citep{Hoeflich02,Kasen06}. Brighter SNe, such as 91T-like objects, tend to have a more prominent secondary $i$-band maximum, and fainter SNe, such as 91bg-like SNe, have a lack of a secondary $i$-band maximum. The latter case is caused by the merging of the primary and secondary maxima due to the quickly receding photosphere and a small \Nifs\ mass \citep[see \eg][]{Hoeflich02,Ashall20}. 

For \SCSN, the lack of a prominent $i$-band secondary maximum is a defining trait \footnote{\red{We note that CSS 140126 has a secondary $i$-band maximum but has a uncertain classification due to a low S/N spectra. However it fits into our sample because it peaks in the $i$-band after the $B$-band.}}. This \red{suggests}, unlike normal SNe~Ia, that there is a lack of recombination of iron group elements in the ejecta at 2~d to 40~d past maximum. Given that \SCSN\ are significantly brighter than sub-luminous SNe~Ia, it is unlikely that the cause of the lack of a secondary $i$-band maximum is the same as in sub-luminous SNe~Ia. In the case of \SCSN, it \red{suggests} that a significant amount of luminosity is not produced by the radioactive decay \Nifs, or that there is full mixing in the ejecta \citep[\eg][]{Kasen06}. However, as is seen in 2002cx-like SNe, if  there were full mixing in the ejecta of \SCSN\ we would expect to see a prominent $H$-band break and even lower ejecta velocities in the photoperic phase \cite[\eg][]{Kromer15,Stritzinger15}. Furthermore, if \SCSN\ were powered predominately by the radioactive decay of \Nifs\ it would be expected that they would have a very distinct secondary $i$-band maximum, such as in normal or 1991T-like SNe~Ia. This is not the case for \SCSN. 
 
\begin{figure*}[tb]
\centering
 \includegraphics[width=0.99\textwidth]{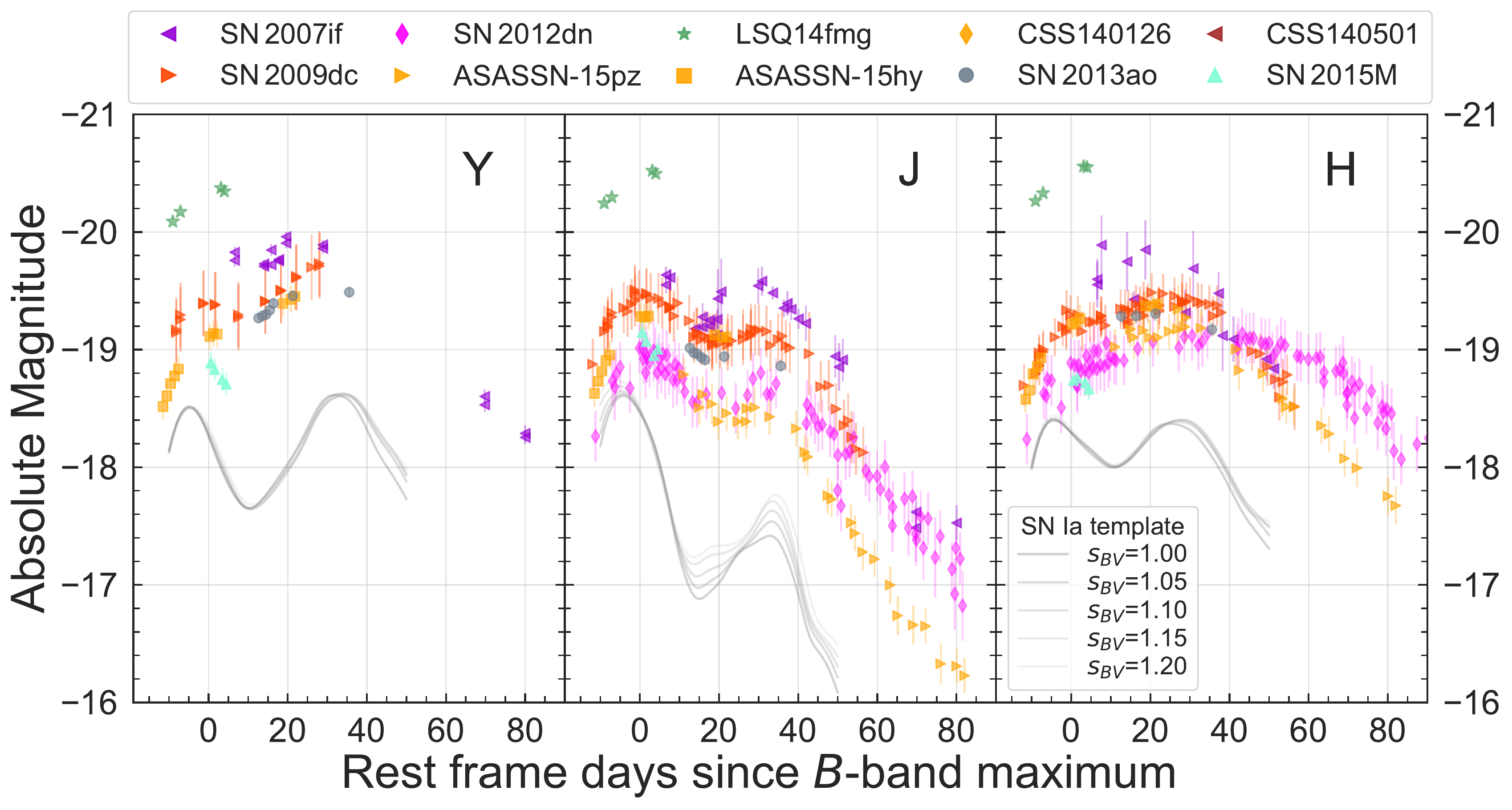} 
\caption{Absolute-magnitude $YJH$ light curves of the \SCSN\ in the sample. The light curves have not been K- or S-corrected. Normal SN~Ia light-curve templates of \sBV=1.00, 1.05, 1.10, 1.15, and 1.20 from \textit{SNooPy}  are plotted in grey colors for comparison, with the peak magnitude matching to the luminosity–decline rate relation from Fig.~4 of \citealt{Burns18}. We chose not to normalize the NIR light curve comparison to peak as the time of the NIR maximum is uncertain for many objects. Overall, the NIR photometric properties of \SCSN\ are distinct from those of normal SNe~Ia. }
\label{fig:compLCNIR}
\end{figure*}

The NIR light curves of \SCSN\ are vastly different than normal SN~Ia (Fig.~\ref{fig:compLCNIR}). Ten of the \SCSN\ in the sample have NIR light curves. They are in general much brighter than the normal population in the NIR. None of the \SCSN\ have a clear secondary maximum in the $Y$ or $H$ bands. The diversity between the \SCSN\ is also large. For all \SCSN, the phase of the NIR primary maxima occurs significantly after that of $B$-band maximum, whereas the NIR primary maxima of the normal population consistently transpires a few days before their $B$-band maxima. For example, the $H$-band light curve of SN~2012dn peaks $\sim$50~d past $B$-band maximum. As a consequence, the NIR, specifically the $H$ band, may be the most effective way to distinguish \SCSN\ from normal SNe. These prolonged NIR light curves of \SCSN\ imply that there could be some additional sources of luminosity \citep{Nagao17,Nagao18}. 

Another intriguing photometric peculiarity of \SCSN\ is their UV luminosity. \SCSN\ are $\sim$2~mag brighter than normal SNe~Ia in the mid-UV  \citep{Brown14}. \citet{Brown14} analyzed the UV properties of SN~2009dc and SN~2012dn, which are located at opposite ends of the \SCSN\ peak luminosity distribution in the optical, and found that both of these objects peak at $\sim-$18~mag in the swift uvm2 band. \citet{Lu21} analyzed the swift uvm2 light curves of all published \SCSN\ and confirmed these results. The cause of this excess flux is unknown. It is not likely to be due to a larger amount of \Nifs\ in the outer layers, as the ionization state of \SCSN\ is generally low, and the $H$-band break is not observed until past +50\,d from maximum light (Section~\ref{sec:nir_spec}).  The high UV flux could be caused by low metallicity of the progenitor, where a low metallicity produces a lack of Fe-group elements in the outer layers and thus a lack of line blanketing \citep{Mazzali14}. It could also be produced by an additional energy source which is not \Nifs\ such as interaction with any surrounding material \citep[e.g.,][]{Nagao17, Hsiao20}.  \red{For a discussion on bolometric light curves see section \ref{sect:bolLC}. }

Among the thirteen objects in the sample, six have early light curves ($< -10~d$) and  constraints on the first epoch.  
\red{The rise times of the two SNe from this work, CSS140501 and SN~2015M, are obtained by fitting the pre-maximum $V$-band photometry with a second order polynomial function, where the data was constrained by the discovery survey’s photometry and last non-detection limits as mentioned in Section~\ref{sec:Discoveryinfo}.
The remaining  the rise times were obtained from literature. The rise time of SN~2007if was obtained from an unfiltered magnitude \citep{Scalzo10}. The rise time of SN~2009dc is given in the $R$-band \citep{Silverman11}, and ASASSN-15hy \citep{Lu21} and ASASSN-15pz \citep{Chen19} are provided in the $V$-band.
} These SNe and their respective rise times are SN~2007if ($24.2 \pm 0.4~d$; \citealt{Scalzo10}), SN~2009dc ($23 \pm 2~d$; \citealt{Silverman11}), ASASSN-15pz ($21.4 \pm 2~d$; \citealt{Chen19}), ASASSN-15hy ($22.5 \pm 4.6$; \citealt{Lu21}), CSS140501 ($20.9 \pm 6.7$; this work), and SN~2015M ($19.8 \pm 4.8$; this work). This implies an average rise time of $22.0 \pm 3.8$~d.

\subsection{Luminosity Width Relation}
\label{sect:LWR}
SNe~Ia follow an intrinsic luminosity width relation (LWR), where
brighter SNe have broader light curves. Two common ways to determine
the broadness and the time scale of the light curve are the parameters
$\DmB$ \citep{Phillips93} or \sBV\ \citep{Burns14}.  $\DmB$ measures the change in $B$-band magnitude between maximum light and 15 rest-frame days past then, and \sBV\ is the time difference between the occurrence of $B$-band maximum and the reddest point in the $(B-V)$ color curve divided by 30~d \citep{Burns14}. 

To establish the location of \SCSN\ on the LWR, the light-curve parameters were measured using the \textit{SNooPy} package \citep{Burns11}.  No light curve templates were used to fit the photometry or their derived parameters. Rather, the $B_{max}$, $\DmB$ and \sBV\ parameters were directly measured from the rest-frame K-corrected light curves interpolated with Gaussian processes. The \textit{SNooPy} \textsc{get\_color} function was used to calculate the color curves. For the color curves, no interpolation between data was performed when multi-band observations on the same night were not available. The color curves were produced with Gaussian processes to obtain the time of the reddest point in the $(B-V)$ color curve relative to $B$-band maximum. This value was divided by 30 to obtain \sBV.
This same technique was used to obtain the color curves in Section \ref{sect:colorcurv}, and to derive the colors at maximum light.  For all photometric measurements, uncertainties were obtained using a bootstrapping technique with 150 iterations. All rest frame, K-corrected, light curves were corrected for galactic and host extinction (where applicable). 
$B_{max}$ was converted to $M_{B}$ using the distance moduli presented in Table \ref{table:prop}. 

\begin{figure*}
\centering
 \includegraphics[width=.99\textwidth]{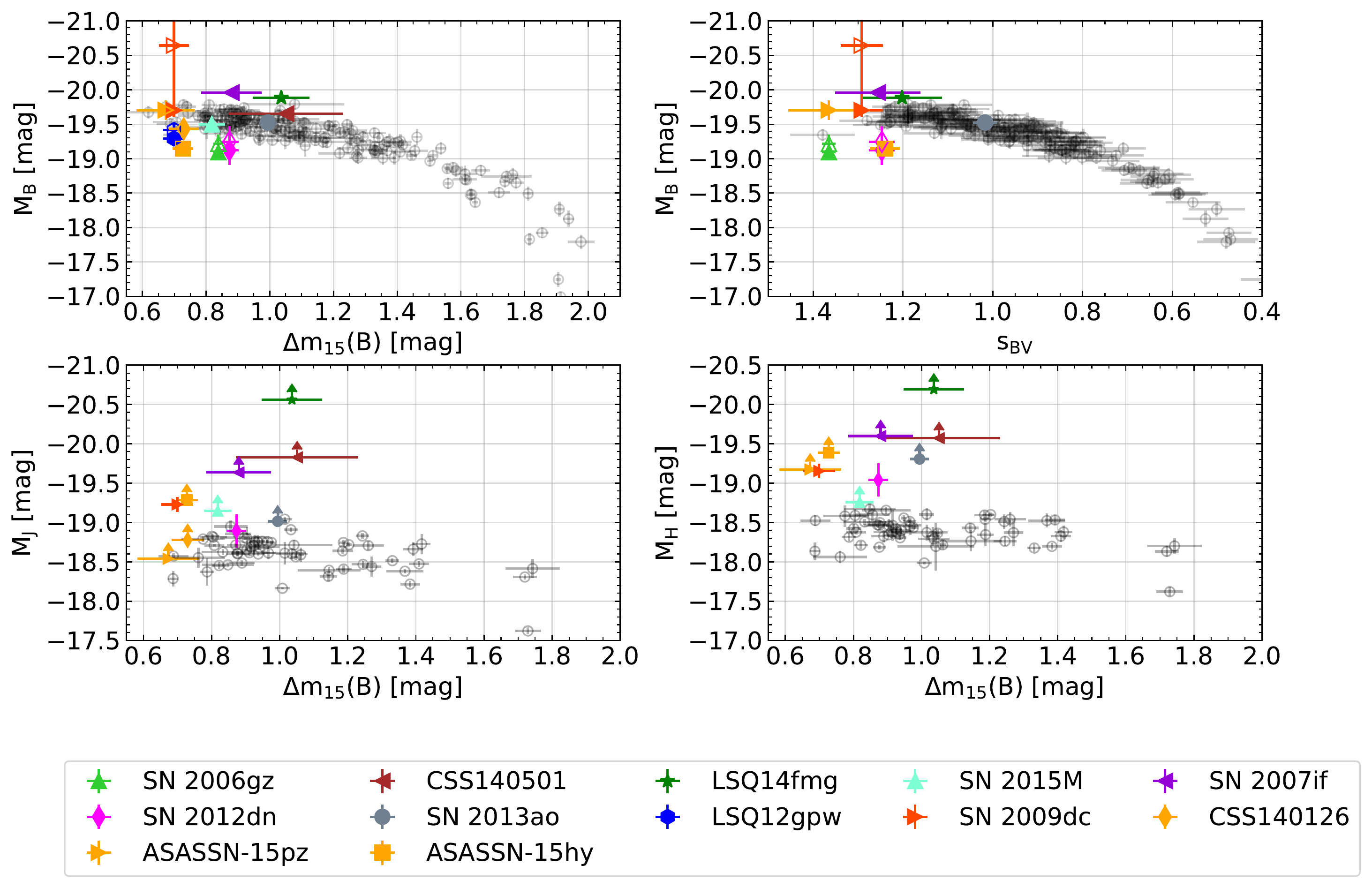}
\caption{The luminosity width relation. \textit{Top left:} The absolute $B$-band magnitude ($M_B$)plotted as a function of $\DmB$. \textit{Top right:} $M_B$ plotted as a function of \sBV. In both top panels the open symbols are corrected for galactic and host-galaxy extinction. \textit{Bottom left:} $M_{J}$ as a function of $\DmB$. \textit{Bottom right:} $M_{H}$ as a function of $\DmB$. For both of the bottom panels, many of the points are lower flux limits as the light curves are still rising during the final photometric observations. The \SCSN\ points in the NIR LWR have not been corrected for host-galaxy extinction as the extinction in the NIR is negligible and the values of host-galaxy extinction are uncertain. In all plots the gray symbols are the LWR relation constructed using SNe~Ia observed by the CSP \citep{Krisciunas17,Phillips19}. }
\label{fig:LWR}
\end{figure*}

The $B$-band LWR as a function of $\DmB$ and \sBV\ is presented in the top panels of  Fig. \ref{fig:LWR}.
The \SCSN\ are all slowly declining with $\DmB<$1.3\,mag and \sBV$>$1.   
\SCSN\ are located below, above and in the same area of the LWR as normal SNe~Ia.
 SN\,2012dn, SN\,2013ao,  ASASSN-15pz, and SN~2015M are all located in the main part of the LWR. 
 SN2006gz, LSQ12gpw and  ASASSN-15hy are less luminous than their $B$-band light curve shape would imply if they were standard SNe~Ia (i.e. they are less luminous than the LWR), whereas, SN\,2007if, SN\,20009dc, and LSQ14fmg are all brighter than the LWR. 
Unlike the previous suggestion of \citet{Taubenberger11}, it is evident with a larger sample that not all \SCSN\ are overluminous. However, host galaxy extinctions of \SCSN\ are highly uncertain and this may affect the location of some SNe on the LWR, although for the majority of points there is a reliable limit from the Na ID EW (Section~\ref{sect:Extinction}).  We note that most of our objects are within the Hubble flow (z$>$0.02) therefore the distance derived to the SN from the hosts are accurate.

The NIR is less affected by extinction, therefore it is a more suitable wavelength range to 
analyze any intrinsic luminosity differences between \SCSN\ and other SNe~Ia. 
We do not attempt to correct the NIR photometry for host extinction.
The bottom panels of Fig.~\ref{fig:LWR} show the LWR in the $J$ and
$H$ bands. Unfortunately, the temporal coverage in the NIR is much
worse than the optical. Therefore, for all SNe except SN~2009dc and
SN~2012dn only lower flux limits  can be set for the peak
luminosity. The lower limits were determined by measuring the
brightest photometric point on the rising light curves.   In the
$J$-band, most of the SNe are over luminous (brighter than $-$19~mag)
except for ASASSN-15pz, CSS140126, and SN~2012dn that have a
luminosity similar to that of normal SNe~Ia. However for  ASASSN-15pz and CSS140126the these values  are lower limits on the luminosity and they could be intrinsically brighter.  Interestingly, the
$H$-band is the most effective wavelength range to distinguish between
\SCSN\ and normal SNe~Ia. All of the \SCSN\ are brighter than normal
SNe~Ia in the $H$ band.
They range from $M_{H}=-18.76$~mag (SN~2015M), to  $M_{H}=-20.19$~mag (LSQ14fmg). The bright $H$-band is one of the few ubiquitous properties of the subclass of \SCSN. \red{The fact that \SCSN\ are not standarizable may suggest that that there is more than one driving parameter driving the explosion. i.e. More than just a range of WD masses. There could be a range of both WD masses and envelope masses which produce \SCSN\, see section \ref{sect:discussion} for a  detailed discussion.}

As shown by \citet{Galbany21}, \SCSN\ are preferentially located in low metallicity and low mass galaxies with high sSFR, and are therefore more common in the high redshift universe. The fact that \SCSN\ are not standardizeable and they do not follow the LWR in the optical or NIR could have direct consequences for dark energy experiments. We have shown here that it is easier to remove \SCSN\ from cosmological experiments with rest frame NIR data. But due to cosmic expansion, this strategy may limit future dark energy experiments to lower redshifts. With only near-maximum light, rest-frame, optical observations, \SCSN\ may bias dark energy experiments. We briefly discuss this in Section~\ref{sect:Hubble}. However, detailed simulations of this is outside the scope of this work.

\subsection{Color curves}
\label{sect:colorcurv}
\begin{figure*}
\centering
 \includegraphics[width=0.99\textwidth]{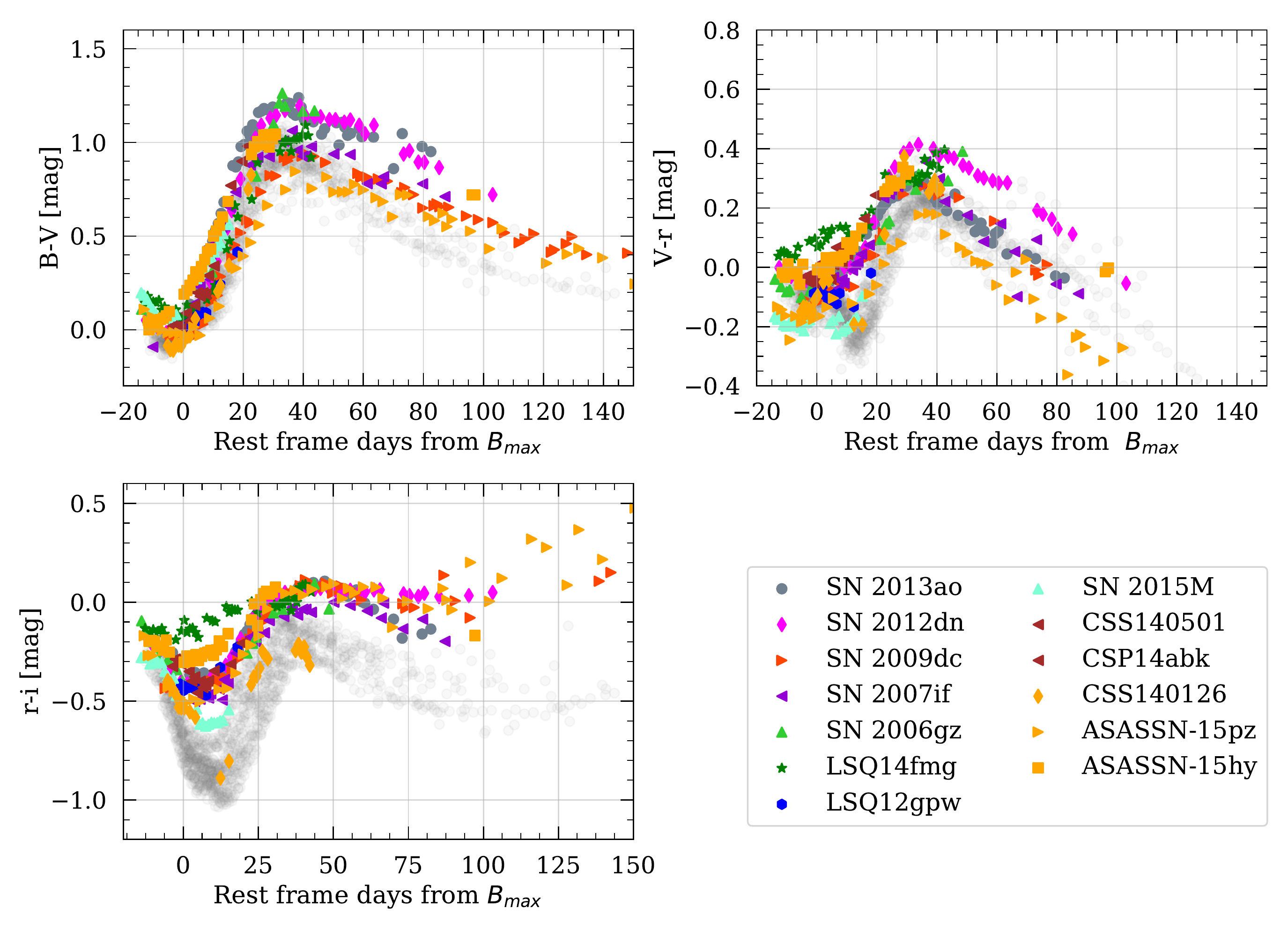}
\caption{The observed color curves of our \SCSN\ sample corrected for Milky Way extinction (solid markers) and compared to a sample of normal SNe~Ia from the CSP (light gray). The normal SNe~Ia were selected so they have $\DmB<1.3$~mag and an $E(B-V)_{host}<$0.15~mag.  The $B-V$ color curves of  \SCSN\ are similar to those of normal SNe~Ia,  yet the $r-i$ color curves  differ significantly. All curves have been corrected for Galactic extinction, but not host galaxy extinction. }
\label{fig:colorLC}
\end{figure*}

The observed color curves of the \SCSN, corrected for Milky Way extinction,  are shown in Fig.~\ref{fig:colorLC}.
At early times, the $(B-V)$ curves are similar to normal SNe~Ia.
They start red and reach their bluest point around maximum light. Between maximum light and +40\,d, the ejecta cool until the reddest epoch is reached, after which the colors turn blue again. However, \SCSN\ do not follow a tight Lira-like law \citep{Phillips99}, and exhibit significant diversity. The reddest points in  $B-V$ color curves cover a larger range ($\sim$0.7-1.3\,mag) in \SCSN\ than in normal SNe~Ia.  After the turnover, the \SCSN\ events show a variety of gradients in the Lira tail.   As discussed by \citet{Lu21}, ASASSN-15hy is unusual regarding its $(B-V)$ color curve. It does not get bluer during the early phase, and the evolution towards redder $(B-V)$ occurs much earlier than other SNe~Ia. The ejecta begin to cool and get redder from $\sim -$10\,d relative to $B$-band maximum, reaching $(B-V) \approx 0.2$\,mag at maximum light. \citet{Lu21} interpreted this to be caused by a lower \Nifs\ mass explosion in the core degenerate scenario.

The $(r-i)$ color curves of \SCSN\ show the largest diversity and generally do not behave like normal SNe~Ia. There is a continuum of properties from LSQ14fmg which gets monotonically redder from the first measurement then stays flat after +25\,d, to SN~2015M and CSS140126 which reach the bluest values. Generally, the \SCSN\ start at similar $(r-i)$ values as normal SNe~Ia, but do not reach such large negative values.  More interestingly, at +50\,d relative to maximum the \SCSN\ $(r-i)$ color curves remain flat and in some cases continue to turn redder. This is caused by light curves at longer wavelengths in \SCSN\ being much broader than normal SNe~Ia. In large all-sky surveys such as ZTF and the Vera Rubin Observatory, analyzing the $(r-i)$ color curve up to +50~d may be one of the most effective ways to distinguish \SCSN\ from the general population of SNe~Ia.

\subsection{\sBV\ vs. \iBmax}
\begin{figure}
\centering
 \includegraphics[width=.5\textwidth]{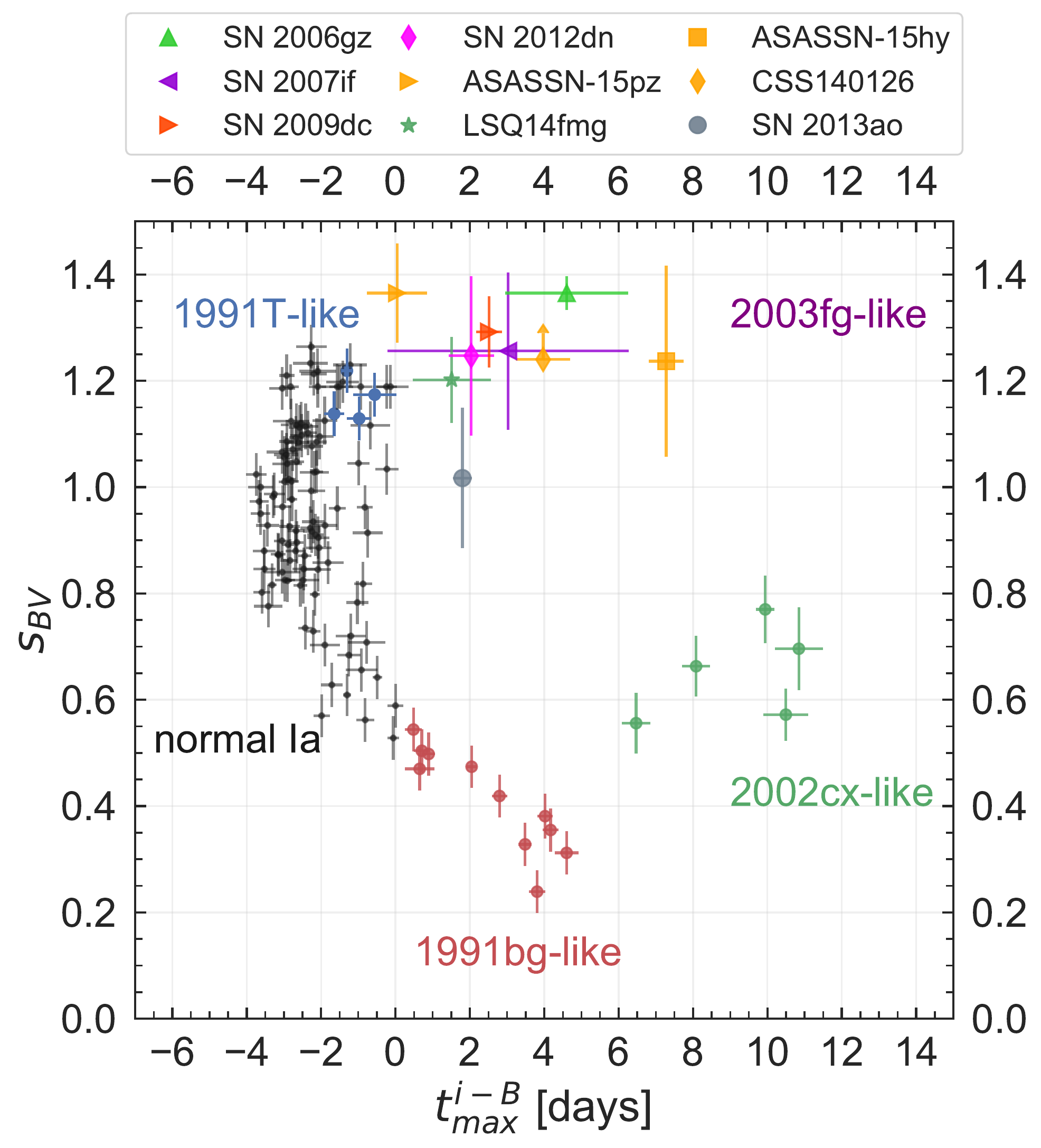}
\caption{The time of the $i$-band maximum relative to the $B$-band maximum vs. \sBV\ with data taken from \citet{Ashall20} and \SCSN\ of this work. These values do not include K-corrections since they are of various subtypes, this is also consistent with the analysis of \citet{Ashall20}. All of the \SCSN\ are located in the top right corner of the figure.  }
\label{fig:sBViband}
\end{figure}

\citet{Ashall20} demonstrated that the timing of the $i$-band primary maximum relative to the $B$-band maximum can be used as a powerful diagnostic to distinguish between sub-types of thermonuclear SNe. It was found that for \SCSN\ the time of the primary $i$-band maximum was later than that of $B$ band. Combining this information with \sBV\ was found to be an excellent way to identify \SCSN.  Fig.~\ref{fig:sBViband} shows the relation from \citet{Ashall20} labeled with the \SCSN\ from this work. 
Interestingly, ASASSN-15pz, which has a hint of an $i$-band secondary inflection,  also has the smallest value of \iBmax. Although note that  CSS140126 has both a strong $i$-band secondary maximum and a  late \iBmax. Conversely, ASASSN-15hy has the largest value of \iBmax\ and no secondary $i$-band maximum. This may indicate a connection between the value of \iBmax\ and the presence of an $i$-band secondary maximum.

\subsection{Bolometric light curves} \label{sect:bolLC}
Pseudo-bolometric light curves were constructed using \red{observed photometry} by employing the the direct method in \textit{SNooPy}. Where needed, the observation gaps in between light curves were interpolated with Gaussian processes,
but no extrapolations were applied outside the time range of individual light curves. Three wavelength regions were selected in order to explore the pseudo-bolometric peak luminosity and flux ratios in the UV, optical, and NIR. 
The light curve of a normal SN~Ia 2007af \citep{Krisciunas17} was also constructed using the same method as above for comparison. 

\begin{figure}[tb]
\centering
 \includegraphics[width=0.48\textwidth]{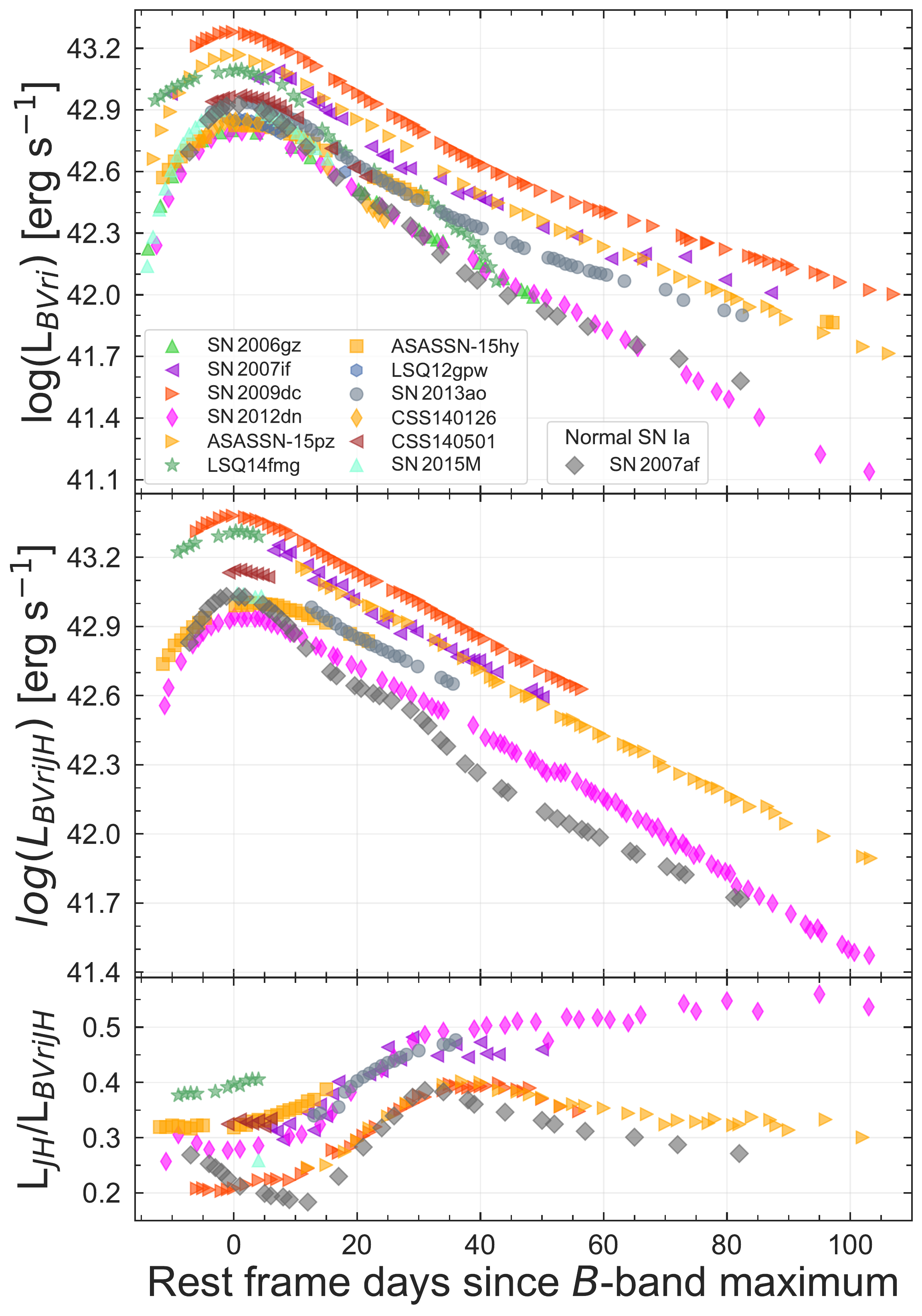} 
\caption{Pseudo-bolometric light curves of \SCSN. The top panel shows the Pseudo-bolometric light curves constructed with optical $BVri$ bands, the middle panel presents those with optical $BVri$ and NIR $JH$ bands, and the bottom panel displays the fraction of the NIR luminosity.}
\label{fig:BVri_JH_BoloLC}
\end{figure}

The $BVri$ ($4200$ to $7300$~\AA) pseudo-bolometric light curves are presented in the top panel of Fig.~\ref{fig:BVri_JH_BoloLC}.
The average peak luminosity in the sample is $L_{\rm{peak}} = 10^{42.97 \pm 0.16}$ erg~s$^{-1}$ for 11 \SCSN\ (SN~2003fg and LSQ12gpw are excluded due to their poor data coverage around the peak). When the NIR ($\sim \lambda 4200 - 16{,}000$~\AA) is also included in the construction of the bolometric light curves \SCSN\ peak at $L_{\rm{peak}} = 10^{43.16 \pm 0.22}$ erg~s$^{-1}$ from the four SNe that have available peaks. 

Interestingly, the fraction of flux in the NIR is generally higher in all \SCSN\ compared to the normal SN~Ia 2007af, as shown in the bottom panel in Fig.~\ref{fig:BVri_JH_BoloLC}. This is consistent with with \SCSN\ having high NIR peak luminosities.
All \SCSN\ gradually increase in NIR flux fraction until one month after the maximum.
It should be noted that SN~2012dn (which has the same luminosity as a
normal SN~Ia) shows the most prolonged and largest NIR contribution
that persists   well past +30~d.

\begin{figure*}[tb]
\centering
 \includegraphics[width=0.8\textwidth]{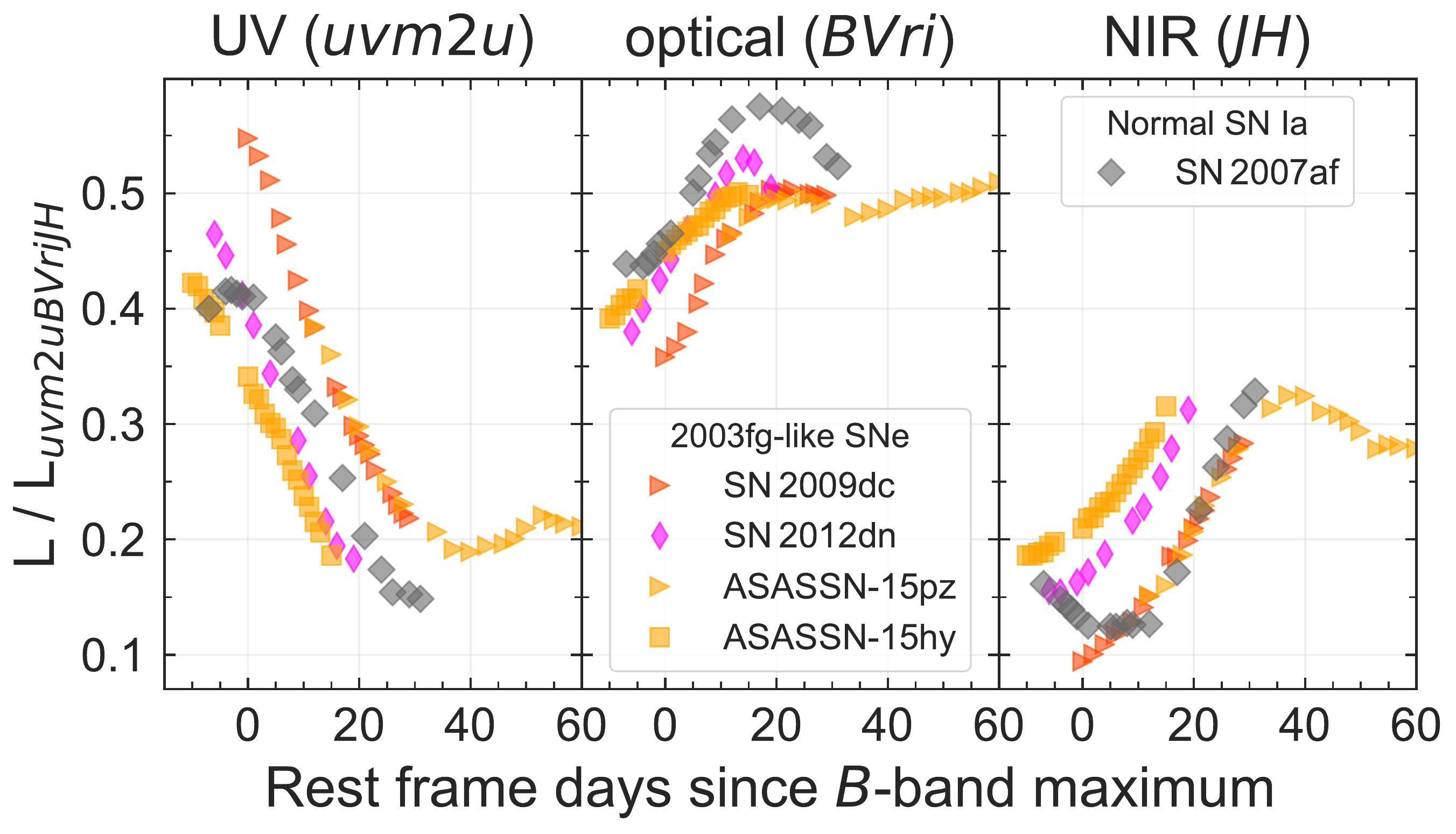} 
\caption{Fractions of pseudo-bolometric luminosity of \SCSN\ in UV ($\sim \lambda 2200 - 4200$), optical ($\sim \lambda 4200 - 7300$) and NIR ($\sim \lambda 7300 - 16{,}000$) regions. Unlike normal SNe~Ia that have the optical region accounting for $\sim$80\% of the luminosity, \SCSN\ all show a significant fraction of luminosity in the UV and NIR. The time evolution of the fractions also shows the flux redistribution from UV to NIR.}
\label{fig:BoloLC_ratios}
\end{figure*}

For four \SCSN\ it was possible to construct UV and NIR ($\sim \lambda 2200 - 16{,}000$~\AA) pseudo-bolometric light curves. Both ASASSN-15hy and SN~2012dn have a peak luminosity of $L_{\rm{peak}} = 10^{43.17 \pm 0.01}$ erg~s$^{-1}$.
Despite the small size of the sample, the effect of the bright UV and NIR as well as the flux redistribution are clearly demonstrated (see  Fig.~\ref{fig:BoloLC_ratios}). 
In \SCSN\ there is no increase in  UV flux at early times, unlike in normal SN~Ia.
The UV fraction of the bolometric flux is already declining when \SCSN\ were first observed, while the UV fraction increases until just before maximum light for normal SNe Ia.
The \SCSN\ also show a low optical ($\sim \lambda 4200 - 7300$) luminosity fraction and high NIR ($\sim \lambda 7300 - 16{,}000$) fraction compared to the normal SN~Ia 2007af.

\subsection{Hubble residuals}
\label{sect:Hubble}
While the exact rate of \SCSN\ is unknown they are extremely rare and make up a very small fraction of SNe~Ia in the local universe. However, they prefer low-mass and high specific star forming host galaxies both of which will increase their fraction in high-redshift SNe surveys. They do not follow the LWR, have broad light curves,  and some \SCSN\ are not over-luminous and overlap normal SNe~Ia in peak brightness.
Furthermore, in the rest frame $B$ and $V$ bands, it is difficult, if not impossible, to distinguish \SCSN\ from normal SNe~Ia.
Therefore, \SCSN\ have the potential to bias dark energy experiments. Although a full simulation of this is beyond the scope of this paper, we determine what the Hubble residuals would be if \SCSN\ were treated as normal SNe~Ia and fit with light-curve fitting tools. 

\begin{figure}
\centering
 \includegraphics[width=.45\textwidth]{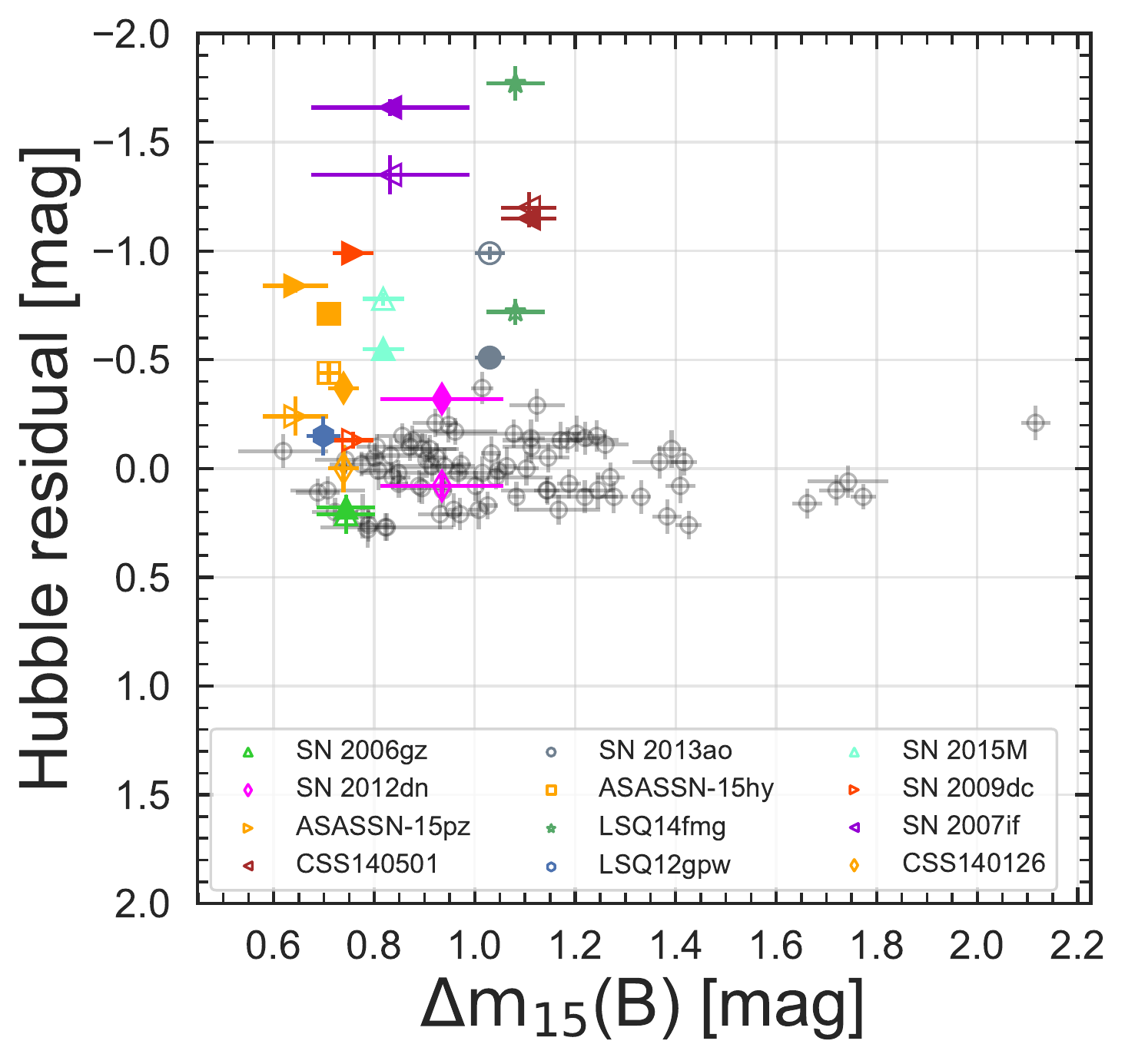}
\caption{Hubble residuals of a selection of normal SNe~Ia from the CSP (black open markers) and of the \SCSN\ fit with both the $B$ and $V$ bands (open symbols) and all available bands (solid symbols). The Hubble residuals are smaller if only the $B$ and $V$ bands are used. \label{fig:hubble}}
\end{figure}

To determine the Hubble residual for each \SCSN, we used the \textit{SNooPy} \textit{EBV\_model2}, with \textit{st} (which is the input setting for \sBV) as the light curve shape parameter.  To `simulate'  the effect of future dark energy experiments such as those from the Nancy Grace Roman Space Telescope, we fit the light curves from +0 to +30~d relative to $B$-band maximum. The light-curve fitting is done twice, once with only the $B$- and $V$- bands and then with all of the bands from UV to NIR.  For both fits, \red{most} of the \SCSN\ have negative Hubble residuals (see Table~\ref{table:hubble}). This implies that when run though light curve fitters, \SCSN\ are too bright for their light curve shape. 
Fig.~\ref{fig:hubble} presents the Hubble residuals as a function of light-curve decline rate. 
The mean residuals are $\Delta\mu(all)=-0.74\pm0.02$~mag and $\Delta\mu(BV)=-0.48\pm0.50$~mag.  The Hubble residuals are generally smaller and the light curves fit better to the templates when only the $B$ and $V$ bands are used. Thus, in the case that only rest-frame $B$ and $V$ bands are observed, \SCSN\ may not be identified and removed in dark energy experiments and will cause a bias. A more detailed simulation is warranted to determine the true extent of this contamination and is beyond the scope of this work.

\begin{deluxetable}{ c c c}
\tablewidth{\textwidth}
\tablecaption{The Hubble residual of the \SCSN\ in the sample. The values were calculated using \textit{SNooPy} \textit{EBV\_model2} and compared to the CMB corrected redshift distance. The cosmological parameters used for this section are  $H_{0}$=73\,km\,s$^{-1}$\,Mpc, $\Omega_{m}$=0.27 $\Omega_{\Lambda}$=0.73.  The Hubble residuals were computed using two cases: 1) $B$ and $V$ bands only and 2) all of the available bands ($uvuBVgrizYHJ$). \label{table:hubble}}
\tablehead{
\colhead{SN}&
\colhead{$\Delta\mu$($BV$)}&
\colhead{$\Delta\mu$($all$)}\\
\colhead{}&
\colhead{Mag}&
\colhead{Mag}}
\startdata
2003fg	&	$\cdots$			&	$\cdots$			\\
2006gz	&	0.21	$\pm$	0.09	&	0.18	$\pm$	0.04	\\
2012dn	&	0.08	$\pm$	0.05	&	$-$0.32	$\pm$	0.02	\\
ASASSN-15pz	&	$-$0.24	$\pm$	0.09	&	$-$0.84	$\pm$	0.03	\\
2007if	&	$-$1.35	$\pm$	0.09	&	$-$1.66	$\pm$	0.04	\\
2009dc	&	$-$0.13	$\pm$	0.04	&	$-$0.99	$\pm$	0.02	\\
LSQ12gpw	&	$-$0.15	$\pm$	0.09	&	$-$0.15	$\pm$	0.08	\\
2013ao	&	$-$0.99	$\pm$	0.03	&	$-$0.51	$\pm$	0.03	\\
CSS140126	&	0.00	$\pm$	0.11	&	$-$0.37	$\pm$	0.03	\\
CSS140501	&	$-$1.20	$\pm$	0.07	&	$-$1.15	$\pm$	0.04	\\
LSQ14fmg	&	$-$0.72	$\pm$	0.06	&	$-$1.77	$\pm$	0.08	\\
2015M	&	$-$0.78	$\pm$	0.03	&	$-$0.55	$\pm$	0.02	\\
ASASSN-15hy	&	$-$0.44	$\pm$	0.05	&	$-$0.71	$\pm$	0.03	\\
\enddata
\end{deluxetable}

\section{Spectroscopic properties}
\label{sect:spec}
In this section, the spectroscopic properties of \SCSN\ are presented. First, we concentrate on optical wavelength spectra and line identifications. Then we discuss the NIR spectra and their line identifications. The velocity and  pseudo equivalent width (pEW) measurements, and properties including the Branch diagram are then presented.

\subsection{Optical wavelength spectra}
\begin{figure*}
\centering
 \includegraphics[width=1.0\textwidth]{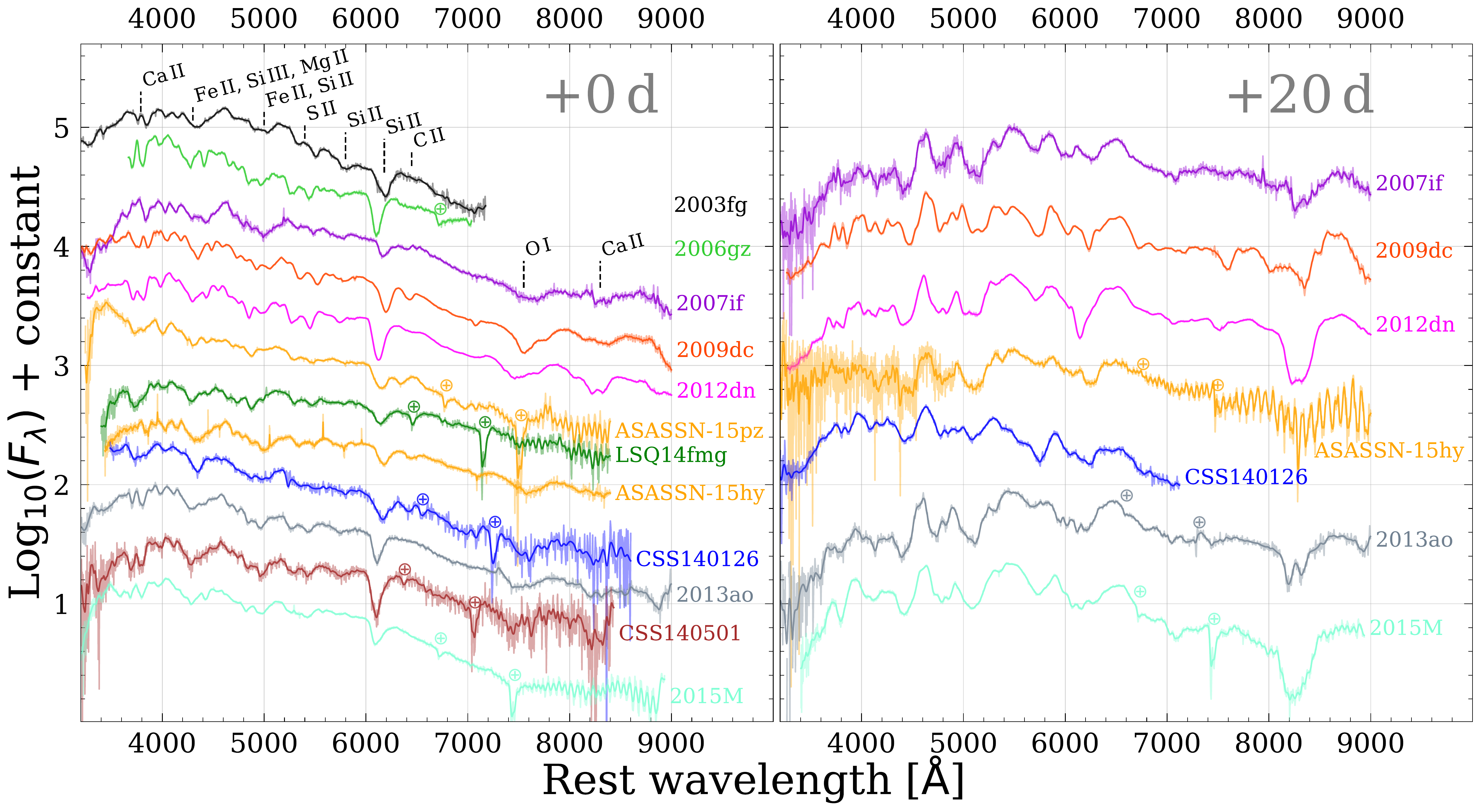}  
\caption{\textit{ Left panel:} Maximum light spectra of all of the \SCSN\ in the sample. \textit{Right panel:} \SCSN\ spectra at +20~d for objects which have spectra at this phase. }
\label{fig:maxspec}
\end{figure*}

All available maximum light and +20~d spectra of \SCSN\ are presented in Fig.~\ref{fig:maxspec}. 
At maximum light, the spectra 
show the standard lines associated with SNe~Ia \citep[\eg][]{Branch06,Ashall18}, see table \ref{table:ions} Many of the \SCSN\ also have strong \CII\ $\lambda$6580 and $\lambda$7234
features persisting through maximum light.  \SCSN\ also have weak \CaII\ features at this phase. 
By +20~d from maximum light the  \CaII\ feature is much stronger, and \red{the spectrum contains} no residual \CII.

\begin{figure*}
\centering
 \includegraphics[width=.99\textwidth]{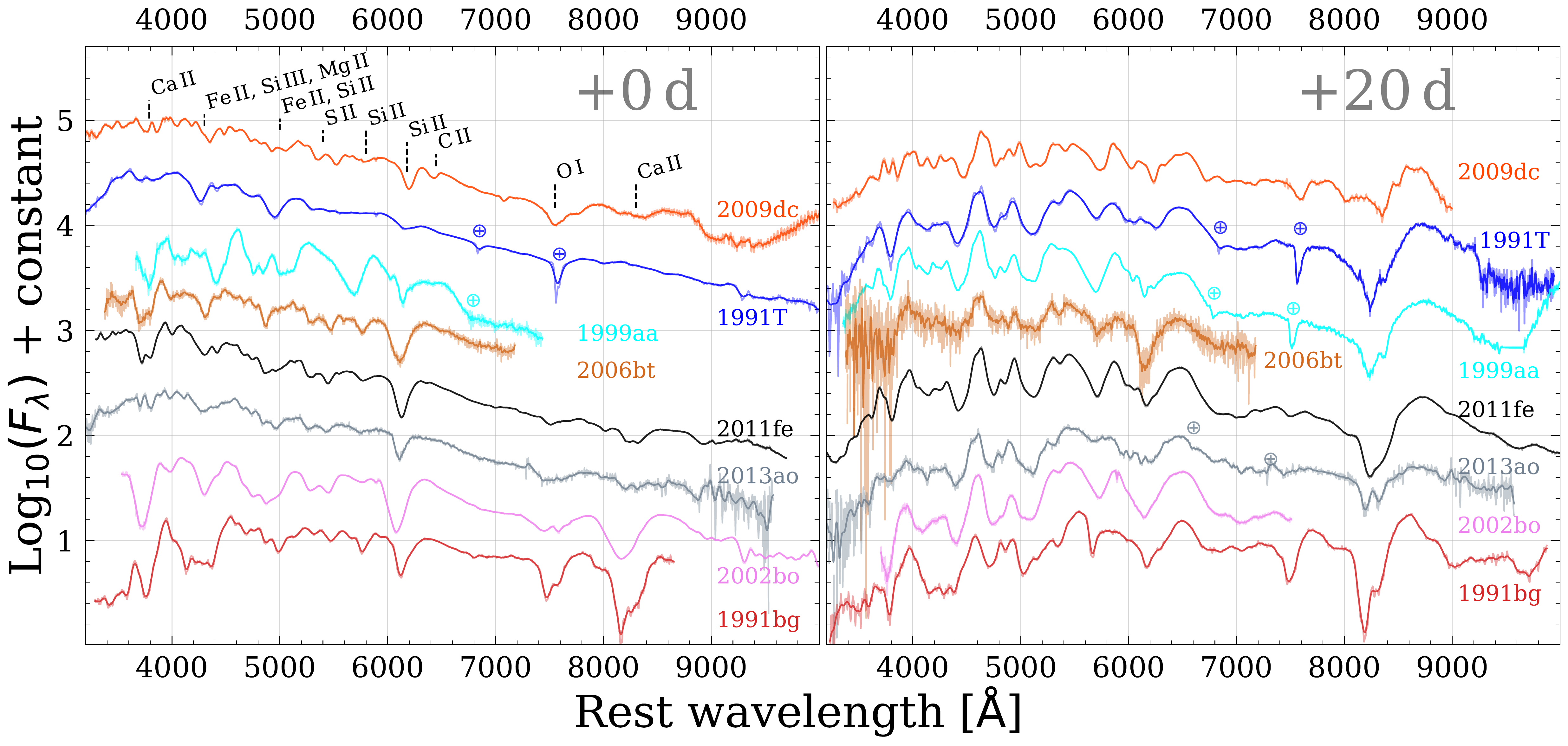}
\caption{Comparison between maximum light (\textit{left}) and +20~d (\textit{right}) spectra of the over-luminous 2003fg-like SN~2009dc and an under-luminous one (SN~2013ao) along with a selection of other SNe~Ia sub-types. }
\label{fig:maxspeccomp}
\end{figure*}

The maximum light and +20~d spectrum of SN\,2009dc and SN\,2013ao are compared to a variety of sub-types of SNe~Ia in Fig.~\ref{fig:maxspeccomp}.  SN\,2009dc and SN\,2013ao  were chosen for this comparison as they are located at the extreme ends of the luminosity parameter space. 
Both SN~2009dc and SN~2013ao appear to have slightly weaker or ``washed out'' spectral features compared to the other SNe.  
However, in terms of ionization state SN~2013ao appears to be most similar to SN\,2006bt \citep{Foley10}, and SN\,2011fe \citep{Mazzali14}. With maximum light spectra alone, it is almost impossible to distinguish between SN\,2013ao and the normal SN\,2011fe, and the only noticeable difference are weaker \CaII\ features in SN~2013ao. On the other hand, SN~2009dc has a similar ionization state to SN~2013ao, but also has strong \CII\ absorption and lower velocities, as is seen with the \SiII\ $\lambda$6355 feature. Although SN\,2009dc appears to be as blue as SN\,1991T, the lack of \FeIII\ and a stronger \SiII\ $\lambda$5972 feature demonstrate that the ionization state in the line forming region of SN~2009dc is lower. 

\citet{Hachinger12} claim these ``washed out'' features and the low ionization state in \SCSN\ are produced by an additional thermal luminosity source, which is H/He deficient. \citet{Hsiao20} suggest that this could be due to interaction with a C-O envelope in the core degenerate scenario. \citet{Taubenberger19} propose that it could be caused by the violent merger of two WDs, although the low continuum polarization makes the latter unlikely \citep{Tanaka10,Cikota19}.

\begin{deluxetable}{ c c }
\tablewidth{\textwidth}
\tablecaption{The main spectra lines identified in the \SCSN\ spectra at maximum light. \label{table:ions}}
\tablehead{
\colhead{Ion}&
\colhead{Wavelength}\\
\colhead{}&
\colhead{$\mathrm{\AA}$}}
\startdata
\CaII\	&	$\lambda\lambda$3968, 3933		\\
\SiII\	&$\lambda$4130		\\
\MgII\ & $\lambda$4481\\
\SiIII\ & $\lambda$4552\\
\FeII\ &$\lambda$5169 \\
\FeIII\ & $\lambda$5156\\
\SII\ &  $\lambda$5453,$\lambda$5606\\
\SiII\ & $\lambda$5972,$\lambda$6355\\
\OI\  &$\lambda$7771\\
\CaII\ & $\lambda\lambda$8498, 8542, 8662\\
\enddata
\end{deluxetable}

At +20~d, the spectra of all sub-types of SNe~Ia are similar. All of the \SCSN\ and comparison SNe~Ia except SN~2013ao have an emission feature in the 5900~\AA\ region. This feature has been attributed to either [\CoIII] 5888~\AA\ \citep{Dessart14} or Na~{\sc i}~D emission \citep{Mazzali08}.
If this feature is attributed to [\CoIII], the lack of this emission
in SN~2013ao may be caused by the lack of \Nifs\ above the
photosphere. This is consistent with the lack of an $H$-band break in the NIR spectra of SN~2013ao. Note however that SN~2009dc, SN~2015M, and ASASN-15hy do not have an $H$-band break at these epochs, but all show this emission feature at 5900~\AA. We thus conclude that the feature is more likely caused by Na~{\sc i}~D emission, which  is not seen in SN~2013ao due to higher temperature and density in the ejecta. Alternatively, differences in the progenitor configuration, including metallicity differences, could produce a reduced Na abundance.

\begin{figure}
\centering
 \includegraphics[width=.48\textwidth]{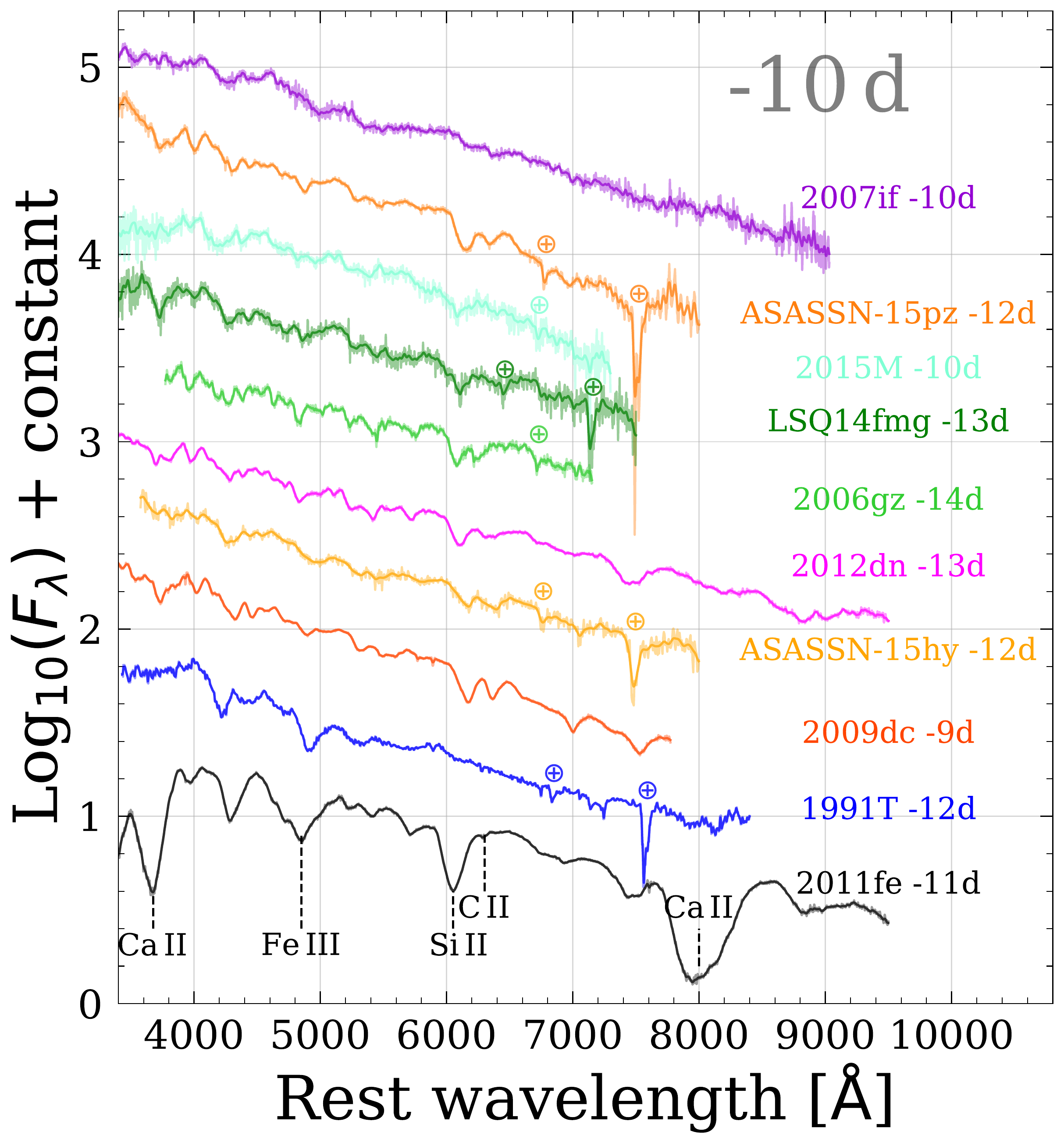}
\caption{Early-time spectroscopic comparison between \SCSN, the normal SN~2011fe, and the over-luminous SN~1991T. All \SCSN\ with adequate data tend to show strong \CII\ and no significant \FeIII.  In contrast,  SN~1991T has strong \FeIII\ and SN~2011fe generally has stronger features. In these phases, it is possible to distinguish \SCSN\ from other sub-types of SNe~Ia. }
\label{fig:m10spec}
\end{figure}

One of the easiest ways to distinguish between \SCSN\ and other
sub-types of SNe~Ia is with spectra at $-10$~d with respect to maximum
light (Fig.~\ref{fig:m10spec}).  At this epoch, \SCSN\ have weak
features and are dominated by continuum, \SiII\ absorption, and have
very weak or no \CaII\ and \FeIII\ features. On the other hand,
SN~2011fe is redder and has strong P Cygni profiles which include a
large amount of intermediate mass elements such as a strong \SiII\
$\lambda$6355 and \CaII\ features. Similar to \SCSN, SN\,1991T has a
hot continuum and no \CaII\ features.  However, SN\,1991T has a strong
\FeIII\ feature at $\sim$4900~\AA\ and no \CII\ absorption, which is
unlike \SCSN. With this in mind, we suggest that the lack of a strong
\FeIII\ absorption in early-time spectra should become  one of the  defining characteristics of \SCSN.

Several hundred days post explosion, 
the ejecta of SNe are optically thin and dominated by forbidden transitions (see \citealt{Taubenberger19}).
The ionization state of these lines provides critical information about the rate of recombination and the density in the center of the explosion.
Normal SNe~Ia have nebular spectra which are dominated by strong [\FeIII] at $\sim$4700~\AA\ and a weaker [\FeII] emission at $\sim$5200~\AA, with an [\FeII]/[\NiII]/[\CaII] emission at $\sim$7300~\AA\ \citep[\eg][]{Graham17}. Luminous SNe~Ia, such as SN~1991T, have a higher ionization state with strong [\FeIII] lines \citep{Cappellaro01}, and sub-luminous SNe~Ia tend to have a stronger [\CaII] and weaker [\FeIII] emission than normal SNe~Ia \citep[\eg][]{Mazzali12,Galbany19}.

Four \SCSN, SN~2006gz \citep{Maeda09},
SN~2007if \citep{Taubenberger13a}, SN~2009dc \citep{Taubenberger13a}, and SN~2012dn \citep{Taubenberger19} have nebular phase spectra. Interestingly, despite being luminous, all of these objects show a low ionization state with weak [\FeIII] emission. They also have strong [\CaII] emission. SN~2012dn (the least luminous of the four) has the strongest [\CaII] emission, significantly stronger than the other \SCSN. SN\,2012dn also shows [\OI] emission at 6300~\AA, which has only been observed in the nebular phase spectra of one other SN~Ia, the sub-luminous SN~2010lp \citep{Taubenberger:2013b}. 
Although sub-luminous and \SCSN\ sit at the opposite ends of the LWR they share similar traits in nebular phase spectra of low-ionization state and strong [\CaII] emission. However, in \SCSN\ the low ionization is likely caused by low ejecta velocities (\red{see section \ref{sect:vel}}), high central densities, and an increased recombination rate; whereas in the sub-luminous SNe~Ia (\eg\ SN~1986G; \citealt{Phillips87,Ashall16b}), the low ionization state is thought to be caused by less heating owing to their smaller \Nifs\ masses. 

\subsection{NIR spectra}
\label{sec:nir_spec}
NIR spectroscopy provides critical information on the physics of SNe~Ia \citep[\eg][]{Kirshner73,Marion09,Hsiao19}.
In the NIR, the photosphere recedes faster than at shorter wavelengths, allowing for deeper parts of the ejecta to be exposed at earlier times. The NIR also contains different ions than the optical, such as \CI\ 1.0693\mic\ and the $H$-band break \red{($\sim$1.4--1.9~\mic)}. 
If \SCSN\ contain a large carbon shell, it would be expected that as the ejecta cool, the ionization state of carbon would transition from singly ionized to neutral. Given that the carbon shell is large and generally dominated by \CII\ which is  seen up to and past maximum light it is expected that there would be strong NIR \CI\ well past maximum light. This is seen in  SN~2015M that shows a distinct \CI\ 1.0693\mic\ absorption at $\sim$11 000~\kms (see Fig.~\ref{fig:specNIR}). \CI\ may also be seen in   ASASSN-15hy (see \citealt{Lu21}).  For the \SCSN\ objects which do not show \CI, 
either the carbon has become optically thin, stays ionized  at all epochs, or the \CI\ line is very weak possibly due to the presence of He in the outer layers as discussed in the Appendix D of \citet{Lu21}.


The $H$-band break is formed from a multiplet of allowed \CoII, \FeII, and \NiII\ emission lines located well above the photosphere \citep{Wheeler98, Hoeflich02}. The strongest and bluest of these lines is \CoII\ 1.57\mic. The $H$-band break appears when the photosphere recedes into the \Nifs\ region. For normal SNe~Ia, this begins a few days after maximum light, and for sub-luminous SNe~Ia, the break emerges slightly later, at $\sim$+8\,d. The later appearance can be interpreted as the photosphere having to recede through more material to reach the \Nifs\ region. The strength of the $H$-band break correlates with light curve shape, where brighter SNe have a stronger break \citep{Hsiao13}. Furthermore, the velocity of the bluest edge (\ved) of the $H$-band region at $10\pm3$\,d can be used to directly measure the edge of the \Nifs\ region in a SN~Ia \citep{Ashall19a}, where more luminous SNe~Ia have larger values of \ved. \ved\ can be used to discriminate between SNe~Ia explosion models \citep{Ashall19b}. 

\begin{figure}
\centering
 \includegraphics[width=.5\textwidth]{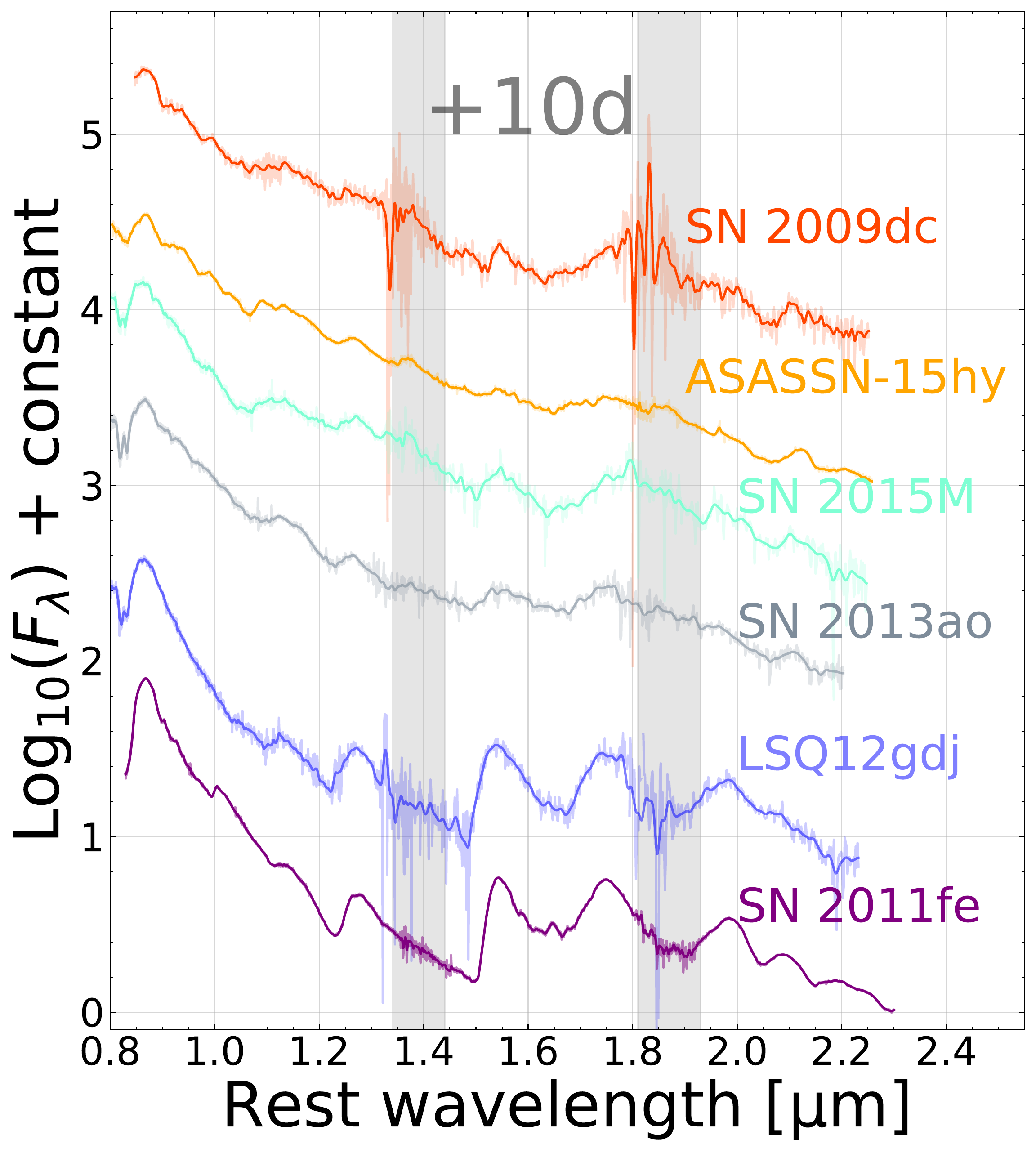}
\caption{The NIR spectra of four \SCSN, as well as the 1991T-like LSQ12gdj and the normal SN~2011fe around +10~d past maximum light. None of the four \SCSN\ show a strong $H$-band break. For presentation purposes the spectra have been interpolated with a gaussian filter having a 3-sigma smoothing length. }
\label{fig:HbreakNIr}
\end{figure}

Fig.~\ref{fig:HbreakNIr} shows the NIR spectra of the four \SCSN\ which have spectra at $10\pm3$\,d, as well as the 1991T-like LSQ12gdj and the normal SN\,2011fe. Unlike normal, sub-luminous, and 1991T-like SNe~Ia, \SCSN\ show a very weak or no $H$-band break by +10~d. 
The lack of an $H$-band break indicates that the photospheres of \SCSN\ have not receded into the \Nifs\ region by this time and that the mass above the \Nifs\ is large. As 1991T-like SNe~Ia have a higher ionization state but still have a strong $H$-band break, an ionization effect can be ruled out as the cause of their lack of $H$-band break.
We note that although sub-luminous SN~Ia have a weak $H$-band break it is intrinsically different from the $H$-band break in  \SCSN. SN~2009dc and ASASSN-15hy have NIR spectral observations that extend to +85 and +80~d past maximum, respectively. In these SNe the break appears at much later epochs. In SN~2009dc, the $H$-band break appears between +24 and +85~d \citep{Taubenberger11}, and in ASASN-15hy, it appears between +30~d and +80~d \citep{Lu21}. The delayed onset of the $H$-band break demonstrates that the \Nifs\ region is in the very inner layers of the ejecta and the photosphere has to recede through a large optically thick envelope before reaching the bulk of the \Nifs. 
\red{With this in mind in the next section we will measure the velocities and pEWs of the early time spectra to determine the chemical composition and structure of \SCSN. }

\subsection{Velocity and pEW fitting method}
\label{sect:vel}
For the optical wavelength spectra, the velocity minima and pEW  of the main spectral features were obtained using the Measure Intricate Spectral Features In Transient Spectra  (\texttt{misfits}\footnote{\url{http://github.com/sholmbo/misfits}}; S.~Holmbo et al., in prep.) code.
To acquire the minimum of a P Cygni absorption feature, the rest frame spectra were smoothed by passing them through a low-pass filter after Fourier transforming them to remove the high-frequency noise, as described in \citet{Marion09}. An error spectrum was computed by obtaining the differences between the observed spectra and the  Fourier transformed smoothed spectrum. The absolute values of the residuals are smoothed with a Gaussian function.  The corresponding Gaussian smoothed version that is scaled contains 68\% of the absolute value of the residual level and is used as the 1-sigma error spectrum.  

The \textit{velocity.gaussians} function was utilized to obtain the minimum of an absorption feature. The boundaries of the wavelength region were manually selected for each feature. A linear continuum and a single Gaussian function were simultaneously fit to the feature, and the best fit was determined by chi-squared minimization. To estimate the uncertainty, a Monte Carlo approach was adopted with 1000 realizations. The realizations were generated by including the flux uncertaintie assuming a normal distribution and the boundary uncertainties assuming a uniform distribution. The minimum wavelengths were converted to velocity using the relativistic Doppler formula and rest wavelength of the feature. The mean and the standard deviation of the velocities measured from the Monte Carlo realization were adopted as the value and the 1$\sigma$ uncertainty of the velocity, respectively.

The pEW of the features was calculated following the prescription of \citet{Garavini07}. The  \textit{width.shallowpew} function was used  in 
\texttt{misfits}, where uncertainties were determined with  the same Monte Carlo method mentioned above. 
The mean value and standard deviation were taken as the pEW and its 1-$\sigma$ uncertainty.


\subsection{Velocity and pEW measurements}
\begin{figure}
\centering
 \includegraphics[width=.5\textwidth]{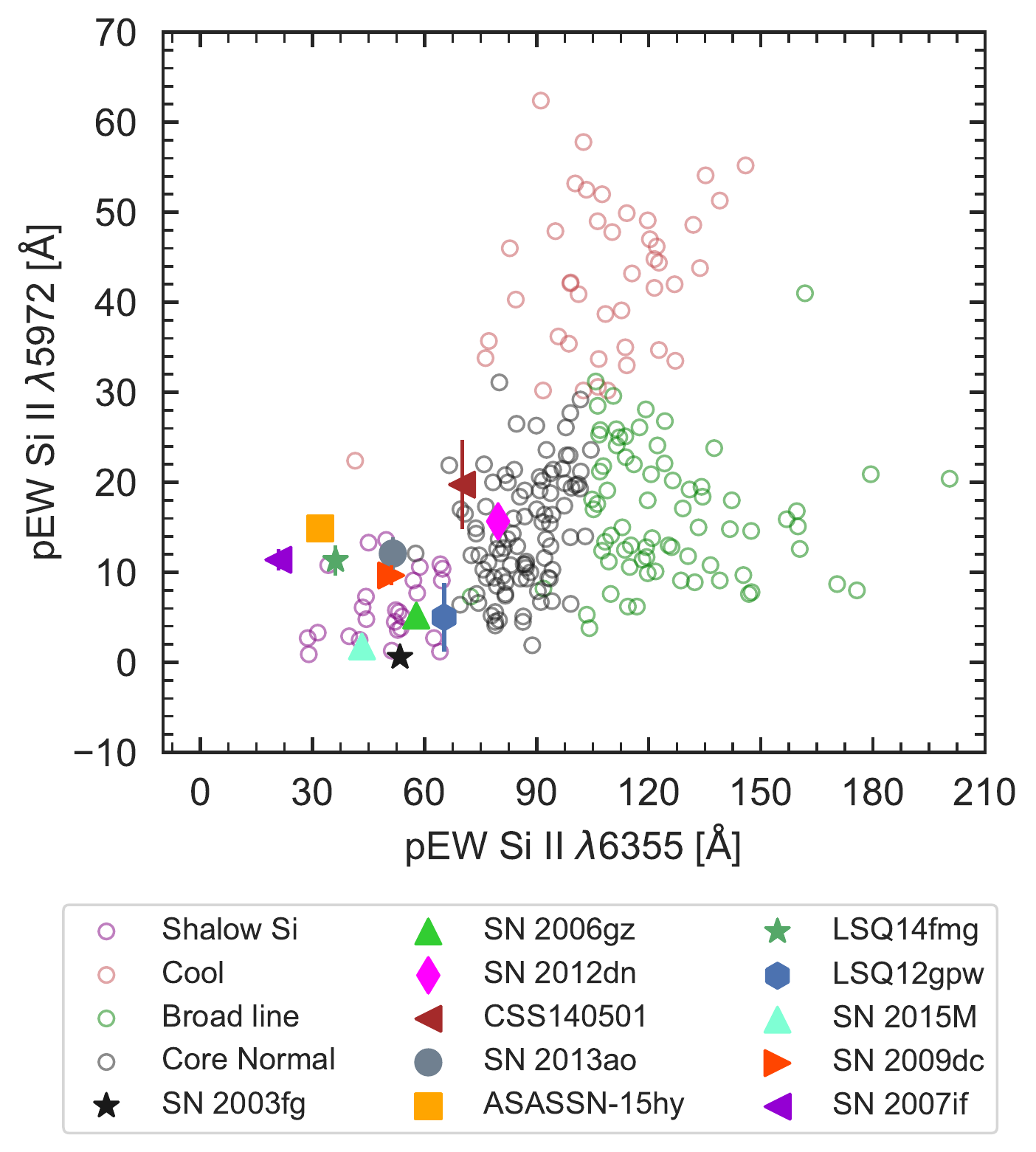}
\caption{The Branch diagram produced using SN~Ia data from \citet{Blondin12} and \citet{Folatelli13}. Most \SCSN\ are located within the same area as the shallow silicon (SS) SNe~Ia.}
\label{fig:branch}
\end{figure}

In this work, the pEW of \SiII\ $\lambda$5972, $\lambda$6355 and \CII\ $\lambda$6580 were measured, as well as the velocity of \SiII\ $\lambda$6355 and \CII\ $\lambda$6580. The pEW of the $\lambda$6355 and $\lambda$5972 \SiII\ features have been shown to be a powerful diagnostic to identify SN~Ia sub-types, where normal SNe~Ia can be separated into four  groups, \citep[core normal, shallow Si, broad line and cool][]{Branch06,Burrow20}.  In this parameter space, most \SCSN\ are located primarily in the shallow-silicon area close to over luminous objects such as 1991T-like objects (see Fig.~\ref{fig:branch}). However, SN\,2012dn and CSS140501 are in a similar area as core normal SNe.
These two SNe also overlap with the normal population in the LWR (Fig.~\ref{fig:LWR}).

\begin{figure*}
\centering
 \includegraphics[width=.99\textwidth]{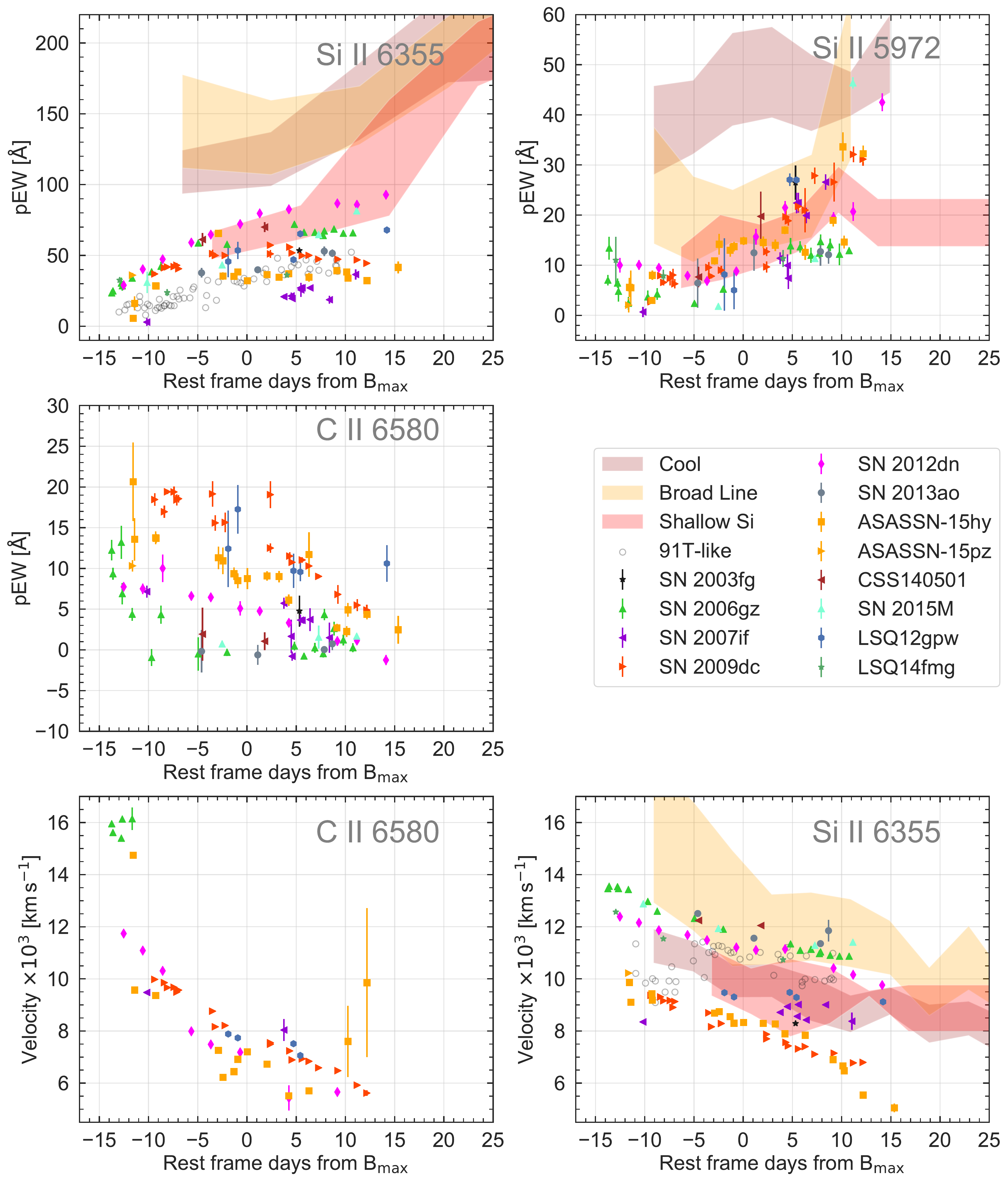}
\caption{pEW  and velocity measurements as a function of phase.
\textit{Top left:} \SiII\ $\lambda$6355 pEW as a function of phase relative to maximum.
\textit{Top right:} \SiII\ $\lambda$5972 pEW as a function of phase relative to maximum.
\textit{Middle left:} \CII\ $\lambda$6580 pEW as a function of phase from maximum.
\textit{Bottom left:} \CII\ $\lambda$6580 velocity as a function of phase relative to  maximum.
\textit{Bottom right:} \SiII\ $\lambda$6355 velocity as a function of phase relative to maximum. }
\label{fig:spec_vel}
\end{figure*}

The bottom panels of Fig.~\ref{fig:spec_vel} 
show the velocity of the \CII\ $\lambda$6580 and \SiII\ $\lambda$6355\ features as a function of time. 
The velocities decrease over time as would be expected from a homologous expansion and a receding photosphere. The velocity spread in \SiII\ $\lambda$6355\ between the fastest and slowest \SCSN\ is 5,000~\kms.
At the earliest epochs, around $-$10~d, the velocities range from $\sim$13,500~\kms\ (SN~2006gz) to $\sim$8,000~\kms\ (SN~2007if), and by maximum light the spread has decreased to 11,500~\kms\ for the fastest expansion and 7,500\kms\ for the slowest. 
Despite the large spread, some \SCSN\ exhibit some of the slowest
velocities of any SN~Ia. The change in velocity from the $-$10\,d to
maximum light roughly indicates the depth of the Si shell. The change in \SiII\ $\lambda$6355 velocity between --10~d to maximum light ranges
from 2,000~\kms\ in SN~2006gz, to 1,000~\kms\ in SN~2012dn, and
0~\kms\ in SN~2007if. 

The \SCSN\ with the fastest \SiII\ velocities are consistent with the ``shallow Silicon" (SS) SNe~Ia from \citet{Folatelli13}.  They are also consistent with the velocities of SN~1991T-like SNe (M.~Phillips et al., in preparation). 
However, the lower velocities and early-time \red{flatter} \SiII\ evolution are unusual compared to normal SNe~Ia. 
In some cases, \SCSN\ may have a very confined intermediate mass element layer in velocity space, and do not show a rapid drop in velocities at the earliest phases. 
This lack of an early drop may be caused by a \red{compression of the Si shell}, due to it running into an envelope \citep{Quimby06}.
In SNe~Ia when the velocity measurements reach a minimum and stay at that value for a prolonged time, it usually requires the photosphere to pass through the base of the layer.
By +10~d past maximum light the Si velocity is still declining. 
However, we note that features are susceptible to ionization changes and the bottom of the \SiII\ layer does not always correspond to the bottom of the Si-region. 
Furthermore, the red side of the \SiII\ feature becomes contaminated by \FeII\ lines after maximum light. For most \SCSN\ this is after +10\,d but for ASASSN-15hy this occurs 2-3\,d past maximum \citep{Lu21}. This can artificially produce a sudden velocity drop between 0 and +10~d.


The velocities of the \CII\ feature range between 10,000 to 16,000~\kms\ at $-$10~d to 8,000~\kms\ at maximum light. For some of the \SCSN\ (LSQ12gpw, SN~2012dn, and ASASSN-15hy) the velocity of the \CII\ feature is lower than that of \SiII. This may be an indication of mixing of the C and Si layers, or it may be a projected velocity effect where the \SiII\ is located well above the photosphere, but the \CII\ is located close to the photosphere, ensuring that most of the absorption is produced from material that is not directly moving towards the observer \citep{Hoeflich90}. 

The top two panels and middle panel of Fig.~\ref{fig:spec_vel} contain the pEW measurement of \SiII\ $\lambda$5972, $\lambda$6355 and \CII\ $\lambda$6580. The \SiII\ $\lambda$6355 feature slowly increases in pEW over time for all objects.
At early times, around $-$10\,d, the pEW of \SiII\ $\lambda$6355
ranges from 5-50~\AA, and rises to 20-90~\AA\ by +10~d. The pEW of the
\SCSN\ cover a larger range in values than 1991T-like SNe, which range from 0-20~\AA\ at early times to 30-50~\AA\ by +10~d, and from shallow Si objects which cover a range of 0-50~\AA\ at maximum light. 

The pEW measurements of the \SiII\ $\lambda$5972 feature range from 0-10~\AA\ at $-$10~d, to 10-30~\AA\ at +10~d and  follow a similar trend to \citet{Branch06} SS SNe. Generally, the increasing pEW of this feature is interpreted as a cooling photosphere and the \SiII\ $\lambda$5972 line getting populated due to the recombination of \SiIII\ \citep[\eg][]{Hachinger08,Ashall18}. 

The pEW of \CII\ $\lambda$6580 is more difficult to measure as it sits on the top of the re-emission  of the \SiII\ $\lambda$6355 feature. The pEW decreases over time for all SNe, except SN~2006gz, SN~2013ao, CSS140501, and SN~2015M which have pEW values consistent with 0. 
The pEW measurments ranges from 5-20~\AA\ at early times and 0-5~\AA\ at +10\,d. Three of the \SCSN\ (SN~2009dc, LSQ12gpw, and ASASSN-15hy) have persistent \CII\ features well past maximum light. Interestingly, these SNe also have the slowest \SiII\ $\lambda$6355 velocities and the broadest light curves. These correlations will be discussed in more detail in the next section.  Note that the pEW of  \CII\ $\lambda$6850 region was measured even if no absorption feature was visible in the spectra, hence some SNe have values consistent with zero.

\section{Parameter study}
\label{sect:correlations}
\begin{figure*}
\centering
 \includegraphics[width=1.0\textwidth]{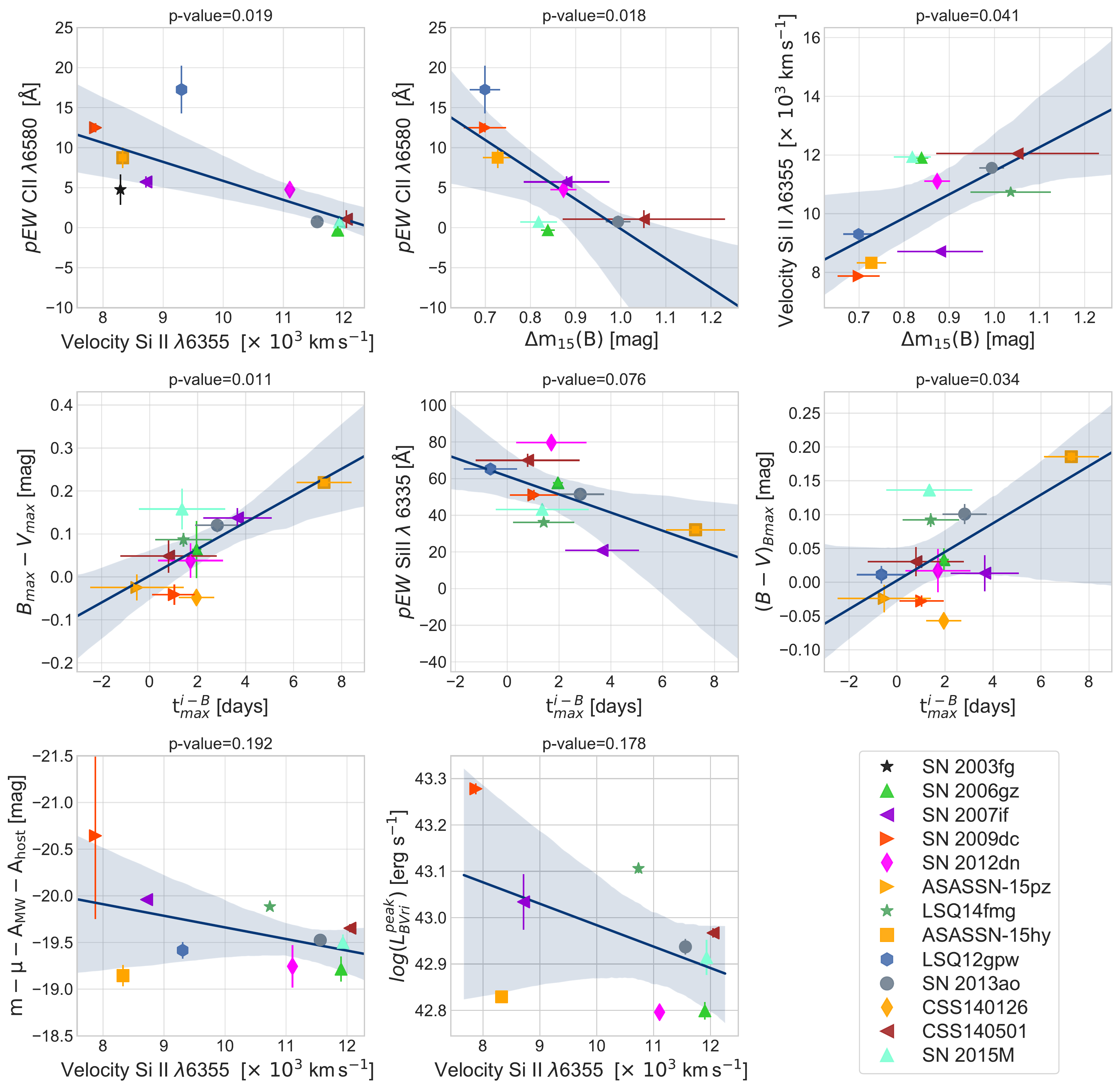}
\caption{Correlation plots between measured parameters of \SCSN. Above each panel the p-value is given. All relationships are statistically significant except for the lower left and lower middle panels between $M_{B}$, L$^{peak}_{BVri}$ and Si velocity. For each panel a line of best fit determined by a least-squares technique is provided along with a 1-sigma uncertainty shaded region. It should be noted that \iBmax\ in this plot has been K-corrected. A full pair-plot of all parameters can be found in Appendix \ref{sec:corelationA}.}
\label{fig:corela}
\end{figure*}

Having measured and presented the main parameters of the \SCSN, we now turn our attention to the correlations between these parameters and what these may imply about the physics of progenitors and explosion mechanisms. Fig.~\ref{fig:corela} presents six of the most significant correlations that were found within our data set, as well as two important non-correlations. For each correlation the least-squared best-fit line  is given along with the 1-sigma uncertainty region. We show the \SiII\ $\lambda$6355 velocity and the \CII\ $pEW$, both obtained within 3 days of $B$-maximum, \iBmax, the color $(B-V)$ at B-band maximum corrected only for Milky Way extinction, the color $(B_{\mathrm{max}}-V_{\mathrm{max}})$, and the $B$-band and pseudo-bolometric peak magnitude and flux corrected for host galaxy extinction. 
The p-value for each fit is provided above each panel.

Arguably the most interesting correlation is between the $pEW$ of \CII\ and the velocity of the \SiII\ $\lambda$6355, both taken at maximum light.
SNe with slower \SiII\ velocities tend to have larger values of \CII\ $pEW$. Slower Si velocities could be produced by a WD exploding inside a carbon-rich envelope. The ejecta would slow down as they run into the envelope \citep{Noebauer16}. In a simplistic picture,
a large envelope mass would produce lower ejecta velocities, as more kinetic energy would be deposited into a more massive envelope. 
A larger envelope mass would also produce longer diffusion times and broader light curves, as is seen in the top middle panel of Fig.~\ref{fig:corela}, where there is a correlation between \CII\ $pEW$ and $\DmB$. \SCSN\ with broader light curves have larger \CII\ features and slower \SiII\ $\lambda$6355 velocities. These three correlations point to a non-degenerate carbon-rich envelope as being the dominant cause of the observed diversity between \SCSN. Interestingly, these correlations are not seen in normal SNe~Ia, demonstrating that normal SN~Ia are not produced via the envelope model. For example, faster declining SNe have the slowest ejecta velocities  \citep[\eg][]{Benetti05,Gall18,Ashall18,Galbany19}.


The unique $i$-band behavior sets \SCSN\ apart from other luminous events. \SCSN\ generally show no secondary $i$-band maximum.
The weak $i$-band secondary requires either full mixing in the ejecta or a lack of recombination of the Fe-group layers above the photosphere \citep{Hoeflich02,Kasen06,Jack15}. In the case of \SCSN, 
the lack of an $H$-band break at +10\,d indicates that the photosphere is not within the \Nifs\ region and therefore the ejecta cannot be fully mixed. Instead, the weak $i$-band secondary maximum may be caused by a lack of recombination of Fe-group elements.  The timing of the $i$-band maximum (\iBmax)\footnote{Note that in Fig.~\ref{fig:corela} \iBmax\ is obtained from the rest frame K-corrected data, whereas in Fig.~\ref{fig:sBViband} the data are not K-corrected to be consistent with the definition from \citet{Ashall20}.} allows for \SCSN\ to be distinguished from other SNe~Ia \citep{Ashall20}.  \SCSN\ with larger values of \iBmax\ tend to have lower values of \SiII\ $\lambda$6355 $pEW$. This is in contradiction with the trend seen between subluminous to normal SN~Ia, where a lower value of \iBmax\ and stronger $i$-band secondary maximum are correlated with a higher ionization state and faster velocities \citep[\eg][]{Kasen06}. This is seen in 1991T-like SNe, but not in \SCSN\ that show a weaker \SiII\ $\lambda$6355 $pEW$ feature and larger values of \iBmax, possibly caused by longer diffusion timescales through the large carbon-rich outer envelope.  

The \iBmax\ parameter is also found to be correlated with the observed $(B-V)$ color at maximum light, and the difference between the $B_{max}$ and $V_{max}$ magnitudes (middle left and middle right panels of Fig.~\ref{fig:corela}). In both cases, the photometry was not corrected for host-galaxy extinction. Redder \SCSN\ have larger values of \iBmax\ indicating cooler ejecta and a larger re-processing of flux toward redder wavelengths. This is consistent with a homologous expansion which is adiabatically cooling.  Line blanketing in the UV can also reprocess flux into redder wavelengths through fluorescence. It causes redder colors in SNe \citep{Mazzali00,Lentz00}. As the line blanketing is due to the presence of heavy elements in the outer layers, differences in the magnitude of the line blanketing (resulting in differences in the UV-optical colors) could be caused by differences in the metallicity of the progenitor or the shape of the outer density profile \citep{Walker12}. \red{We note that the three correlations in the middle panels of Fig. \ref{fig:corela} may be largely driven by ASASSN-15hy.}

It should be noted, however, that there is no statistically significant correlation between the peak  ($B$-band or bolometric) luminosity (both corrected and not corrected for host galaxy extinction) and the \SiII\ velocity at maximum light. This is inconsistent with predictions from the super-$M_{Ch}$ scenario \citep{Howell06}.
The lack of a correlation here implies that more than just the mass of the exploding WD drives the luminous display. 


Overall the correlations in Fig.~\ref{fig:corela} are consistent with a degenerate core exploding inside a carbon-rich envelope. This could occur in the core-degenerate scenario. We discuss this further in Section~\ref{sect:discussion}.

\section{Discussion}
\label{sect:discussion}


Here we place into context our findings relative to three leading models of \SCSN. 
One leading model consists of the disruption of a   C-O WD  that exceeds the $M_{Ch}$ limit due to rapid rotation and/or high magnetic fields \citep{Yoon05,Das13}. 
Alternatively the merger of two WDs could produce   SNe~Ia  exceeding the $M_{Ch}$-mass limit \citep[e.g.,][]{Scalzo10}. 
Finally, another viable model may be the disruption of a C-O degenerate core within a dense circumstellar material  environment  \citep{Hachinger12,Noebauer16}. This is also referred to as an envelope model \citep{Hoeflich:Khokhlov:96}. Such an explosion could be associated with an explosion of a  degenerate core of an asymptotic giant branch (AGB) star (i.e. in the core degenerate scenario; \citealt{Hsiao20, Lu21}), or from a C-O WD explosion with surrounding circumstellar dust \citep{Nagao17,Nagao18}.
We discuss each of these models below.


\subsection{Super-M$_{Ch}$ WD}
A single WD may exceed the $M_{ch}$ limit due to rapid rotation or high magnetic fields \citep{Yoon05,Das13}. The mass limit of such models is thought to be 1.8\Msun \citep{Yoon05}. A C/O WD exploding at such masses may be able to produce enough \Nifs\ to power the extreme luminosities observed in some \SCSN. However, due to this scenario requiring a detonation as the explosion mechanism, it has problems producing the large amounts of unburnt carbon, and intermediate mass elements, as well as the low ionization observed in the maximum light spectra \citep{Hoeflich:Khokhlov:96}.  

In the Super-M$_{Ch}$ WD scenario it was predicted that an increased total WD mass should result in a larger \Nifs\ mass and a larger binding energy which would result in higher diffusion time scales, higher luminosities, lower kinetic energies, and lower ejecta velocities \citep{Howell06}. However, in our sample there is no correlation between \SiII\ $\lambda$6355 velocity and peak $B$-band or bolometric magnitude\footnote{Note that the distances to the \SCSN\ are well determined and not the cause of this lack of correlation.} (as a proxy for binding energy and \Nifs\ mass), or between $\DmB$ and  $B$-band or bolometric magnitude (both as a proxy for WD mass). Both of these correlations would be expected if the driving parameter amongst \SCSN\ was ejecta mass.  Furthermore, to produce the luminosity of the most luminous \SCSN, such as SN~2003fg, requires an ejecta mass of 2.1\Msun\ \citep{Howell06}. This is above the mass limit (1.8\Msun) of a single super Super-M$_{Ch}$ WD \citep{Yoon05}.

\subsection{Dynamically merging WDs}
 In the double-degenerate scenario, two WDs may dynamically merge and produce a large \Nifs\ mass. The advantage of this scenario is that the total summed mass of the two WDs can exceed  1.8\Msun. The low levels of continuum polarization in the two \SCSN\ that have data, SN~2007if and SN~2009dc \citep{Tanaka10, Cikota19}, make dynamical mergers an unlikely avenue to produce these \SCSN, as such models are highly aspherical \citep{Bulla16}. 
As there are many varieties of WD mass which may merge, as well as off center \Nifs\ distributions from  dynamical merger models,  it would also be expected that dynamical mergers do not produce the correlations seen in the data.

\subsection{Envelope model}
The data and correlations presented in this work (Fig.~\ref{fig:corela}) are consistent with the hydrogen (and possibly helium) free envelope model.  In such model, a C/O WD explodes within a non-degenerate C-rich envelope (e.g., \citealt{Hoeflich:Khokhlov:96}). For a given WD mass, a more massive envelope would produce stronger carbon lines, lower Si velocities, and longer diffusion time scales. This is evident in the correlations between \SiII\ $\lambda$6355 velocity  and \CII\ 6580\AA\ pEW (which is a proxy for envelope mass) and $\DmB$ (which is a proxy for diffusion time).  The more massive carbon envelope would also produce the covering mass above the \Nifs\ region which would naturally explain the observed very late onset of the $H$-band break. Furthermore, there are multiple factors affecting the luminosity.  For a given WD mass, varying the envelope mass would produce a correlation between the expansion velocity and luminosity. In the case of a more massive envelope, the exploding WD would have more mass to deposit its energy into and decrease its speed, and this deposited energy would be converted into luminosity. Thus, both envelope mass and the \Nifs\ mass contribute to the observed luminosity. Additional factors may be the flame propagation speed (e.g., deflagration or detonation) as discussed in \citet{Lu21}. Such a model may also be referred to as the deflagration-core-degenerate scenario. In the envelope scenario, reprocessing of the flux from the optical to NIR in the envelope would also produce the high NIR flux observed. 

%


A viable progenitor scenario within the envelope model configuration is the core degenerate scenario.  The core degenerate scenario is the explosion of the degenerate C/O core in the center of an AGB star. The signature of a superwind detected in the observations of LSQ14fmg provides a compelling link to an AGB progenitor \citep{Hsiao20}. This class of models provides results which match the observational properties of both LSQ14fmg \citep{Hsiao20} and ASASSN-15hy \citep{Lu21}.

Finally, in the core degenerate scenario there should be significant
X-ray luminosity \citep{Lu21}. This should be searched for  in future nearby events.  Another prediction of the core degenerate scenario is the formation of CO in the high-density and low-temperature non-degenerate envelope. It has been proposed that active CO formation manifests as the observed rapid decline in the optical light curve at various phases in SN~2009dc, SN~2012dn, LSQ14fmg, and CSS140126 \citep{Hsiao20}. The timing of this drop is dictated by the envelope's ability to cool and is correlated with the envelope mass as indicated by the minimum of the \SiII\ $\lambda$6355 velocity \citep{Quimby06,Hsiao20}. Faster expanding ejecta cool faster. Another prediction from the core-degenerate scenario is an interaction with previous superwind episodes of the AGB star which may occur between 1 to 10 years after the SN explosion \citep{Hsiao20}. Observationally this appear as a UV late time re-brightening \citep[\eg][]{Graham19}.

One important parameter for core-degenerate models is low metallicity ($Z < Z_{\odot}^{-4}$) \citep{Lu21}, which is also seen at the local environment of \SCSN\ \citep{Galbany21}. These low metallicities may come from population I or II stars. \red{Contrary to normal SNe~Ia, \SCSN\ show no increase in UV flux ratio at early times.}  This is possibly caused by low metallicity of the progenitor and reduced line blanketing in the outer ejecta.

\section{Conclusion}
\label{sect:conclusion}
This paper presents a homogeneous sample of nine \SCSN\ observed by the Carnegie Supernova Project I \& II, which are analyzed in addition with 4 objects from the literature. This is the most complete \SCSN\ dataset to date.

Photometrically not all \SCSN\ are over-luminous. In fact in the optical ($B$ and $V$ bands), they populate the main part of the LWR with  absolute $B$-band magnitudes between $\sim-$19 to $\sim-$21~mag. \SCSN\ begin to differ from normal SNe~Ia in the redder bands. In the $i$ band, \SCSN\ peak after time of $B$-band maximum and have weak secondary maxima. In the NIR bands, \SCSN\ are  unique and are at least 1~mag brighter than normal SNe~Ia with the same optical light curve shape. Furthermore, their rise in the $H$ band can be up to 40~d longer than in the $B$ band.

Light-curve fitters determine that \SCSN\ have negative Hubble residuals; i.e. they are too bright for their light curve shape.
As \SCSN\ preferentially explode in low mass, low metallicity and high specific star-forming galaxies they are more prevalent in the high-redshift universe. \textit{Therefore, due to the similarity between normal and \SCSN\ in the bluer bands ($B$\& $V$), future high-redshift cosmological surveys should ensure they obtain rest-frame NIR observations in order to minimize bias introduced by the contamination
of \SCSN.} As this may not always be possible, it is important to carefully study 2003fg-like SNe to fully understand the bias they will cause in SN cosmology.




Optical spectra of \SCSN\ are similar to that of normal SNe~Ia, but most have strong carbon absorption well past maximum light, as well as low velocity gradients before maximum light. In the NIR, \SCSN\ do not show a distinct $H$-band break at $\sim$10~d. In \SCSN, this $H$-band break is not visible until beyond 70~d past maximum light. 

With our large sample of \SCSN\ we  find that the  ubiquitous characteristics of all  \SCSN\ are:
\begin{itemize}
    \item A broad optical light curve shape ($\DmB<1.3$~mag).
    \item The primary $i$-band peaks after the phase of $B$-band maximum.
    \item A lack of strong \FeIII\ features in the  early spectra
    \item A peak $H$-band absolute magnitude brighter than $-$19~mag.
    \item Carbon absorption at early times ($-$10~d from maximum light).
    \item No clear $H$-band break at +10~d from maximum light.
\end{itemize}
These criteria should be  used in future studies to determine if a SN is truly 2003fg-like.

In \SCSN\ the luminous long-rising NIR light curves may be caused by the reprocessing of flux to the NIR as the result of an explosion inside a massive envelope. The lack of  an early $H$-band break also demonstrates that the photosphere is not within the \Nifs\ region until a much later epoch. 
These observations provide direct evidence that there is a significant amount of ejecta above the \Nifs\ region.

A number of unique and interesting correlations were found within our dataset. 
There are strong correlations between the pEW of the \CII\ feature at maximum light, the \SiII\ velocity at maximum light, and $\DmB$. \SCSN\ with larger \CII\ pEWs have slower \SiII\ velocities at maximum light and broader light curves. These correlations are fully consistent with an envelope model where a C/O degenerate star explodes within an envelope. 
In such a configuration for a given degenerate core mass a larger envelope mass would produce slower Si velocities and larger diffusion time scales. 
Given that there are no H or He lines in \SCSN\ spectra it is likely that this envelope is carbon/oxygen dominated.   One promising progenitor scenario and explosion mechanism is the core degenerate scenario \citep{Kashi11,Hsiao20,Lu21}.

The data presented here provide a new critical piece of information in determining the source of diversity of \SCSN\ and the nature of SNe~Ia in general. It is clear that \SCSN\ are far more diverse than previously thought. Only with high-precision observations extending from  optical through NIR wavelengths can the physics be clearly understood. It is apparent that simply changing the mass of the exploding WD will not produce all of the observational characteristics. Our data are consistent with a degenerate core exploding within a carbon-rich envelope, with the core degenerate scenario providing one of the strongest paths to produce 2003fg-like SN events.

\vspace{0.5cm}
\begin{acknowledgements}
The authors would like to thank Vanessa D\'{i}az for helping with data visualization. 
CA and BJS are supported by NASA grant 80NSSC19K1717 and NSF grants AST-1920392 and AST-1911074.
M.S. and F.T. are supported by grants from the Villum FONDEN (28021) and the Independent Research Fund Denmark (8021-00170B).
E.B. was supported in part by NASA grant 80NSSC20K0538.
 N.B.S. acknowledges support from  the Texas A\&M University Mitchell/Heep/Munnerlyn Chair in Observational Astronomy.
L.G. was funded by the European Union's Horizon 2020 research and innovation programme under the Marie Sk\l{}odowska-Curie grant agreement No. 839090.
The CSP has been funded by the NSF under grants AST-0306969, AST-0607438, AST-1008343, AST-1613426, AST-1613455, and AST-1613472, and in part by  a Sapere Aude Level 2 grant funded by the Danish Agency for Science and Technology and Innovation  (PI M.S.).
Time domain research by D.J.S. is supported by NSF grants AST-1821987, 1813466, \& 1908972, and by the Heising-Simons Foundation under grant \#2020-1864. 
Based on observations made with the Nordic Optical Telescope, owned in collaboration by the University of Turku and Aarhus University, and operated jointly by Aarhus University, the University of Turku and the University of Oslo, representing Denmark, Finland and Norway, the University of Iceland and Stockholm University at the Observatorio del Roque de los Muchachos, La Palma, Spain, of the Instituto de Astrofisica de Canarias.
\end{acknowledgements}

\facilities:  Magellan, du Pont,  Swope, Nordic Optical Telescope 

\software IRAF \citep{Tody86,Tody93}, SNooPy \citep{Burns11}, Astropy \citep{Astropy13,Astropy18}, IDL Astronomy user's library \citep{Landsman95} and misfits (S. Holmbo et al., in prep.). 

\clearpage

\bibliographystyle{aasjournal}
\bibliography{hbreak,new}

\appendix
\setcounter{figure}{0}
\setcounter{table}{0}
\renewcommand{\thefigure}{A\arabic{figure}}
\renewcommand{\theHfigure}{A\arabic{figure}}
\renewcommand{\thetable}{A\arabic{table}}
\renewcommand{\theHtable}{A\arabic{table}}

\section{Observational Details}
\subsection{Discovery Information}
\label{sec:Discoveryinfo}

The discovery information of the five \SCSN\ with CSP-II data that are not previously published is outlined here:

LSQ12gpw was discovered by the La Silla-QUEST Low Redshift Supernova Survey \citep[LSQ;][]{Baltay13}. The earliest images where the object is present were obtained on 2012 November 16.1 and 16.2 UT yielding 20.9 and 20.8 mag, respectively. The image of last non-detection was taken on 2012 November 14.2 UT with a limiting magnitude of 21.5 mag. The classification spectrum was taken on 2012 December 6.25 UT with EFOSC2 on the ESO New Technology Telescope as part of the Public ESO Spectroscopic Survey for Transient Objects \citep[PESSTO;][]{Smartt15}. The spectrum resembles that of SN~2006gz \citep{Maguire12}. The rising light curves of LSQ12gpw have been examined in detail by \citet{Firth15} and \citet{Jiang18}. The Swope light curves have also been published by \citet{Walker15}. Here, we present the same dataset but calibrated using the procedures outlined in \citet{Phillips19}.

SN~2013ao was discovered by the Catalina Real-Time Transient Survey \citep[CRTS;][]{Drake09} and was originally designated CSS130315:114445-203140 and SSS130304:114445-203141. The SN was discovered in an image taken on 2013 March 4.77 UT yielding a discovery magnitude of 17.0 mag. There is no strong constraint on the explosion date from CRTS as the previous image of the object was taken more than 8 months prior \citep{Drake13}. The classification spectrum was taken on 2013 March 6.21 UT by PESSTO and resembles that of pre-maximum normal SNe~Ia \citep{Inserra13}.

CSS140126:120307-010132 or CSS140126 for short was also discovered by CRTS using multiple images obtained on 2014 January 2.5 UT, yielding a discovery magnitude of 18.8 mag. There is no strong constraint on the explosion date from CRTS as the previous image of the object was taken more than 6 months prior. The classification spectrum was taken on 2014 January 4.32 UT by PESSTO and resembles that of 1991T-like SNe~Ia \citep{Fleury14}. 

CSS140501:170414+174839 or CSS140501 for short was also discovered by the CRTS using multiple images obtained on 2014 May 1.5 UT, yielding a discovery magnitude of 18.7 mag. The SN is also present in multiple images taken on 2014 April 23.5 UT, corresponding to $-16.7$~d relative to $B$-band maximum, at 19.7 mag. The last non-detection is based on multiple images taken on 2014 April 7.5 UT with a limiting magnitude of 20.7 mag. The classification spectrum was taken on 2014 May 5.32 UT by PESSTO and was reported to match the spectra of several SNe~Ia before maximum light \citep{Benitez14}. 

SN~2015M (KISS15n) was discovered by the Kiso Supernova Survey \citep[KISS;][]{Morokuma14} using an image obtained on 2015 May 10.54 UT yielding a discovery magnitude of $g=18.3$ mag. No transient was detected in an image observed by the Intermediate Palomar Transient Factory \citep{Masci17} on 2015 May 6.30 UT with a limiting magnitude of $g=19.4$ mag. ASAS-SN \citep{Shappee14, Kochanek17} provides an additional non-detection on 2015 May 7.44 UT with a limiting magnitude of $V=16.85$ mag. Two classification spectra were obtained using the SPRAT instrument on the Liverpool Telescope on 2015 May 17.08 UT and the Andalucia Faint Object Spectrograph and Camera on the Nordic Optical Telescope on 2015 May 16.12 UT. Both spectra resemble those of pre-maximum 1991T-like objects \citep{kiss15ndisc}. While the object is located toward the Coma Cluster, it is unclear whether it is a cluster member. The supernova redshift from the classification spectra is roughly consistent with that of the Coma Cluster.  Images obtained using the Advanced Camera for Surveys (ACS) on the Hubble Space Telescope as part of the ACS Coma Cluster Treasury Survey show the presumed host galaxy at the position of the SN \citep[COMAi13032.301p275841.02;][]{Hammer10}.

\subsection{CSP Photometry and Spectroscopy}
\label{sec:CSP_phot_spec}
\begin{figure*}[htb!]
\centering
\gridline{\fig{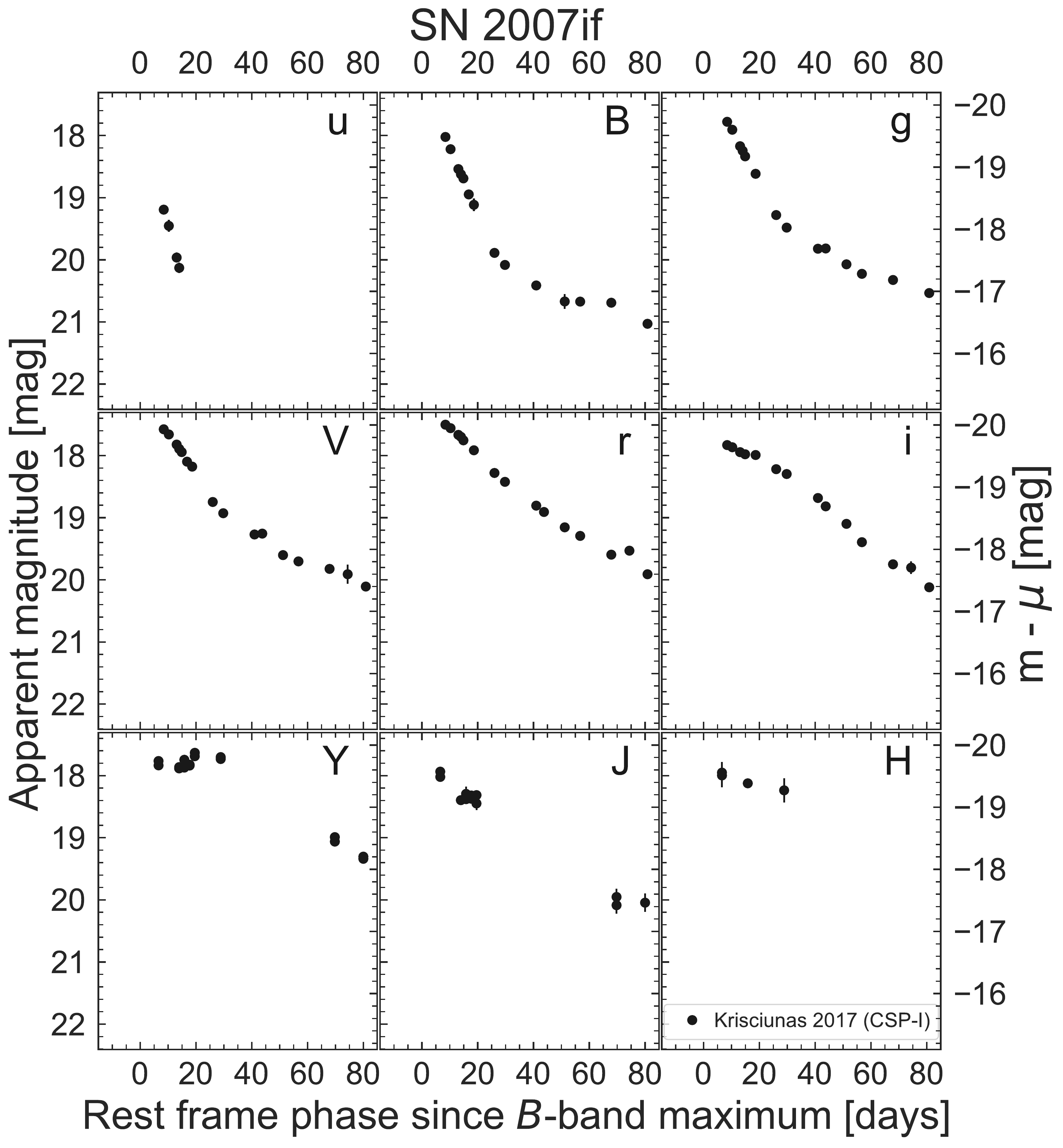}{0.33\textwidth}{}
          \fig{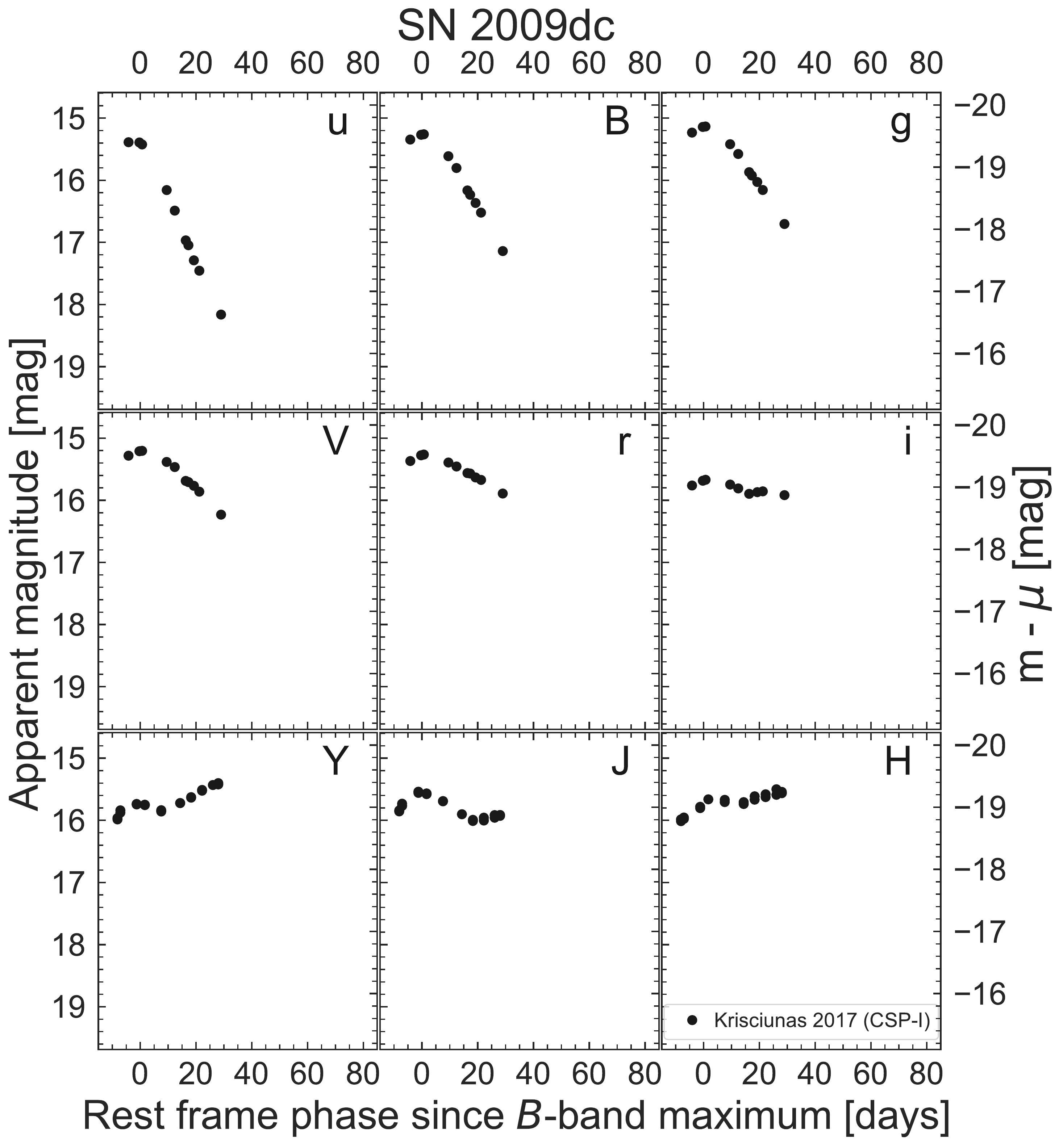}{0.33\textwidth}{}
          \fig{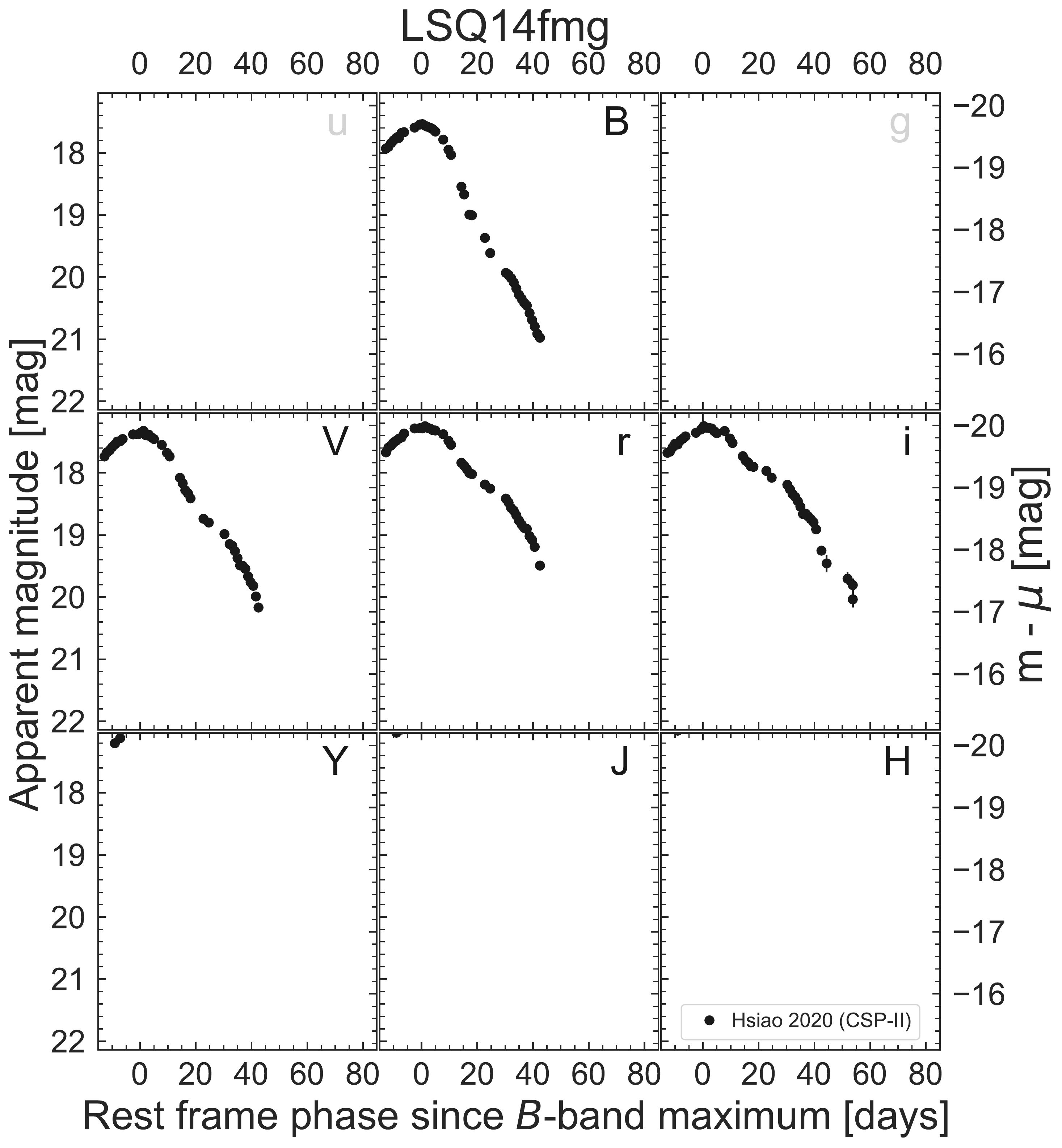}{0.33\textwidth}{}}
\gridline{\fig{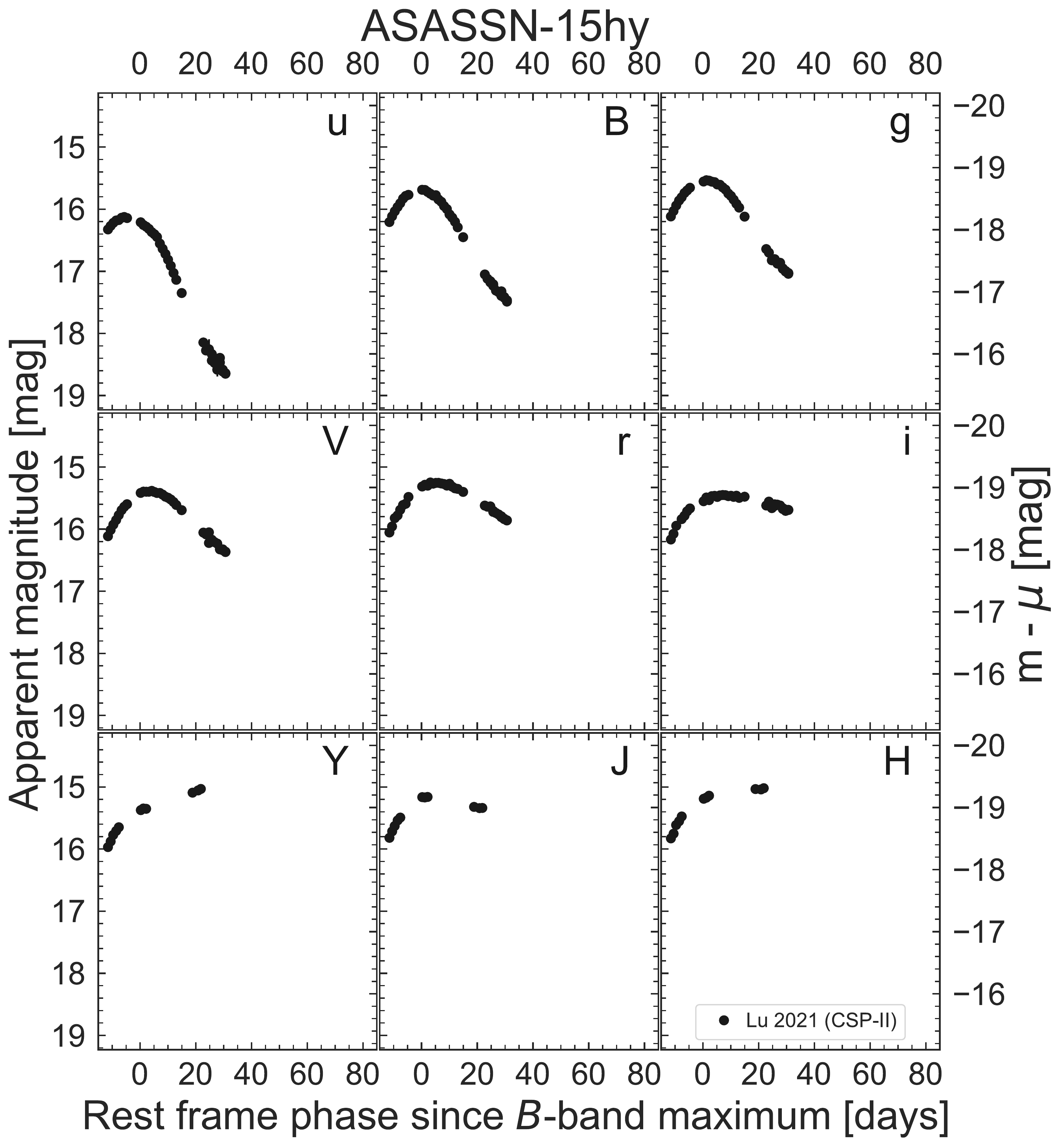}{0.33\textwidth}{}
          \fig{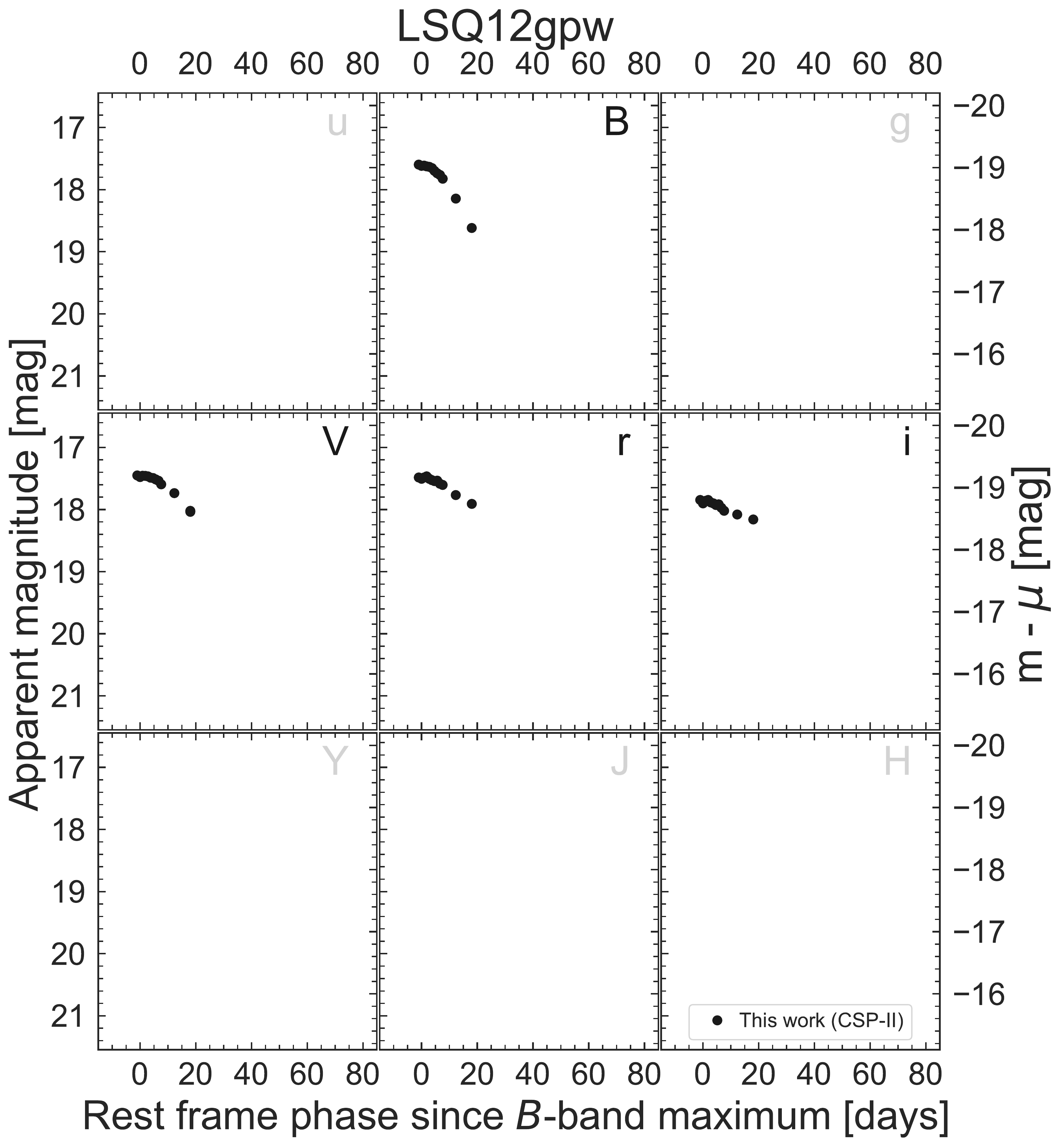}{0.33\textwidth}{}
          \fig{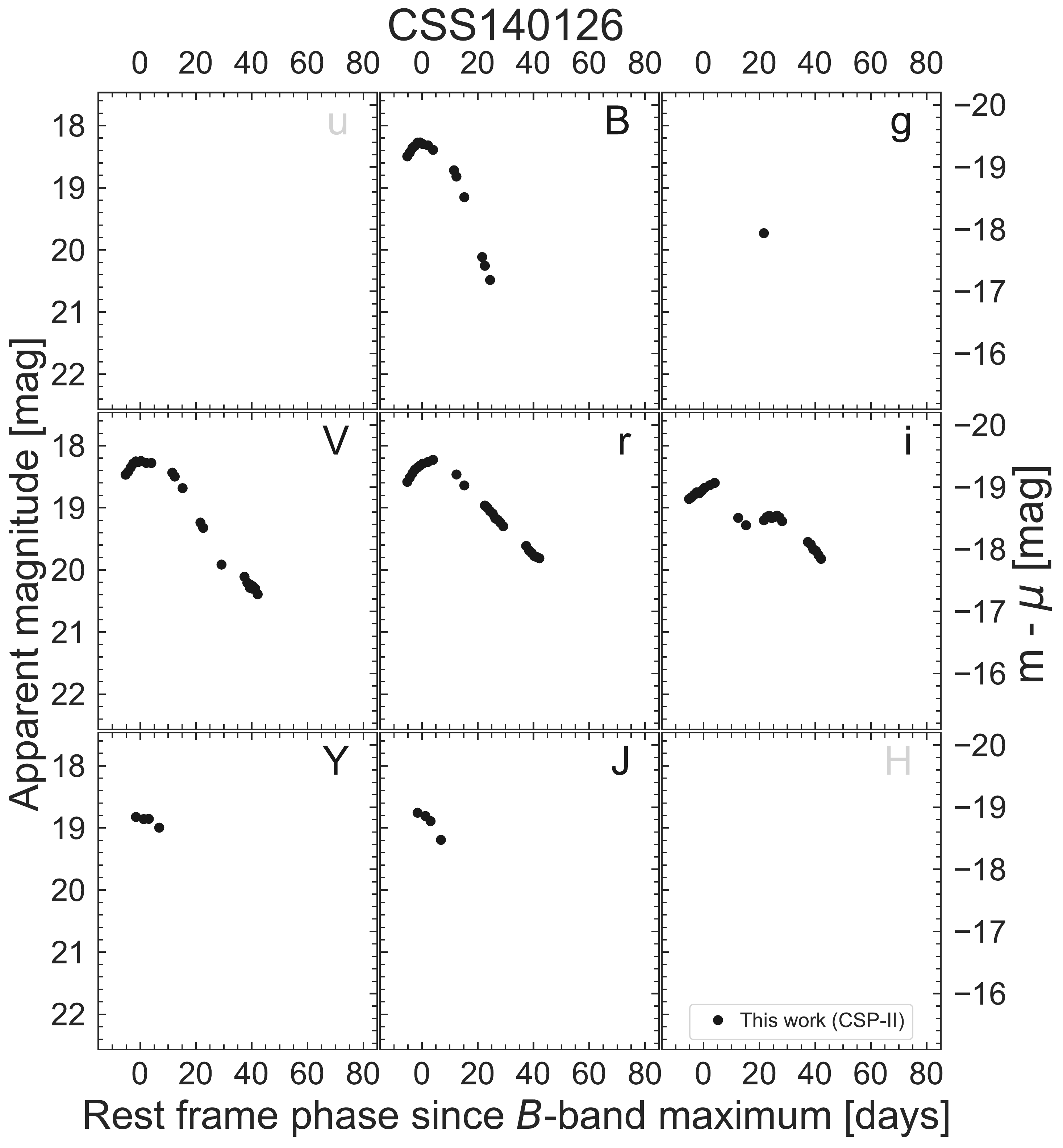}{0.33\textwidth}{}}
\gridline{\fig{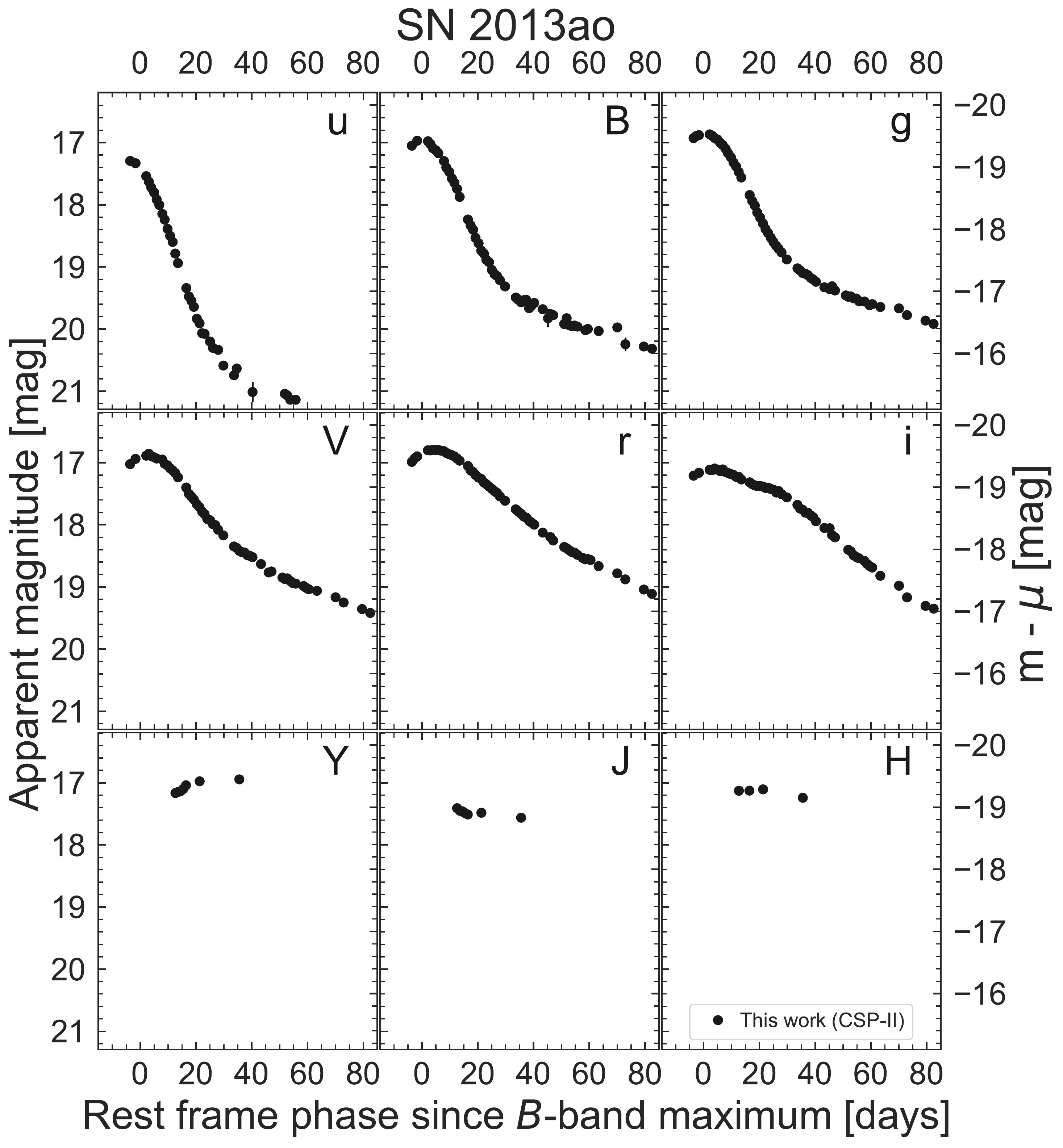}{0.33\textwidth}{}
          \fig{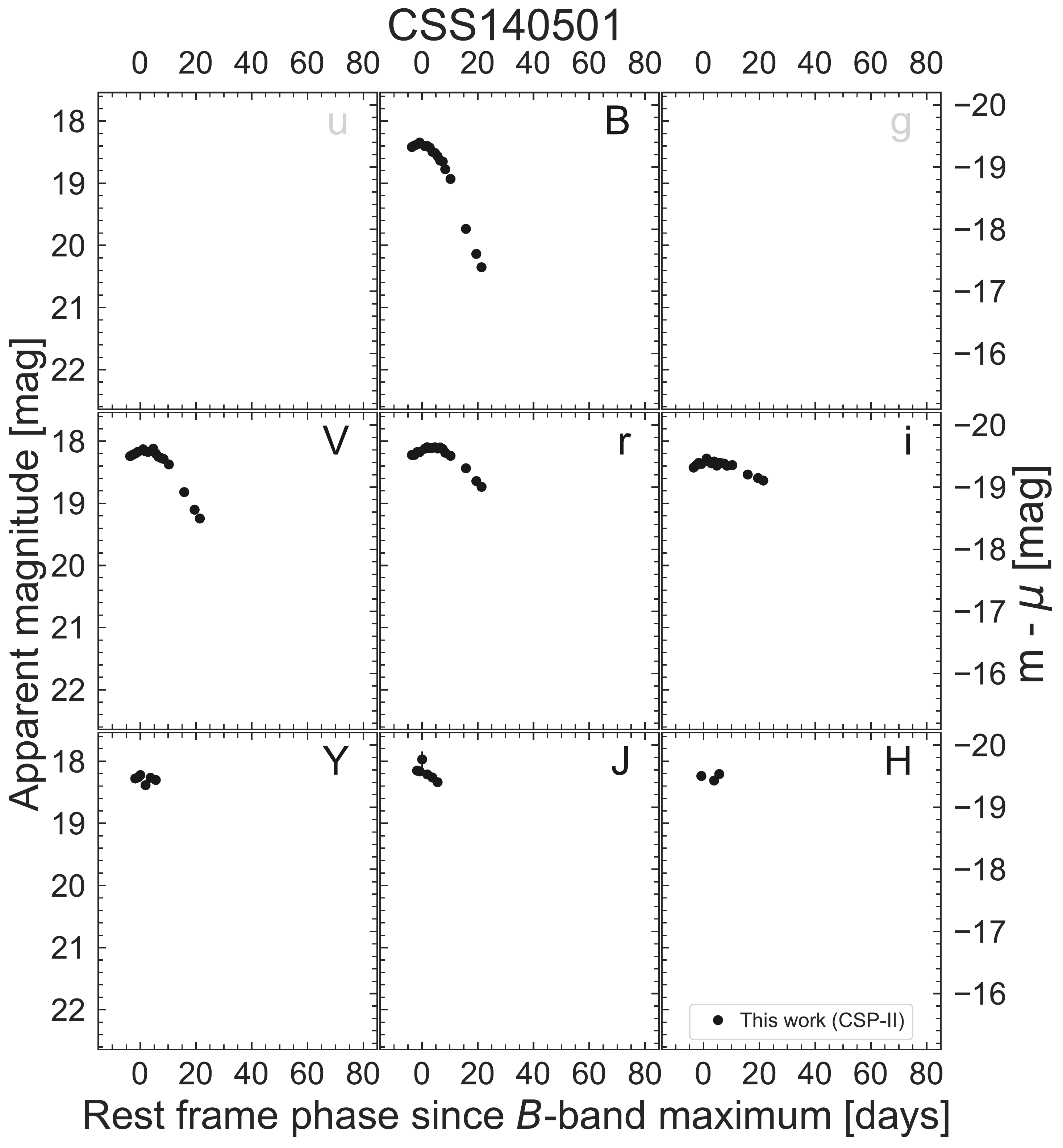}{0.33\textwidth}{}
          \fig{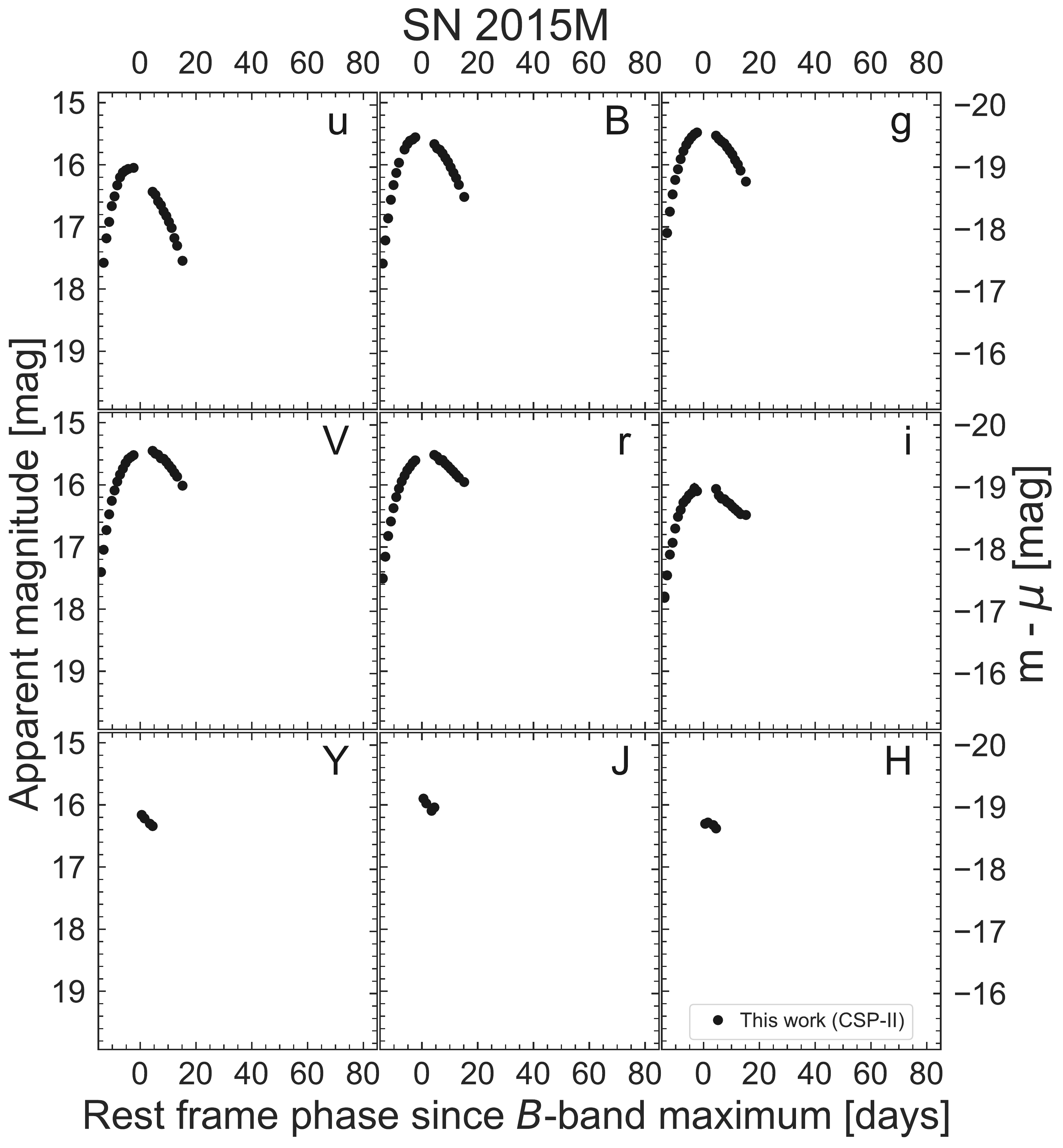}{0.33\textwidth}{}}         
\caption{Individual light curves of the nine \SCSN\ followed by the CSP. If the data were previously published elsewhere it is stated in the figure caption. }
\label{fig:LC_individuals}
\end{figure*}

The photometry of LSQ12gpw, SN~2013ao, CSS140126, CSS140501 and SN~2015M is tabulated in Table \ref{table:phot_lsq12gpw}, \ref{table:phot_sn2013ao}, \ref{table:phot_CSP13abs}, \ref{table:phot_CSP114abk} and \ref{table:phot_KISS15n} respectively. 
Together with four previously published \SCSN\ followed by CSP-I and CSP-II, the light curves are plotted individually in Fig. \ref{fig:LC_individuals}. 
All magnitudes are presented in the CSP natural system and have not been K- or S-corrected. 

Twenty-four optical and six NIR spectroscopic observations of nine 2003fg-like SNe were obtained by the CSP I \& II. The log of these unpublished CSP spectra can be seen in Table~\ref{table:specobs}.

\section{Correlations}
\label{sec:corelationA}
In Fig. \ref{fig:pairplot_correlations}, pair plots of all the parameters previously shown in the text is presented. Section~\ref{sect:correlations} highlights the ones have significant correlations.


\begin{deluxetable}{c c c c c c} [hb!]
\tablecolumns{6}
\tabletypesize{\footnotesize}
\tablecaption{CSP natural system photometry for LSQ12gpw. \label{table:phot_lsq12gpw}} 
\tablehead{
\colhead{MJD} & \colhead{Phase\tablenotemark{a}} & \colhead{$B$} & \colhead{$V$} & \colhead{$r$} & \colhead{$i$}\\
\colhead{Days} & \colhead{Days} & \colhead{Mag} & \colhead{Mag} & \colhead{Mag} & \colhead{Mag}}
\startdata
56268.2&$-$1.0&17.60(0.01)&17.45(0.02)&17.48(0.02)&17.85(0.03)\\
56269.3&0.0&17.62(0.01)&17.48(0.02)&17.50(0.02)&17.90(0.03)\\
56270.2&0.9&17.61(0.02)&17.46(0.02)&17.48(0.02)&17.86(0.03)\\
56271.2&1.9&17.63(0.02)&17.46(0.02)&17.47(0.03)&17.85(0.04)\\
56272.2&2.8&17.63(0.02)&17.47(0.02)&17.51(0.02)&17.89(0.03)\\
56273.2&3.8&17.65(0.02)&17.49(0.02)&17.53(0.03)&17.90(0.03)\\
56274.1&4.6&17.70(0.02)&17.49(0.02)&17.54(0.03)&17.93(0.03)\\
56275.2&5.7&17.74(0.02)&17.52(0.02)&17.54(0.03)&17.92(0.04)\\
56276.2&6.6&17.76(0.02)&17.54(0.02)&17.58(0.03)&17.97(0.03)\\
56277.2&7.6&17.82(0.02)&17.59(0.02)&17.60(0.03)&18.02(0.03)\\
56282.2&12.3&18.14(0.02)&17.74(0.02)&17.77(0.02)&18.08(0.03)\\
56288.2&18.0&18.62(0.03)&18.03(0.03)&17.91(0.02)&18.16(0.03)\\
\enddata
\tablenotetext{a}{Rest frame days relative to the time of $B$-band maximum given in Table \ref{table:propphot}.}
\end{deluxetable}

\startlongtable
\begin{deluxetable*}{ccccccccccc}
\tabletypesize{\footnotesize}
\tablecaption{CSP natural system photometry for SN~2013ao. \label{table:phot_sn2013ao}} 
\tablehead{
\colhead{MJD} & \colhead{Phase\tablenotemark{a}} & \colhead{$u$} &\colhead{$B$} & \colhead{$g$} &\colhead{$V$} & \colhead{$r$} & \colhead{$i$} & \colhead{$Y$} & \colhead{$J$} & \colhead{$H$}\\
\colhead{Days} & \colhead{Days} & \colhead{Mag} & \colhead{Mag} & \colhead{Mag} & \colhead{Mag}& \colhead{Mag} & \colhead{Mag} & \colhead{Mag} & \colhead{Mag}& \colhead{Mag}}
\startdata
56358.3&$-$3.6&17.29(0.02)&17.05(0.01)&16.93(0.01)&17.02(0.02)&16.99(0.01)&17.21(0.01)&$\cdots$&$\cdots$&$\cdots$\\
56359.3&$-$2.6&$\cdots$&$\cdots$&16.89(0.01)&$\cdots$&16.93(0.01)&$\cdots$&$\cdots$&$\cdots$&$\cdots$\\
56360.3&$-$1.7&17.33(0.02)&16.97(0.02)&16.88(0.01)&16.94(0.03)&16.90(0.01)&17.16(0.01)&$\cdots$&$\cdots$&$\cdots$\\
56364.3&2.2&17.54(0.02)&16.98(0.01)&16.87(0.01)&16.89(0.01)&16.80(0.01)&17.11(0.01)&$\cdots$&$\cdots$&$\cdots$\\
56365.2&3.0&17.63(0.02)&17.03(0.02)&16.89(0.01)&16.86(0.03)&16.80(0.01)&17.12(0.01)&$\cdots$&$\cdots$&$\cdots$\\
56366.2&4.0&17.72(0.02)&17.09(0.01)&16.92(0.01)&16.89(0.01)&16.79(0.01)&17.09(0.01)&$\cdots$&$\cdots$&$\cdots$\\
56367.2&5.0&17.80(0.02)&17.12(0.01)&16.95(0.01)&16.91(0.01)&16.80(0.01)&17.11(0.01)&$\cdots$&$\cdots$&$\cdots$\\
56368.2&5.9&17.92(0.01)&17.17(0.01)&17.00(0.01)&16.93(0.01)&16.80(0.01)&17.14(0.01)&$\cdots$&$\cdots$&$\cdots$\\
56369.2&6.9&18.00(0.02)&$\cdots$&17.04(0.01)&$\cdots$&16.81(0.01)&17.11(0.01)&$\cdots$&$\cdots$&$\cdots$\\
56370.3&7.9&18.15(0.02)&17.30(0.02)&17.10(0.01)&16.95(0.03)&16.82(0.01)&17.15(0.01)&$\cdots$&$\cdots$&$\cdots$\\
56371.2&8.8&18.24(0.02)&17.40(0.01)&17.17(0.01)&17.02(0.01)&16.84(0.01)&17.17(0.01)&$\cdots$&$\cdots$&$\cdots$\\
56372.2&9.8&18.39(0.02)&17.47(0.02)&17.24(0.01)&17.04(0.01)&16.86(0.01)&17.18(0.01)&$\cdots$&$\cdots$&$\cdots$\\
56373.2&10.7&18.50(0.02)&17.58(0.01)&17.32(0.01)&17.09(0.01)&16.88(0.01)&17.20(0.01)&$\cdots$&$\cdots$&$\cdots$\\
56374.2&11.7&18.60(0.02)&17.65(0.01)&17.38(0.01)&17.12(0.01)&16.90(0.01)&17.23(0.01)&$\cdots$&$\cdots$&$\cdots$\\
56375.2&12.6&18.78(0.02)&17.74(0.02)&17.47(0.01)&17.17(0.03)&16.93(0.02)&17.23(0.02)&17.17(0.02)&17.41(0.02)&17.13(0.04)\\
56376.2&13.6&18.94(0.03)&17.87(0.02)&17.56(0.01)&17.24(0.02)&16.97(0.01)&17.27(0.01)&17.15(0.01)&17.45(0.04)&$\cdots$\\
56377.2&14.5&$\cdots$&$\cdots$&$\cdots$&$\cdots$&$\cdots$&$\cdots$&17.14(0.01)&17.46(0.03)&$\cdots$\\
56378.2&15.5&$\cdots$&$\cdots$&$\cdots$&$\cdots$&$\cdots$&$\cdots$&17.10(0.01)&17.49(0.03)&$\cdots$\\
56379.2&16.6&19.34(0.06)&18.24(0.03)&17.84(0.02)&17.40(0.04)&17.05(0.01)&17.32(0.02)&17.04(0.01)&17.51(0.03)&17.13(0.04)\\
56380.2&17.4&19.48(0.04)&18.33(0.02)&17.94(0.02)&17.50(0.03)&17.12(0.01)&17.35(0.02)&$\cdots$&$\cdots$&$\cdots$\\
56381.2&18.4&19.55(0.04)&18.40(0.02)&18.01(0.01)&17.54(0.03)&17.15(0.02)&17.37(0.02)&$\cdots$&$\cdots$&$\cdots$\\
56382.1&19.2&19.64(0.04)&18.53(0.02)&18.12(0.01)&17.59(0.01)&17.20(0.01)&17.38(0.01)&$\cdots$&$\cdots$&$\cdots$\\
56383.2&20.3&19.83(0.04)&18.62(0.02)&18.21(0.01)&17.67(0.02)&17.24(0.01)&17.38(0.01)&$\cdots$&$\cdots$&$\cdots$\\
56384.2&21.2&19.91(0.04)&18.74(0.02)&18.30(0.01)&17.72(0.01)&17.26(0.01)&17.38(0.01)&16.98(0.01)&17.48(0.04)&17.11(0.04)\\
56385.2&22.2&20.07(0.03)&18.79(0.02)&18.39(0.01)&17.79(0.02)&17.32(0.01)&17.41(0.01)&$\cdots$&$\cdots$&$\cdots$\\
56386.1&23.1&20.08(0.03)&18.89(0.02)&18.45(0.01)&17.83(0.01)&17.35(0.01)&17.40(0.01)&$\cdots$&$\cdots$&$\cdots$\\
56387.1&24.0&$\cdots$&18.92(0.02)&18.53(0.01)&17.91(0.01)&17.38(0.01)&17.42(0.01)&$\cdots$&$\cdots$&$\cdots$\\
56388.2&25.1&20.20(0.03)&19.05(0.02)&18.60(0.01)&17.93(0.03)&17.42(0.01)&17.43(0.01)&$\cdots$&$\cdots$&$\cdots$\\
56389.2&26.0&20.30(0.04)&19.12(0.02)&18.66(0.01)&17.99(0.02)&17.46(0.01)&17.48(0.01)&$\cdots$&$\cdots$&$\cdots$\\
56390.1&26.9&$\cdots$&19.14(0.02)&18.71(0.01)&18.00(0.03)&17.48(0.01)&17.46(0.02)&$\cdots$&$\cdots$&$\cdots$\\
56391.2&28.0&20.34(0.05)&19.21(0.02)&18.77(0.01)&18.08(0.01)&17.54(0.01)&17.51(0.01)&$\cdots$&$\cdots$&$\cdots$\\
56393.2&29.8&20.59(0.07)&19.31(0.04)&18.88(0.02)&18.17(0.02)&17.61(0.01)&17.56(0.01)&$\cdots$&$\cdots$&$\cdots$\\
56397.2&33.7&20.74(0.05)&19.49(0.03)&19.02(0.02)&18.35(0.02)&17.75(0.01)&17.68(0.02)&$\cdots$&$\cdots$&$\cdots$\\
56398.1&34.6&20.64(0.06)&19.53(0.02)&19.06(0.01)&18.37(0.02)&17.79(0.01)&17.74(0.01)&$\cdots$&$\cdots$&$\cdots$\\
56399.1&35.5&$\cdots$&19.57(0.02)&19.10(0.01)&18.42(0.02)&17.82(0.01)&17.76(0.01)&16.95(0.01)&17.56(0.03)&17.24(0.04)\\
56400.1&36.5&$\cdots$&19.54(0.02)&19.11(0.01)&18.44(0.02)&17.87(0.01)&17.80(0.01)&$\cdots$&$\cdots$&$\cdots$\\
56401.0&37.3&$\cdots$&19.53(0.03)&19.13(0.02)&18.44(0.03)&17.88(0.01)&17.81(0.02)&$\cdots$&$\cdots$&$\cdots$\\
56402.2&38.4&$\cdots$&19.66(0.04)&19.18(0.02)&18.49(0.03)&17.94(0.01)&17.85(0.01)&$\cdots$&$\cdots$&$\cdots$\\
56403.0&39.3&$\cdots$&19.62(0.04)&19.20(0.02)&18.50(0.02)&17.96(0.02)&17.88(0.02)&$\cdots$&$\cdots$&$\cdots$\\
56404.0&40.2&21.01(0.16)&19.58(0.04)&19.24(0.02)&18.52(0.02)&18.00(0.01)&17.94(0.02)&$\cdots$&$\cdots$&$\cdots$\\
56407.2&43.3&$\cdots$&19.68(0.08)&19.33(0.04)&18.63(0.03)&18.12(0.02)&18.05(0.02)&$\cdots$&$\cdots$&$\cdots$\\
56409.2&45.2&$\cdots$&19.83(0.15)&19.36(0.05)&$\cdots$&$\cdots$&18.06(0.02)&$\cdots$&$\cdots$&$\cdots$\\
56410.1&46.1&$\cdots$&19.76(0.05)&19.31(0.02)&18.77(0.02)&18.20(0.01)&18.16(0.01)&$\cdots$&$\cdots$&$\cdots$\\
56411.2&47.0&$\cdots$&19.78(0.04)&19.38(0.02)&18.75(0.02)&18.25(0.01)&18.20(0.01)&$\cdots$&$\cdots$&$\cdots$\\
56415.2&51.0&$\cdots$&19.92(0.04)&19.46(0.02)&18.85(0.02)&18.36(0.01)&$\cdots$&$\cdots$&$\cdots$&$\cdots$\\
56416.2&51.8&21.04(0.07)&19.83(0.03)&19.48(0.01)&18.87(0.03)&18.38(0.01)&18.40(0.02)&$\cdots$&$\cdots$&$\cdots$\\
56417.1&52.8&21.07(0.07)&19.94(0.02)&19.47(0.01)&18.86(0.03)&18.41(0.01)&18.43(0.01)&$\cdots$&$\cdots$&$\cdots$\\
56418.2&53.8&21.14(0.08)&19.95(0.02)&19.51(0.01)&18.90(0.03)&18.44(0.01)&18.49(0.02)&$\cdots$&$\cdots$&$\cdots$\\
56419.2&54.8&$\cdots$&19.94(0.02)&19.51(0.01)&18.94(0.02)&18.45(0.01)&18.52(0.01)&$\cdots$&$\cdots$&$\cdots$\\
56420.2&55.7&21.14(0.07)&19.96(0.03)&19.55(0.01)&18.95(0.03)&18.48(0.02)&18.54(0.02)&$\cdots$&$\cdots$&$\cdots$\\
56422.2&57.7&$\cdots$&$\cdots$&19.56(0.02)&$\cdots$&18.53(0.01)&18.58(0.01)&$\cdots$&$\cdots$&$\cdots$\\
56423.2&58.6&$\cdots$&20.02(0.03)&$\cdots$&18.98(0.03)&18.56(0.01)&18.63(0.02)&$\cdots$&$\cdots$&$\cdots$\\
56424.1&59.5&$\cdots$&20.00(0.03)&19.61(0.01)&19.01(0.03)&18.55(0.01)&18.66(0.02)&$\cdots$&$\cdots$&$\cdots$\\
56425.1&60.4&$\cdots$&$\cdots$&19.60(0.02)&19.04(0.04)&18.57(0.02)&18.69(0.02)&$\cdots$&$\cdots$&$\cdots$\\
56428.2&63.4&$\cdots$&20.03(0.02)&19.64(0.01)&19.06(0.02)&18.66(0.01)&18.82(0.02)&$\cdots$&$\cdots$&$\cdots$\\
56435.1&70.0&$\cdots$&19.97(0.07)&19.67(0.04)&19.17(0.04)&18.78(0.02)&18.98(0.03)&$\cdots$&$\cdots$&$\cdots$\\
56438.2&73.0&$\cdots$&20.24(0.11)&19.78(0.04)&19.25(0.04)&18.88(0.02)&19.17(0.03)&$\cdots$&$\cdots$&$\cdots$\\
56445.0&79.6&$\cdots$&20.28(0.03)&19.86(0.02)&19.36(0.03)&19.04(0.01)&19.30(0.02)&$\cdots$&$\cdots$&$\cdots$\\
56448.1&82.5&$\cdots$&20.32(0.03)&19.92(0.02)&19.42(0.03)&19.11(0.01)&19.35(0.02)&$\cdots$&$\cdots$&$\cdots$\\
\enddata
\tablenotetext{a}{Rest frame days relative to the time of $B$-band maximum given in Table \ref{table:propphot}.}
\end{deluxetable*}

\begin{deluxetable*}{c c c c c c c c c}
\tablewidth{0pt}
\tablecolumns{6}
\tablecaption{CSP natural system photometry for CSS140126. \label{table:phot_CSP13abs}} 
\tablehead{
\colhead{MJD} & \colhead{Phase\tablenotemark{a}} & \colhead{$B$} & \colhead{$g$} & \colhead{$V$} & \colhead{$r$} & \colhead{$i$} & \colhead{$Y$} & \colhead{$J$}\\
\colhead{Days} & \colhead{Days} & \colhead{Mag} & \colhead{Mag} & \colhead{Mag} & \colhead{Mag}& \colhead{Mag} &\colhead{Mag}& \colhead{Mag}}
\startdata
56662.3&$-$5.3&18.50(0.01)&$\cdots$&18.47(0.01)&18.58(0.01)&18.86(0.02)&$\cdots$&$\cdots$\\
56663.3&$-$4.3&18.43(0.01)&$\cdots$&18.42(0.01)&18.51(0.01)&18.83(0.02)&$\cdots$&$\cdots$\\
56664.3&$-$3.4&18.36(0.01)&$\cdots$&18.35(0.01)&18.45(0.01)&18.79(0.02)&$\cdots$&$\cdots$\\
56665.3&$-$2.5&18.33(0.01)&$\cdots$&18.29(0.01)&18.39(0.01)&18.75(0.02)&$\cdots$&$\cdots$\\
56666.4&$-$1.6&18.27(0.01)&$\cdots$&18.25(0.01)&18.35(0.01)&18.77(0.01)&18.82(0.02)&18.76(0.02)\\
56667.3&$-$0.6&18.27(0.01)&$\cdots$&18.26(0.01)&18.32(0.01)&18.73(0.01)&$\cdots$&$\cdots$\\
56668.3&0.3&18.30(0.01)&$\cdots$&18.25(0.01)&18.29(0.01)&18.68(0.01)&$\cdots$&$\cdots$\\
56669.3&1.2&$\cdots$&$\cdots$&$\cdots$&$\cdots$&$\cdots$&18.86(0.02)&18.81(0.02)\\
56670.3&2.2&18.32(0.01)&$\cdots$&18.28(0.01)&18.26(0.01)&18.64(0.01)&$\cdots$&$\cdots$\\
56671.3&3.1&$\cdots$&$\cdots$&$\cdots$&$\cdots$&$\cdots$&18.86(0.02)&18.89(0.02)\\
56672.3&4.0&18.39(0.02)&$\cdots$&18.28(0.02)&18.23(0.01)&18.60(0.02)&$\cdots$&$\cdots$\\
56675.3&6.8&$\cdots$&$\cdots$&$\cdots$&$\cdots$&$\cdots$&19.00(0.03)&19.19(0.04)\\
56680.4&11.5&18.72(0.02)&$\cdots$&18.43(0.02)&$\cdots$&$\cdots$&$\cdots$&$\cdots$\\
56681.3&12.4&18.82(0.02)&$\cdots$&18.50(0.01)&18.46(0.01)&19.16(0.02)&$\cdots$&$\cdots$\\
56684.4&15.2&19.15(0.01)&$\cdots$&18.68(0.01)&18.64(0.01)&19.28(0.02)&$\cdots$&$\cdots$\\
56691.2&21.6&20.11(0.02)&19.73(0.01)&19.24(0.01)&$\cdots$&19.20(0.02)&$\cdots$&$\cdots$\\
56692.3&22.6&20.26(0.02)&$\cdots$&19.32(0.01)&18.96(0.01)&19.15(0.02)&$\cdots$&$\cdots$\\
56693.3&23.5&$\cdots$&$\cdots$&$\cdots$&19.00(0.02)&19.13(0.02)&$\cdots$&$\cdots$\\
56694.3&24.4&20.48(0.03)&$\cdots$&$\cdots$&19.06(0.01)&19.16(0.02)&$\cdots$&$\cdots$\\
56695.3&25.4&$\cdots$&$\cdots$&$\cdots$&19.09(0.02)&19.15(0.02)&$\cdots$&$\cdots$\\
56696.3&26.3&$\cdots$&$\cdots$&$\cdots$&19.17(0.01)&19.13(0.02)&$\cdots$&$\cdots$\\
56697.3&27.2&$\cdots$&$\cdots$&$\cdots$&19.20(0.01)&19.15(0.02)&$\cdots$&$\cdots$\\
56698.3&28.1&$\cdots$&$\cdots$&$\cdots$&19.24(0.01)&19.22(0.02)&$\cdots$&$\cdots$\\
56699.4&29.2&$\cdots$&$\cdots$&19.91(0.02)&19.30(0.02)&$\cdots$&$\cdots$&$\cdots$\\
56708.3&37.4&$\cdots$&$\cdots$&20.11(0.04)&19.61(0.03)&19.55(0.03)&$\cdots$&$\cdots$\\
56709.4&38.5&$\cdots$&$\cdots$&20.21(0.04)&19.69(0.02)&19.59(0.02)&$\cdots$&$\cdots$\\
56710.4&39.3&$\cdots$&$\cdots$&20.26(0.04)&19.72(0.02)&19.67(0.03)&$\cdots$&$\cdots$\\
56711.4&40.3&$\cdots$&$\cdots$&20.28(0.04)&19.78(0.02)&19.69(0.02)&$\cdots$&$\cdots$\\
56712.4&41.2&$\cdots$&$\cdots$&20.30(0.03)&19.79(0.02)&19.76(0.03)&$\cdots$&$\cdots$\\
56713.4&42.2&$\cdots$&$\cdots$&20.39(0.03)&19.81(0.02)&19.82(0.03)&$\cdots$&$\cdots$\\
\enddata
\tablenotetext{a}{Rest frame days relative to the time of $B$-band maximum given in Table \ref{table:propphot}.}
\end{deluxetable*}

\begin{deluxetable*}{c c c c c c c c c}
\tablewidth{0pt}
\tablecolumns{6}
\tablecaption{CSP natural system photometry for CSS140501. \label{table:phot_CSP114abk}} 
\tablehead{
\colhead{MJD} & \colhead{Phase\tablenotemark{a}} & \colhead{$B$} & \colhead{$V$} & \colhead{$r$} & \colhead{$i$} & \colhead{$Y$} & \colhead{$J$} & \colhead{$H$}\\
\colhead{Days}& \colhead{Days} &\colhead{Mag}& \colhead{Mag} &\colhead{Mag}& \colhead{Mag} &\colhead{Mag}& \colhead{Mag} &\colhead{Mag}}
\startdata
56783.3&$-$3.6&18.42(0.02)&18.24(0.02)&18.22(0.02)&18.42(0.02)&$\cdots$&$\cdots$&$\cdots$\\
56784.3&$-$2.7&18.39(0.01)&18.22(0.01)&18.22(0.01)&18.39(0.02)&$\cdots$&$\cdots$&$\cdots$\\
56785.3&$-$1.8&18.38(0.01)&18.20(0.02)&18.18(0.01)&18.35(0.02)&18.28(0.01)&18.15(0.02)&$\cdots$\\
56786.3&$-$0.8&18.34(0.01)&18.17(0.02)&18.18(0.02)&18.37(0.02)&18.26(0.01)&18.16(0.02)&18.24(0.04)\\
56787.3&0.1&$\cdots$&$\cdots$&$\cdots$&$\cdots$&18.22(0.01)&17.97(0.13)&$\cdots$\\
56788.4&1.0&18.40(0.02)&18.13(0.02)&18.12(0.02)&18.28(0.04)&$\cdots$&$\cdots$&$\cdots$\\
56789.2&1.9&18.40(0.02)&18.16(0.02)&18.10(0.01)&18.33(0.02)&18.39(0.02)&18.21(0.02)&$\cdots$\\
56790.2&2.8&18.43(0.03)&18.17(0.02)&18.11(0.02)&18.36(0.02)&$\cdots$&$\cdots$&$\cdots$\\
56791.2&3.8&18.49(0.04)&18.16(0.03)&18.11(0.02)&18.33(0.02)&18.27(0.01)&18.26(0.02)&18.31(0.04)\\
56792.3&4.7&18.51(0.04)&18.12(0.03)&18.10(0.02)&18.39(0.03)&$\cdots$&$\cdots$&$\cdots$\\
56793.2&5.6&18.57(0.03)&18.20(0.03)&18.12(0.02)&18.35(0.02)&18.30(0.02)&18.34(0.02)&18.21(0.04)\\
56794.3&6.6&18.63(0.03)&18.25(0.02)&18.10(0.02)&18.36(0.02)&$\cdots$&$\cdots$&$\cdots$\\
56795.3&7.5&18.65(0.02)&18.27(0.02)&18.13(0.02)&18.36(0.02)&$\cdots$&$\cdots$&$\cdots$\\
56796.2&8.4&18.77(0.02)&18.28(0.02)&18.19(0.02)&18.40(0.02)&$\cdots$&$\cdots$&$\cdots$\\
56798.3&10.3&18.93(0.04)&18.38(0.02)&18.24(0.02)&18.39(0.04)&$\cdots$&$\cdots$&$\cdots$\\
56804.2&15.8&19.73(0.02)&18.82(0.01)&18.44(0.01)&18.54(0.01)&$\cdots$&$\cdots$&$\cdots$\\
56808.3&19.5&20.14(0.02)&19.10(0.02)&18.64(0.01)&18.59(0.01)&$\cdots$&$\cdots$&$\cdots$\\
56810.3&21.4&20.35(0.02)&19.24(0.01)&18.74(0.01)&18.63(0.02)&$\cdots$&$\cdots$&$\cdots$\\
\enddata
\tablenotetext{a}{Rest frame days relative to the time of $B$-band maximum given in Table \ref{table:propphot}.}
\end{deluxetable*}

\begin{deluxetable*}{ccccccccccc}
\tablewidth{0pt}
\tablecaption{CSP natural system photometry for SN~2015M. \label{table:phot_KISS15n}} 
\tablehead{
\colhead{MJD} & \colhead{Phase\tablenotemark{a}} & \colhead{$u$} &\colhead{$B$} & \colhead{$g$} &\colhead{$V$} & \colhead{$r$} & \colhead{$i$} & \colhead{$Y$} & \colhead{$J$} & \colhead{$H$}\\
\colhead{Days}& \colhead{Days} &\colhead{Mag}& \colhead{Mag} &\colhead{Mag}& \colhead{Mag} &\colhead{Mag}& \colhead{Mag} &\colhead{Mag}& \colhead{Mag} &\colhead{Mag}}
\startdata
57154.1&$-$14.1&$\cdots$&17.59(0.02)&$\cdots$&17.40(0.02)&17.51(0.02)&17.81(0.03)&$\cdots$&$\cdots$&$\cdots$\\
57155.1&$-$13.1&17.58(0.04)&17.22(0.01)&17.10(0.01)&17.05(0.01)&17.16(0.01)&17.46(0.02)&$\cdots$&$\cdots$&$\cdots$\\
57156.1&$-$12.1&17.18(0.02)&16.86(0.01)&16.75(0.01)&16.73(0.01)&16.82(0.01)&17.12(0.02)&$\cdots$&$\cdots$&$\cdots$\\
57157.1&$-$11.1&16.92(0.02)&16.56(0.01)&16.48(0.01)&16.47(0.01)&16.59(0.01)&16.93(0.04)&$\cdots$&$\cdots$&$\cdots$\\
57158.1&$-$10.2&16.66(0.01)&16.33(0.01)&16.24(0.01)&16.26(0.01)&16.38(0.01)&16.70(0.01)&$\cdots$&$\cdots$&$\cdots$\\
57159.1&$-$9.2&16.51(0.02)&16.13(0.01)&16.07(0.01)&16.09(0.01)&16.20(0.01)&16.51(0.01)&$\cdots$&$\cdots$&$\cdots$\\
57160.1&$-$8.2&16.33(0.01)&15.97(0.01)&15.91(0.01)&15.95(0.01)&16.06(0.01)&16.40(0.01)&$\cdots$&$\cdots$&$\cdots$\\
57161.1&$-$7.2&16.20(0.01)&$\cdots$&15.78(0.01)&15.84(0.01)&15.94(0.01)&16.28(0.01)&$\cdots$&$\cdots$&$\cdots$\\
57162.1&$-$6.3&16.12(0.01)&15.75(0.01)&15.68(0.00)&15.74(0.01)&15.85(0.01)&16.23(0.01)&$\cdots$&$\cdots$&$\cdots$\\
57163.1&$-$5.3&16.09(0.01)&15.67(0.01)&15.61(0.00)&15.65(0.01)&15.77(0.01)&16.17(0.01)&$\cdots$&$\cdots$&$\cdots$\\
57164.1&$-$4.3&16.07(0.01)&15.61(0.01)&15.55(0.01)&15.58(0.01)&15.71(0.01)&16.13(0.01)&$\cdots$&$\cdots$&$\cdots$\\
57165.1&$-$3.3&$\cdots$&15.59(0.01)&15.51(0.00)&15.55(0.01)&15.65(0.01)&16.06(0.09)&$\cdots$&$\cdots$&$\cdots$\\
57166.1&$-$2.3&16.05(0.01)&15.56(0.00)&15.48(0.00)&15.52(0.01)&15.61(0.00)&16.10(0.01)&$\cdots$&$\cdots$&$\cdots$\\
57169.1&0.6&$\cdots$&$\cdots$&$\cdots$&$\cdots$&$\cdots$&$\cdots$&16.16(0.04)&15.90(0.03)&16.30(0.03)\\
57170.1&1.6&$\cdots$&$\cdots$&$\cdots$&$\cdots$&$\cdots$&$\cdots$&16.22(0.04)&15.98(0.03)&16.28(0.03)\\
57172.1&3.5&$\cdots$&$\cdots$&$\cdots$&$\cdots$&$\cdots$&$\cdots$&16.30(0.05)&16.10(0.03)&16.32(0.04)\\
57173.1&4.5&16.43(0.02)&15.67(0.01)&15.53(0.01)&15.45(0.01)&15.52(0.01)&16.07(0.02)&16.34(0.04)&16.04(0.03)&16.38(0.04)\\
57174.1&5.5&16.48(0.02)&15.74(0.01)&15.58(0.01)&15.50(0.01)&15.55(0.01)&16.17(0.01)&$\cdots$&$\cdots$&$\cdots$\\
57175.1&6.5&16.59(0.03)&15.76(0.01)&15.62(0.01)&15.52(0.01)&15.61(0.01)&16.22(0.01)&$\cdots$&$\cdots$&$\cdots$\\
57176.1&7.4&16.65(0.02)&15.82(0.01)&15.66(0.01)&15.57(0.01)&15.60(0.01)&16.23(0.01)&$\cdots$&$\cdots$&$\cdots$\\
57177.1&8.4&16.75(0.02)&15.89(0.01)&15.72(0.01)&15.58(0.01)&15.66(0.01)&16.28(0.01)&$\cdots$&$\cdots$&$\cdots$\\
57178.1&9.4&16.82(0.02)&15.95(0.01)&15.78(0.01)&15.63(0.01)&15.70(0.01)&16.30(0.01)&$\cdots$&$\cdots$&$\cdots$\\
57179.0&10.3&16.92(0.02)&16.04(0.01)&15.84(0.00)&15.69(0.01)&15.74(0.01)&16.35(0.01)&$\cdots$&$\cdots$&$\cdots$\\
57180.1&11.3&17.02(0.02)&16.13(0.01)&15.92(0.01)&15.74(0.01)&15.79(0.00)&16.39(0.01)&$\cdots$&$\cdots$&$\cdots$\\
57181.0&12.2&17.18(0.02)&16.21(0.01)&15.99(0.00)&15.81(0.01)&15.84(0.01)&16.43(0.01)&$\cdots$&$\cdots$&$\cdots$\\
57182.0&13.2&17.30(0.02)&16.32(0.01)&16.09(0.00)&15.87(0.01)&15.88(0.00)&16.47(0.01)&$\cdots$&$\cdots$&$\cdots$\\
57184.0&15.2&17.55(0.02)&16.52(0.01)&16.27(0.01)&16.01(0.01)&15.96(0.01)&16.48(0.01)&$\cdots$&$\cdots$&$\cdots$\\
\enddata
\tablenotetext{a}{Rest frame days relative to the time of $B$-band maximum given in Table \ref{table:propphot}.}
\end{deluxetable*}



\begin{deluxetable*}{c c c c c c c}
\tablewidth{0pt}
\tablecolumns{7}
\tablecaption{\label{table:specobs} Journal of Spectroscopic Observations.}
\tablehead{
\colhead{SN} & 
\colhead{JD$-$2450000} &	
\colhead{Phase\tablenotemark{a}} & 
\colhead{Telescope} & 
\colhead{Instrument}  \\ 
\colhead{} &
\colhead{(Days)} & 
\colhead{(Days)} &
\colhead{} &
\colhead{} }
\startdata
\multicolumn{5}{c}{\bf Optical}\\ 
\hline
\hline
2007if	&	4355.86	&	+5.48	&	du Pont	&	B \& C	\\
2007if	&	4361.83	&	+11.04	&	du Pont	&	B \& C	\\
2007if	&	4365.83	&	+14.77	&	Clay	&	LDSS3	\\
2007if	&	4377.69	&	+25.81	&	ESO 3.6m	&	EFOSC2	\\
2007if	&	4381.67	&	+29.51	&	Baade	&	IMACS	\\
2009dc	&	4938.82	&	$-$8.08	&	Clay	&	LDSS3	\\
2009dc	&	4939.79	&	$-$7.13	&	du Pont	&	B \& C	\\
2009dc	&	4943.78	&	$-$3.22	&	du Pont	&	B \& C	\\
2009dc	&	4944.80	&	$-$2.22	&	du Pont	&	B \& C	\\
2009dc	&	4951.75	&	+4.58	&	Clay	&	LDSS3	\\
2009dc	&	4952.80	&	+5.61	&	Clay	&	LDSS3	\\
2009dc	&	4965.69	&	+18.23	&	Baade	&	IMACS	\\
2009dc	&	4974.70	&	+27.05	&	du Pont	&	B \& C	\\
2009dc	&	4982.65	&	+34.83	&	du Pont	&	B \& C	\\
LSQ12gpw	&	6275.44	&	+5.42	&	NOT	&	ALFOSC	\\
2013ao	&	6388.48	&	+8.67	&	NOT	&	ALFOSC	\\
2013ao	&	6388.48	&	+24.87	&	SALT	&	RSS	\\
2013ao	&	6407.50	&	+43.10	&	NOT	&	ALFOSC	\\
CSS140501\tablenotemark{b}	&	6798.64	&	$-$0.08	&	NOT	&	ALFOSC	\\
2015M	&	7158.62	&	$-$10.15	&	NOT	&	ALFOSC	\\
2015M	&	7166.42	&	$-$2.52	    &	NOT	&	ALFOSC	\\
2015M	&	7176.45	&	+7.28	    &	NOT	&	ALFOSC	\\
2015M	&	7180.40	&	+11.14	    &	NOT	&	ALFOSC	\\
2015M	&	7191.50	&	+21.99	    &	NOT	&	ALFOSC	\\
\hline
\multicolumn{5}{c}{\bf NIR}\\ 
\hline
2013ao&6371.14&+8.74&Baade&FIRE\\
2013ao&6376.22&+13.03&Baade&FIRE\\
2013ao&6383.18&+19.66&Baade&FIRE\\
2013ao&6385.28&+21.66&Baade&FIRE\\
2013ao&6431.15&+65.32&Baade&FIRE\\
2015M&7175.52&+6.37&Baade&FIRE\\
\enddata
\tablenotetext{a}{Rest frame days relative to the time of  $B$-band maximum given in Table \ref{table:propphot}.}
\tablenotetext{b}{CSS140501-170414+174839.}
\end{deluxetable*}

\restartappendixnumbering 
\renewcommand{\thefigure}{B\arabic{figure}}
\renewcommand{\theHfigure}{B\arabic{figure}}

\begin{figure*}[htb!]
\centering
 \includegraphics[width=.99\textwidth]{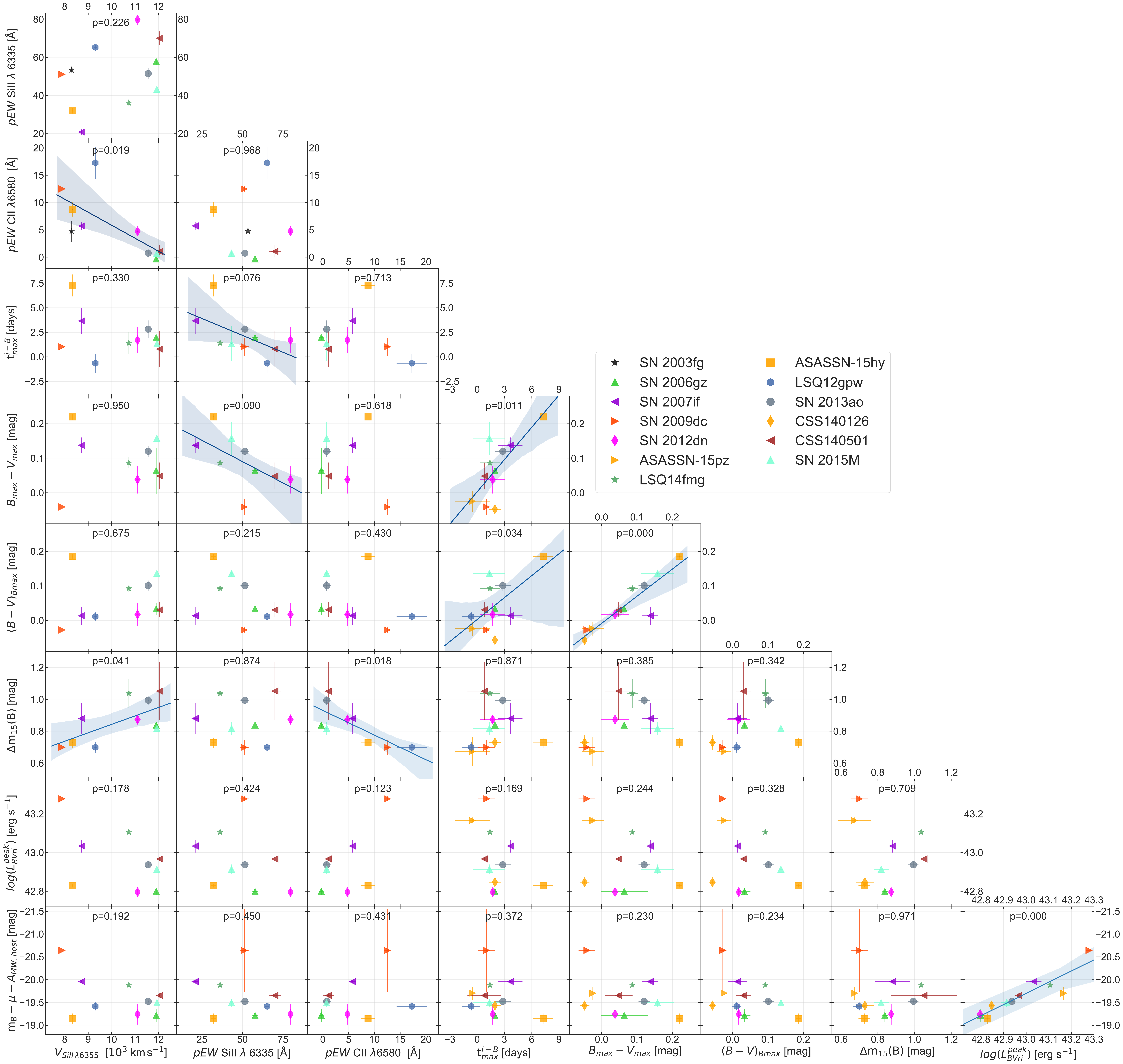}  
\caption{A pair-plot of measured parameters of \SCSN, on top of each panel the p-value is given. Potentially significant correlations with p $\leq$ 0.1 are presented by a best-fit line determined by a least square technique along with 95$\%$ confidence intervals in each panel. }
\label{fig:pairplot_correlations}
\end{figure*}

\end{document}